{}
{}

\documentclass[11pt,a4paper]{article}

\usepackage{graphicx} 
\usepackage{float}
\usepackage{color}
\usepackage[toc,page]{appendix} 
\usepackage{enumerate}
\usepackage{latexsym,amsthm,amssymb,amsfonts,amsbsy,amsmath}
\usepackage{amsmath}
\usepackage{dsfont}
\usepackage{amsfonts}
\usepackage{amssymb}
\usepackage{mathrsfs}
\usepackage{bbm}
\usepackage{mathtools}
\usepackage{cite}

\usepackage{amsthm}
\usepackage{bm} 
\usepackage{bbm} 
\usepackage{slashed} 
\usepackage{cancel}
\usepackage{xcolor}
\usepackage{tikz}
\usetikzlibrary{arrows,matrix,snakes}
\usetikzlibrary{decorations.markings,shapes.geometric,shapes.misc} 
\usepackage{tikz-cd}

\usepackage{pdflscape}
\usepackage{scalefnt}

\definecolor{lightred}{RGB}{255,127,127}
\definecolor{lightgreen}{RGB}{127,255,127}
\definecolor{lightblue}{RGB}{127,127,255}
\definecolor{linkcolor}{rgb}{0,0,0.6}
\usepackage[ pdftex,colorlinks=true,
pdfstartview=FitV,
linkcolor= linkcolor,
citecolor= linkcolor,
urlcolor= linkcolor,
hyperindex=true,
hyperfigures=false]
{hyperref}

\numberwithin{equation}{section}

\usepackage{tabularx} 

\usepackage{fancyhdr} 

\theoremstyle{plain}

\newcommand{\vp}{\varphi}

\newcommand{\R}{\mathbb{R}}

\newcommand{\p}{\partial}
\newcommand{\g}{\mathfrak{g}}
\newcommand{\Ec}{\mathcal{E}}
\newcommand{\Eh}{\widehat{\mathcal{E}}}
\newcommand{\df}{\mathfrak{d}}
\newcommand{\kf}{\mathfrak{k}}
\newcommand{\hf}{\mathfrak{h}}

\newcommand{\ps}[2]{\left\langle#1,#2\right\rangle}
\newcommand{\psb}[2]{\bigl\langle#1,#2\bigr\rangle}
\newcommand{\psB}[2]{\Bigl\langle#1,#2\Bigr\rangle}
\newcommand{\psd}{\langle\cdot,\cdot\rangle}
\newcommand{\pse}{\langle\cdot,\Ec\cdot\rangle}
\newcommand{\s}{\sigma}
\newcommand{\F}[2]{F_{#1}^{{\color{white}#1}#2}}
\newcommand{\Jc}{\mathcal{J}}

\newcommand{\Kc}{\mathcal{K}}
\newcommand{\ti}[1]{_{\bm{\underline{#1}}}}
\newcommand{\Id}{\text{Id}}
\newcommand{\Cd}{\mathsf{C}^{\mathfrak{d}}_{\bm{\underline{12}}}}
\newcommand{\Cg}{\mathsf{C}^{\mathfrak{g}}_{\bm{\underline{12}}}}
\newcommand{\Pc}{\mathcal{P}}
\newcommand{\Hc}{\mathcal{H}}

\newcommand{\dd}{\text{d}}
\newcommand{\Ker}{\text{Ker}}
\newcommand{\Wd}[1]{I_{\text{W}\hspace{-1pt}\text{Z},D}\left[#1\right]}
\newcommand{\Wg}[1]{I_{\text{W}\hspace{-1pt}\text{Z},G}\left[#1\right]}
\newcommand{\ad}{\text{ad}}
\newcommand{\Ad}{\text{Ad}}
\newcommand{\KD}{K\backslash D}
\newcommand{\KDH}{K \backslash D / H}
\newcommand{\Tp}{\null^{\mathsf{T}}}
\newcommand{\Tr}{\text{Tr}}
\newcommand{\cL}{{(\rm L)}}
\newcommand{\cR}{{(\rm R)}}
\newcommand{\cLR}{{(\rm L/R)}}
\newcommand{\hv}{h^{\!\vee}}
\newcommand{\diag}{\text{diag}}

\newcommand{\Oc}{\mathcal{O}}
\newcommand{\Ac}{\mathcal{A}}
\newcommand{\Wc}{\mathcal{W}}
\newcommand{\Pexp}{\text{P}\overleftarrow{\text{exp}}}
\newcommand{\Cdh}{\mathsf{C}^{D\mathfrak{h}}_{\bm{\underline{12}}}}
\newcommand{\Cpm}{\mathsf{C}^{\pm}_{\bm{\underline{12}}}}
\newcommand{\Cp}{\mathsf{C}^{+}_{\bm{\underline{12}}}}
\newcommand{\Cm}{\mathsf{C}^{-}_{\bm{\underline{12}}}}
\newcommand{\ab}{\bar{a}}
\newcommand{\bb}{\bar{b}}
\newcommand{\cb}{\bar{c}}
\newcommand{\Ub}{\overline{U}}
\newcommand{\rhob}{\overline{\rho}}
\newcommand{\Z}{\mathbb{Z}}
\newcommand{\su}{\mathfrak{su}}
\newcommand{\ri}{{\rm i}}
\newcommand{\td}{{\tt t}}
\newcommand{\cD}{{(\rm D)}}
\newcommand{\psib}{{\overline{\psi}\hspace{1pt}}\null}
\newcommand{\slf}{\mathfrak{sl}}
\newcommand{\Lc}{\mathcal{L}}
\newcommand{\C}{\mathbb{C}}
\newcommand{\CP}{\mathbb{CP}^1}

\DeclareFontEncoding{LS1}{}{}
\DeclareFontSubstitution{LS1}{stix}{m}{n}
\DeclareSymbolFont{stixsymbols}{LS1}{stixscr}{m}{n}
\SetSymbolFont{stixsymbols}{bold}{LS1}{stixscr}{b}{n}
\DeclareMathSymbol{\kay}{\mathalpha}{stixsymbols}{"6B}

\definecolor{myGreen}{rgb}{0.0,0.4,0.0}

\makeatletter
\let\@keywords\@empty
\let\@subject\@empty
\providecommand{\keywords}[1]{\gdef\@keywords{#1}}
\providecommand{\subject}[1]{\gdef\@subject{#1}}
\def\thetitle{\@title}
\def\theauthor{\@author}
\def\thesubject{\@subject}
\def\thedate{\@date}
\def\thekeywords{\@keywords}
\makeatother
\AtBeginDocument{
\hypersetup{pdftitle={\thetitle}}
\hypersetup{pdfauthor={\theauthor}}
\hypersetup{pdfsubject={\thesubject}}
\hypersetup{pdfkeywords={\thekeywords}}}

\title{On a class of conformal \texorpdfstring{$\bm\Ec$}{E}-models and\\ their chiral Poisson algebras}

\author{Sylvain Lacroix}

\begin{document}

\begin{center}

\vspace*{2cm}

\begingroup\Large\bfseries\thetitle\par\endgroup

\vspace{1.5cm}

\begingroup
\large Sylvain Lacroix
\endgroup

\vspace{1cm}

\begingroup
\textit{Institute for Theoretical Studies, ETH Z\"urich \\ Clausiusstrasse 47, 8092 Z\"urich, Switzerland} \vspace{11pt} \\
\href{mailto:sylvain.lacroix@eth-its.ethz.ch}{\texttt{sylvain.lacroix@eth-its.ethz.ch}}

\endgroup

\end{center}

\vspace{2cm}

\begin{abstract}
In this paper, we study conformal points among the class of $\Ec$-models. The latter are $\s$-models formulated in terms of a current Poisson algebra, whose Lie-theoretic definition allows for a purely algebraic description of their dynamics and their 1-loop RG-flow. We use these results to formulate a simple algebraic condition on the defining data of such a model which ensures its 1-loop conformal invariance and the decoupling of its observables into two chiral Poisson algebras, describing the classical left- and right-moving fields of the theory. In the case of so-called non-degenerate $\Ec$-models, these chiral sectors form two current algebras and the model takes the form of a WZW theory once realised as a $\s$-model. The case of degenerate $\Ec$-models, in which a subalgebra of the current algebra is gauged, is more involved: the conformal condition yields a wider class of theories, which includes gauged WZW models but also other examples, seemingly different, which however sometimes turn out to be related to gauged WZW models based on other Lie algebras. For this class, we build non-local chiral fields of parafermionic-type as well as higher-spin local ones, forming classical $\Wc$-algebras. In particular, we find an explicit and efficient algorithm to build these local chiral fields. These results (and their potential generalisations discussed at the end of the paper) open the way for the quantisation of a large class of conformal $\Ec$-models using the standard operator formalism of two-dimensional CFT.
\end{abstract}

\newpage

\setcounter{tocdepth}{3}
\tableofcontents

\newpage

\section{Introduction}

$\s$-models form an important class of two-dimensional field theories, with applications in various domains, including high-energy physics, string theory and condensed matter physics. Among this class, a particular role is played by conformal $\s$-models, for which various exact results can be derived, exploiting the powerful constraints imposed by the conformal symmetry~\cite{Belavin:1984vu}. Given a classical $\s$-model characterised by the choice of its target space manifold, equipped with a pseudo-Riemannian metric and a torsion tensor, a natural starting point for the study of its quantisation and its potential conformal invariance is to look at its Renormalisation Group (RG) flow. The latter can be studied through a perturbative expansion in the Planck constant and translates to a geometric flow of the metric and torsion~\cite{Friedan:1980jf,Fradkin:1985ys,Callan:1985ia}. In this language, a conformal $\s$-model then corresponds to a fixed-point of the RG-flow, encoding the quantum scale invariance of the theory.

The next step towards the quantisation of such a model in the language of two-dimensional CFT is the identification of its algebras of extended conformal symmetry. These structures can already be traced back at the classical level, where they take the form of two decoupled chiral sectors in the dynamics of the theory. To be more explicit, let $(t,x)$ be the space-time coordinates of the $\s$-model and $\Ac$ be its algebra of classical observables, equipped with the canonical Poisson bracket. The classical algebras of extended conformal symmetry will then correspond to two subalgebras $\Ac^\cL$ and $\Ac^\cR$ in $\Ac$, formed respectively by local left-moving fields and local right-moving fields, \textit{i.e.} fields whose dynamics depends only on the light-cone coordinates $t+x$ or $t-x$.\footnote{Here and throughout the article, we consider relativistic $\s$-models based on a Lorentzian worldsheet. In the Euclidean case, the left- and right-moving degrees of freedom would be replaced by holomorphic and anti-holomorphic fields.} These subalgebras are closed under Poisson brackets and mutually Poisson-commuting. They contain at least the chiral components of the energy-momentum tensor, generating the classical Virasoro symmetry of the model, but often consists of a larger set of fields, hence the name of ``extended'' conformal symmetry. Typical examples of such structures are affine Lie algebras (appearing in Wess-Zumino-Witten models~\cite{Witten:1983ar,Witten:1983tw}) and higher-spin $\Wc$-algebras (initially introduced in~\cite{Zamolodchikov:1985wn}). In addition to the local chiral fields in $\Ac^\cL$ and $\Ac^\cR$, we note that the theory can also possess non-local ones, which are slightly more complicated to manipulate but still play a useful role in the study of the model. These non-local chiral degrees of freedom can for instance take the form of parafermionic fields~\cite{Lepowsky:1984,Fateev:1985mm,Ninomiya:1986dp,Gepner:1987sm,Karabali:1989dk,Bardakci:1990lbc,Bardakci:1990ad}, which appear in gauged Wess-Zumino-Witten models and which will be relevant for the present paper.

The setup described in the previous paragraph paves the way for an exact operator quantisation of the theory. Indeed, in many cases, the chiral Poisson algebras $\Ac^\cL$ and $\Ac^\cR$ can be quantised into so-called Vertex Operator Algebras, formed by non-commutative chiral fields and describing the extended conformal symmetry of the quantum model. This framework provides a powerful organising principle for the systematic analysis of the model and the exact derivation of some of its quantum properties (for instance, the Hilbert space describing the states of the model can be studied using the representation theory of these operator algebras). \\

The main goal of this paper is to initiate such a ``conformal quantisation programme'' for a class of theories called \textit{conformal $\Ec$-models}. The concept of $\Ec$-model (not necessarily conformal) was initially introduced by Klim\v{c}\'{i}k and \v{S}evera in~\cite{Klimcik:1995ux,Klimcik:1995dy,Klimcik:1996nq,Klimcik:1996np}. It essentially describes $\s$-models whose Hamiltonian degrees of freedom can be gathered into a Lie-algebra valued field satisfying a current Poisson bracket and whose dynamics is governed by a Hamiltonian quadratic in this current. In particular, this formalism makes manifest certain properties of duality of these $\s$-models (more precisely the so-called Poisson-Lie T-duality), which represented the main motivation for their introduction. In this paper, we will argue that this Lie-theoretic formulation also provides a powerful tool to identify conformal occurrences among the class of $\Ec$-models and to describe their chiral algebras at the classical level, thus opening the way for a quantisation of these theories in the standard operator formalism of two-dimensional CFT. To keep the length of the paper contained, we will not address these quantum aspects here: we hope to return to these questions in future works.\\

There exist two different types of $\Ec$-models, called non-degenerate and degenerate. Although we will ultimately be interested in the degenerate ones, which give rise to a richer class of conformal theories, let us first consider the non-degenerate case, which is simpler to study. A non-degenerate $\Ec$-model~\cite{Klimcik:1995ux,Klimcik:1995dy,Klimcik:1996nq} is more easily described using the Hamiltonian formalism. In this context, the main protagonist of the theory is a field $\Jc(t,x)$, valued in a certain Lie algebra $\df$. The latter is referred to as the double algebra and is equipped with a split ad-invariant symmetric bilinear form $\psd$. The main property satisfied by this field is its Poisson bracket, taking the form of a so-called Poisson current algebra. To describe it, we decompose the $\df$-valued field $\Jc(t,x)$ into components $\lbrace \Jc_A(t,x) \rbrace_{A=1}^{\dim\df}$, corresponding to a basis $\lbrace T_A \rbrace_{A=1}^{\dim\df}$ of $\df$. The current Poisson algebra obeyed by $\Jc_A(t,x)$ then reads
\begin{equation}\label{Eq:IntroPbJ}
\bigl\lbrace \Jc_A(t,x), \Jc_B(t,y) \bigr\rbrace = \F{AB}C\,\Jc_C(t,x)\,\delta(x-y) - \eta_{AB}\,\p_x\delta(x-y)\,,
\end{equation}
where $\delta(x-y)$ is the Dirac-distribution, $\F{AB}C$ are the structure constants of the Lie algebra $\df$ and $\eta_{AB}=\ps{T_A}{T_B}$ are the entries of the bilinear form $\psd$ in the basis $\lbrace T_A \rbrace_{A=1}^{\dim\df}$. The dynamics $\p_t=\lbrace \Hc,\cdot\rbrace$ of the theory is defined by the choice of a quadratic local Hamiltonian\vspace{-2pt}
\begin{equation}\label{Eq:IntroH}
\Hc = \frac{1}{2} \int\, \psb{\,\Jc(t,x)\,}{\,\Ec\bigl(\Jc(t,x)\bigr)\,} \, \dd x\,, \vspace{-2pt}
\end{equation}
characterised by the choice of a symmetric operator $\Ec : \df \to \df$, giving its name to the $\Ec$-model. To ensure the relativistic invariance of the theory, this operator is required to be an involution, \textit{i.e.} to satisfy $\Ec^2=\Id$. The choice of $\Ec$ is then equivalent to the data of its eigenspaces decomposition $\df = V_+ \oplus V_-$, with $V_\pm=\Ker(\Ec \mp \Id)$. The eigenspaces $V_+$ and $V_-$ are naturally associated with the two light-cone directions $t+x$ and $t-x$ of the $\s$-model worldsheet: this decomposition is thus particularly suited to describe the chiral properties of non-degenerate $\Ec$-models. Moreover, it turns out to also allow a remarkably simple and purely Lie-algebraic formulation~\cite{Valent:2009nv,Sfetsos:2009dj,Avramis:2009xi,Sfetsos:2009vt} of the 1-loop RG-flow of these theories, in terms of the operator $\Ec$ or equivalently of the eigenspaces $V_\pm$.

The approach that we wish to advertise in the present paper is to use this algebraic formulation for the identification and the study of conformal $\Ec$-models and their operator quantisation. We illustrate this idea by considering the case where the eigenspaces $V_\pm$ are themselves subalgebras of $\df$, \textit{i.e.} are closed under the Lie bracket. Such a condition was previously considered in the reference~\cite{Klimcik:1996hp}: here, we will refer to it as the \textit{strong conformal condition}. We will show that it ensures the following properties:\vspace{-3pt}
\begin{enumerate}[(i)]\setlength\itemsep{0.2pt}
\item the 1-loop RG-flow~\cite{Valent:2009nv,Sfetsos:2009dj,Avramis:2009xi,Sfetsos:2009vt} of the model is trivial, \textit{i.e.} we have 1-loop conformal invariance ;
\item the dynamics of the theory is such that the field $\Jc$ decomposes into a left-moving component valued in $V_+$ and a right-moving component valued in $V_-$ ;
\item these chiral components satisfy two decoupled current Poisson algebras.
\end{enumerate}
Through this condition, we thus succeeded in identifying conformal non-degenerate $\Ec$-models and describing their chiral Poisson algebras. This is reminiscent of the structure of the Wess-Zumino-Witten (WZW) conformal field theory~\cite{Witten:1983ar,Witten:1983tw}, whose chiral fields are famously described by two current algebras. In fact, it was argued in~\cite{Klimcik:1996hp} that non-degenerate $\Ec$-models satisfying the strong conformal condition essentially take the form of WZW theories once realised as $\s$-models in the Lagrangian formulation. This condition thus lead to a rather limited class of conformal theories and is presented in this paper mainly as a proof of concept and as a warm-up for more general investigations. Indeed, we expect that it is in fact not the most general condition ensuring the conformal invariance of the model (which is why we call it ``strong''): we will mention some preliminary results and perspectives on these more general cases in the concluding section \ref{Sec:Conclusion} and hope to return to a more thorough analysis of this question in the future. In the present article, we mostly explore another direction for generalisations, in the context of degenerate $\Ec$-models, which has been less explored. As we will see, this leads to a seemingly wider class of conformal theories and to richer structures in the description of their chiral sectors.\\

Degenerate $\Ec$-models, sometimes also referred to as dressing cosets in the literature, were introduced by Klim\v{c}\'{i}k and \v{S}evera in the article~\cite{Klimcik:1996np} (see also~\cite{Sfetsos:1999zm,Squellari:2011dg} for related constructions) -- here, we will mainly follow the presentation of the more recent works~\cite{Klimcik:2019kkf,Klimcik:2021bjy,Klimcik:2021bqm}. Similarly to the non-degenerate $\Ec$-models, they are defined in terms of a $\df$-valued current $\Jc(t,x)$, satisfying the Poisson bracket \eqref{Eq:IntroPbJ}. The main difference in the degenerate case is that the components of $\Jc$ are not all physical fields of the model: indeed, the theory possesses a gauge symmetry, eliminating some of these degrees of freedom. This symmetry is encoded in the choice of a gauge group $H$, whose Lie algebra is a subalgebra $\hf$ of $\df$, which is required to be isotropic with respect to the bilinear form $\psd$. The physical observables of the degenerate $\Ec$-model are then obtained from the Hamiltonian reduction of the current algebra of $\Jc$ with respect to the subalgebra corresponding to $\hf$ (as explained in the main text, this is done through the imposition of a first-class constraint, which generates the $H$-gauge symmetry).

The Hamiltonian of the theory also takes a quadratic form \eqref{Eq:IntroH}, but with $\Ec$ replaced by a ``degenerate'' operator $\Eh : \df \to \df$, satisfying slightly different properties. In particular, one requires that $\Eh(\hf) = \lbrace 0 \rbrace$ (\textit{i.e.} $\Eh$ is degenerate and its kernel contains $\hf$) and that $\Eh$ commutes with the adjoint action of $H$ on $\df$: these properties ensure that the Hamiltonian is invariant under the $H$-gauge symmetry of the model, as expected. Together with the relativistic invariance of the theory, this setup induces a natural decomposition $\df=\hf \oplus \hf' \oplus V_+ \oplus V_-$ of the double algebra. Here, similarly to the non-degenerate case, $V_\pm = \Ker(\Eh \mp \Id)$ are the eigenspaces of $\Eh$ with eigenvalues $\pm 1$, which are naturally associated with the light-cone directions $t\pm x$ of the model. The difference comes with the additional presence, in the decomposition $\df=\hf \oplus \hf' \oplus V_+ \oplus V_-$, of the gauge subalgebra $\hf=\Ker(\Eh)$ and the space $\hf'$, which can be interpreted as the dual space of $\hf$. Although these additional features make the degenerate $\Ec$-models slightly more complicated than their non-degenerate cousins, this Lie-theoretic formulation still allows a simple characterisation of their dynamics and their 1-loop RG-flow~\cite{Severa:2018pag}.

The main subject of the paper is to study the characteristics of these degenerate $\Ec$-models when one imposes what we call the \textit{strong conformal condition}, namely that $\df_\pm = V_\pm \oplus \hf \oplus \hf'$ are subalgebras of $\df$. In the absence of a gauge symmetry, \textit{i.e.} when $\hf$ and $\hf'$ are trivial, this condition reduces to the one considered above for non-degenerate models. In the main text of this paper, we will prove that it ensures the 1-loop conformal invariance of the theory and the presence of chiral fields in its classical dynamics. More precisely, we will find that the components of the current $\Jc$ in the subspaces $V_\pm$ are not chiral, as in the non-degenerate case, but in fact are chiral ``up to a $H$-gauge transformation''. We will then obtain a $V_+$--valued left-moving field and a $V_-$--valued right-moving field by dressing these components with the adjoint action of a well-chosen non-local field in $H$. We thus get non-local gauge-invariant chiral degrees of freedom, which take the form of so-called parafermionic fields~\cite{Lepowsky:1984,Fateev:1985mm,Ninomiya:1986dp,Gepner:1987sm,Karabali:1989dk,Bardakci:1990lbc,Bardakci:1990ad}. In particular, we show that these fields satisfy two decoupled parafermionic Poisson algebras, which can be interpreted as the Hamiltonian reduction of the $\df_\pm$--current algebras with respect to the $H$-gauge symmetry. In addition to these non-local quantities, we also construct local chiral fields, by considering well-chosen gauge-invariant combinations of the current components (alternatively, these can be thought of as particular combinations of the chiral parafermions which turn out to be local quantities). These fields form the classical algebras of extended conformal symmetry of the model, which we show take the form of higher-spin $\Wc$-algebras. Moreover, we propose an efficient algorithm for the systematic construction of these local chiral fields. Finally, we show that under additional technical assumptions on the decomposition $\df=\hf \oplus \hf' \oplus V_+ \oplus V_-$, the conformal $\s$-models built from the above formalism are realisations of the celebrated Goddard-Kent-Olive (GKO) construction of coset CFTs~\cite{Goddard:1984vk,Goddard:1986ee}.\\

Another part of the article is dedicated to the study of explicit examples of degenerate $\Ec$-models satisfying this strong conformal condition. As one can expect, a prototypical familly of such theories is given by the gauged Wess-Zumino-Witten models, which are $\s$-models whose target space is a Lie group $G$ quotiented by the adjoint action of a certain subgroup $H_0$. These theories were reinterpreted as degenerate $\Ec$-models in~\cite{Klimcik:1999ax}: here, we show that they satisfy the strong conformal condition and thus fit into the above framework (for any choice of $G$ and $H_0$). We further provide a thorough analysis of the $SU(2)/U(1)$ case, for which we give a very explicit description of the parafermionic fields and the first few local fields in the $\Wc$-algebra. We recover this way various results established in the literature and illustrate on a simple example the general construction sketched above.

A natural question is whether this construction also goes beyond the class of gauged WZW theories. As a second example, we build a conformal $\Ec$-model based on a solvable Lie group, leading to a $\s$-model with a simple two-dimensional target space and which in particular does not fit into the GKO scheme. To the the best of our knowledge, this target space is not obtainable directly from a gauged WZW theory. Finally, we consider the conformal point of the Klim\v{c}\'{i}k (or Bi-Yang-Baxter) model: this is a $\s$-model with target space a Lie group $G$~\cite{Klimcik:2008eq,Klimcik:2014bta}, whose conformal limit was studied recently in~\cite{Fateev:1996ea,Bazhanov:2018xzh,Kotousov:2022azm}\footnote{These works were parts of the motivation for the present article. In particular, the construction of the $\Wc$-algebra for conformal $\Ec$-models presented here can be seen as a generalisation of a result obtained in~\cite{Kotousov:2022azm} for the Klim\v{c}\'{i}k model.}, in particular for $G=SU(2)$. Here, we will focus on the analogous case $G=SL(2)$, to avoid certain issues with reality conditions. This example will allow us to raise some interesting open questions, subtleties and potential limitations of the construction. We also consider a one-parameter deformation of this theory, which defines a more complicated but still conformal $\s$-model. Surprisingly, although this conformal theory initially takes a form which seems quite distinct from that of a gauged WZW model, the target space we obtain is in the end related to an axially gauged WZW theory on $(SL(2,\R)\times \R) / \R$~\cite{Witten:1991yr,Horne:1991gn}. In particular, the gauge algebra of this WZW theory is different from the one initially considered to obtain the Klim\v{c}\'{i}k model. It is not clear to us whether this is an accidental low-dimensional isomorphism or the sign of a deeper connection.\\

The motivation behind this paper is two-folds. On the one-side, we want to argue that the formalism of $\Ec$-models allows for a natural and systematic construction of a large class of conformal $\s$-models and the study of their properties, mainly based on Lie-theoretic considerations. On the other hand, we hope that this approach will also be beneficial for the study of the $\Ec$-models themselves and their applications. For instance, these models play a growing role in the investigation of dualities, symmetries and integrable structures in $\s$-models, exploiting their underlying algebraic formulation. These aspects have been extensively studied at the classical level, where they are now well understood. In contrast, these questions are still relatively unexplored at the quantum level (except for instance in relation to perturbative RG-flow approaches). We hope that the powerful techniques of two-dimensional CFT and the rigorous and well-controlled framework it provides will offer a natural and useful entry point for the investigation of these subjects. We will come back on these perspectives in the concluding section \ref{Sec:Conclusion} (and refer to this part for further explanations and references).\\

The plan of the article is the following. Section \ref{Sec:ND} is devoted to non-degenerate $\Ec$-models. We start with a review of their construction and properties in Subsection \ref{SubSec:EModelsN}. We then introduce the strong conformal condition in Subsection \ref{SubSec:ConfN} and show that it implies the 1-loop conformal invariance of these models and the decomposition of their current algebra into two decoupled chiral subalgebras. Finally, we explain the relation between this construction and WZW models in Subsection \ref{SubSec:WZW}.

The main body of the paper is the Section \ref{Sec:D}, which deals with degenerate $\Ec$-models and their conformal occurrences. We first review the general theory of degenerate $\Ec$-models in Subsection \ref{SubSec:EModels} and later study the strong conformal condition and its consequences in Subsection \ref{SubSec:Conf}. More precisely, we prove that it implies the 1-loop conformal invariance in Subsubsection \ref{SubSec:Beta}. We then study the classical non-local chiral fields of these theories and their parafermionic Poisson algebras in Subsubsection \ref{SubSec:Para}, as well as their local chiral fields, forming higher-spin $\Wc$-algebras, in Subsubsection \ref{SubSec:W}. Finally, we explain the relation with the GKO coset construction in Subsubsection \ref{SubSec:GKO}.

Section \ref{Sec:Examples} concerns the study of explicit examples of degenerate $\Ec$-models satisfying the strong conformal condition. We start with general gauged WZW models in Subsection \ref{SubSec:gWZW} and further specialise to the $SU(2)/U(1)$ case, which we explore in detail in Subsection \ref{SubSec:Su2u1}. We then study an example based on a solvable Lie group in Subsection \ref{SubSec:Diam} and finish with the conformal point of the $SL(2)$ Klim\v{c}\'{i}k model in Subsection \ref{SubSec:BYB}.

We conclude and discuss various perspectives in Section \ref{Sec:Conclusion}. Finally, some technical results and computations are gathered in Appendix \ref{App:Dirac}.

\section{Conformal non-degenerate \texorpdfstring{$\bm\Ec$}{E}-models}
\label{Sec:ND}

\subsection[Reminder about non-degenerate \texorpdfstring{$\Ec$}{E}-models]{Reminder about non-degenerate \texorpdfstring{$\bm\Ec$}{E}-models}
\label{SubSec:EModelsN}

The goal of this subsection is to recall the definition and main properties of non-degenerate $\Ec$-models, which form a class of two-dimensional Hamiltonian field theories closely related to $\s$-models, introduced by Klim\v{c}\'{i}k and \v{S}evera in~\cite{Klimcik:1995ux,Klimcik:1995dy,Klimcik:1996nq}. The first half of the subsection concerns the formulation of these models in terms of current Poisson algebras, while the second half reviews their concrete realisations as $\s$-models.

\subsubsection{Hamiltonian formulation in terms of a current Poisson algebra}
\label{SubSec:HamN}

\paragraph{The double Lie algebra.} One of the key ingredient defining a non-degenerate $\Ec$-model is the \textit{double algebra} $\df$, which is a real Lie algebra of even dimension $2d$, with Lie bracket $[\cdot,\cdot]$ and a non-degenerate ad-invariant split symmetric bilinear form $\psd$. The ad-invariance property is the requirement that
\begin{equation}
\ps{[X,Y]}{Z} = \ps{X}{[Y,Z]}\,, \qquad \forall\,X,Y,Z\,\in\,\df\,,
\end{equation}
whereas the splitness condition means that $\psd$ has signature $(d,d)$.

Throughout this paper, $\lbrace T_A \rbrace_{A=1}^{2d}$ will denote a basis of $\df$, while $\F{AB}C$ and $\eta_{AB}$ will denote the structure constants and bilinear form entries in this basis, defined by
\begin{equation}
\bigl[ T_A, T_B \bigr] =\F{AB}C \, T_C \qquad \text{ and } \qquad \ps{T_A}{T_B} = \eta_{AB}\,,
\end{equation}
where we used the Einstein convention of summation over repeated indices in the first equation. Since $\psd$ is non-degenerate, the matrix $\bigl(\eta_{AB}\bigr)_{A,B=1}^{2d}$ is invertible: we will denote its inverse by $\bigl(\eta^{AB}\bigr)_{A,B=1}^{2d}$. From now on, we will use $\eta_{AB}$ and $\eta^{AB}$ to lower and raise indices in $\df$. In particular, we introduce the dual basis
\begin{equation}
T^A = \eta^{AB}\,T_B\,,
\end{equation}
which then satisfies $\ps{T^A}{T_B}=\delta^A_{\;\,B}$.

\paragraph{The current Poisson algebra.} Another crucial ingredient in the definition of the non-degenerate $\Ec$-model is the \textit{current} $\Jc(x)$. It is a field valued in the double algebra $\df$, which depends on the space variable $x$ of the two-dimensional Hamiltonian theory under construction. The main property of this field is its Poisson bracket, which takes the form of a \textit{current Poisson algebra}. To describe it, let us first decompose $\Jc(x) = \Jc_A(x)\,T^A$ along the dual basis $\lbrace T^A \rbrace_{A=1}^{2d}$ introduced in the previous paragraph. The Poisson bracket of the components $\Jc_A$ then reads
\begin{equation}\label{Eq:PbJA}
\bigl\lbrace \Jc_A(x), \Jc_B(y) \bigr\rbrace = \F{AB}C\,\Jc_C(x)\,\delta(x-y) - \eta_{AB}\,\p_x\delta(x-y)\,,
\end{equation}
where $\delta(x-y)$ denotes the Dirac distribution. Although we have written the bracket of $\Jc$ in terms of its components $\Jc_A$ for explicitness, let us note that this bracket can be rewritten in a manifestly basis-independent way. For that, we will use the standard tensorial notations: for $X \in \df$, we introduce the elements $X\ti{1}=X \otimes \Id$ and $X\ti{2}=\Id\otimes X$ of the tensor product algebra $U(\df) \otimes U(\df)$. Moreover, we define
\begin{equation}
\bigl\lbrace \Jc(x)\ti{1}, \Jc(y)\ti{2} \bigr\rbrace = \bigl\lbrace \Jc_A(x), \Jc_B(y) \bigr\rbrace\,T^A \otimes T^B\,,
\end{equation}
which is valued in $U(\df) \otimes U(\df)$ and encodes the Poisson brackets of all the components of $\Jc$ into one unique object. The equation \eqref{Eq:PbJA} can then be rewritten in the following basis-independent way:
\begin{equation}\label{Eq:PbJ}
\bigl\lbrace \Jc(x)\ti{1}, \Jc(y)\ti{2} \bigr\rbrace  = \bigl[ \Cd, \Jc(x)\ti{1} \bigr] \,\delta(x-y) - \Cd\,\p_x\delta(x-y)\,,
\end{equation}
where $\Cd \in\df\otimes\df$ is the split quadratic Casimir of the algebra $\df$, defined as
\begin{equation}\label{Eq:Cas}
\Cd = \eta_{AB} \; T^A \otimes T^B\,.
\end{equation}
The skew-symmetry of the bracket \eqref{Eq:PbJ} is ensured by the identity
\begin{equation}
\bigl[ \Cd, X\ti{1} + X\ti{2} \bigr] = 0\, , \qquad \forall\,X\in\df\,,
\end{equation}
which follows from the ad-invariance of $\df$. Let us also note that the Casimir satisfies the following additional property:
\begin{equation}\label{Eq:CasId}
\ps{\Cd}{X\ti{1}}\ti{1} = X\,, \qquad \forall \, X\in\df\,,
\end{equation} 
which follows from the definition \eqref{Eq:Cas} of $\Cd$ in terms of the bilinear form.\\

Let us end this paragraph with a brief comment. Above, we have introduced the current $\Jc(x)$ as a field depending on the space variable $x$ of the theory. In the following, we will see $x$ as belonging to the circle $\mathbb{S}^1 \simeq \R / 2\pi \mathbb{Z}$ and will require the field $\Jc(x)$ to satisfy the periodic boundary condition $\Jc(x+2\pi) = \Jc(x)$. We note however that most of the results discussed in this paper are independent of this choice: one could formally replace the circle by the real line $\mathbb{R}=(-\infty,+\infty)$ and define a meaningful consistent $\Ec$-model in this setup (typically, one would then require the field $\Jc(x)$ to decrease sufficiently fast at spatial infinity).

\paragraph{The momentum.} Let us introduce the quantity
\begin{equation}\label{Eq:P}
\Pc =  \frac{1}{2} \int_0^{2\pi} \ps{\Jc(x)}{\Jc(x)}\, \dd x = \frac{1}{2}  \int_0^{2\pi} \eta^{AB} \, \Jc_A(x)\,\Jc_B(x)\,\dd x\,.
\end{equation}
Starting from the Poisson bracket \eqref{Eq:PbJ} and using the ad-invariance of $\psd$ and the identity \eqref{Eq:CasId}, one easily checks that
\begin{equation}
\bigl\lbrace \Pc, \Jc(x) \bigr\rbrace = \p_x\,\Jc(x)\,.
\end{equation}
The Poisson bracket of the observable $\Pc$ therefore generates the spatial derivative on the current $\Jc(x)$. We thus identify $\Pc$ as the momentum of the $\Ec$-model.

\paragraph{The Hamiltonian and $\bm\Ec$-operator.} The final ingredient necessary to define the non-degenerate $\Ec$-model is the choice of a Hamiltonian $\Hc$, which determines the dynamics of any observable $\Oc$ of the theory by the time-evolution equation
\begin{equation}
\p_t \Oc = \lbrace \Hc, \Oc \rbrace\,.
\end{equation}
In analogy with the momentum \eqref{Eq:P}, we will build this Hamiltonian as a quadratic combination of the current components:
\begin{equation}\label{Eq:Hn}
\Hc =  \frac{1}{2} \int_0^{2\pi} \bigl\langle{\Jc(x)},{\Ec(\Jc(x))}\bigr\rangle\, \dd x = \frac{1}{2}  \int_0^{2\pi} \Ec^{AB} \, \Jc_A(x)\,\Jc_B(x)\,\dd x\,.
\end{equation}
Here, $\Ec : \df \to \df$ denotes a linear operator on the double algebra $\df$ and $\Ec^{AB} = \ps{T^A}{\Ec\bigl(T^B\bigr)}$. Without loss of generality, we can always take $\Ec$ to be symmetric with respect to the bilinear form $\psd$ (or equivalently require that $\Ec^{AB}=\Ec^{BA}$): under this condition, the operator $\Ec$ is uniquely determined.\\

As we have now defined the dynamics $\p_t$ of the theory, we can investigate its space-time symmetries. It is clear by construction that the model is invariant under space-time translations of the coordinates $(t,x)$, with the Hamiltonian $\Hc$ and the momentum $\Pc$ being the associated conserved generators (which Poisson-commute). A natural further requirement is to ask that the resulting two-dimensional theory is relativistic, \textit{i.e.} invariant under infinitesimal Lorentz boosts of $(t,x)$. This is not the case for a generic Hamiltonian of the form \eqref{Eq:Hn} but in fact translates into a quite simple condition on the operator $\Ec$, namely that it is an involution, \textit{i.e.}
\begin{equation}
\Ec^2 = \Id\,.
\end{equation}
In the rest of this section, we will restrict ourselves to the study of relativistic non-degenerate $\Ec$-models and will thus assume $\Ec^2 = \Id$. Note in particular that this implies the orthogonal decomposition
\begin{equation}\label{Eq:SumV}
\df = V_+ \overset{\perp}{\oplus} V_-, \qquad \text{ where } \qquad V_\pm = \Ker(\Ec \mp \Id)
\end{equation}
are the eigenspaces of $\Ec$. From now on we will suppose that $V_+$ and $V_-$ have the same dimension, which is then equal to $\dim V_\pm = \frac{1}{2} \dim\df = d$. We also introduce
\begin{equation}
\pi_\pm = \frac{1}{2}  \bigl( \Id \pm \Ec \bigr) \;\; : \;\; \df \longrightarrow V_\pm\,,
\end{equation}
which are the orthogonal projectors corresponding to the decomposition $\df=V_+ \oplus V_-$.

Let us finally note that the operator $\Ec:\df\to\df$ is often required in the literature to be such that the symmetric bilinear form $\ps{\cdot}{\Ec\cdot}$ on $\df$ is definite positive, in contrast with the initial bilinear form $\psd$ which has split signature. This simply ensures that the Hamiltonian \eqref{Eq:Hn} takes positive values on all configurations of the field $\Jc(x)$ and thus is bounded from below. Although this is a natural requirement, we will in general not impose this additional condition as it will not be necessary from a technical point of view and as it will for instance allow us to also include $\s$-models with non-euclidian target spaces in the construction. Let us finally note for completeness that in terms of the eigenspaces $V_\pm$ of $\Ec$ introduced above, this positivity condition is equivalent to $\psd$ being definite positive when restricted to $V_+$ and definite negative when restricted to $V_-$.

\paragraph{Equations of motion.} Having defined the Hamiltonian $\Hc$ of the theory, let us now describe its dynamics by writing down the equations of motion of the current $\Jc$. This is easily done by computing the Poisson bracket of $\Hc$ with $\Jc$, using the current algebra \eqref{Eq:PbJ} and the identity \eqref{Eq:CasId}. In the end, one finds
\begin{equation}\label{Eq:EoMn}
\p_t \Jc = \p_x \Ec(\Jc) + \bigl[ \Jc, \Ec(\Jc) \bigr]\,.
\end{equation}
Note that this dynamics takes the form of the zero curvature equation
\begin{equation}\label{Eq:ZceN}
\bigl[ \p_x + \Jc, \p_t + \Ec(\Jc) \bigr] = 0\,
\end{equation}
for a $2$-dimensional connection with spatial component $\Jc$ and temporal component $\Ec(\Jc)$.\\

It will be important for later investigations to rewrite the equations of motion of the theory using light-cone coordinates $x^\pm$ and their derivatives $\p_\pm$, defined by
\begin{equation}
x^\pm = t\pm x \qquad \text{ and } \qquad \p_\pm = \frac{1}{2} (\p_t \pm \p_x)\,.
\end{equation}
Let us also recall the direct sum \eqref{Eq:SumV} of the double algebra in terms of the eigenspaces of $\Ec$ and the corresponding projectors $\pi_\pm : \df \to V_\pm$. We then decompose the current as
\begin{equation}
\Jc = \Jc_+ - \Jc_-, \qquad \text{ with } \qquad \Jc_\pm  = \pm \pi_\pm \Jc \in V_\pm\,.
\end{equation}
The minus sign in $\Jc_-$ has been introduced so that the equation \eqref{Eq:ZceN} becomes $[\p_++\Jc_+,\p_-+\Jc_-]=0$. Using the fact that $\Jc_\pm\in V_\pm$, one can separate this equation of motion into two independent parts:
\begin{subequations}\label{Eq:EoMnJpm}
\begin{align}
&\p_- \Jc_+ = + \pi_+ \bigl[ \Jc_+, \Jc_- \bigr]\,, \\
&\p_+ \Jc_- = - \pi_- \bigl[ \Jc_+, \Jc_- \bigr]\,.
\end{align}
\end{subequations}

\paragraph{Light-cone Hamiltonian.} For completeness, let us finally point out that the light-cone Hamiltonians $\Pc_\pm = \frac{1}{2} (\Hc \pm \Pc)$, which generate the light-cone derivatives $\p_\pm = \lbrace \Pc_\pm, \cdot \rbrace$, take the simple form
\begin{equation}\label{Eq:Ppm}
\Pc_\pm = \pm \frac{1}{2} \int_0^{2\pi} \dd x \; \ps{\Jc_\pm(x)}{\Jc_\pm(x)}\,
\end{equation}
in terms of the $V_\pm$-valued components $\Jc_\pm$ of the current $\Jc$.

\subsubsection[Lagrangian formulation and \texorpdfstring{$\s$}{sigma}-models]{Lagrangian formulation and \texorpdfstring{$\bm \s$}{sigma}-models}\label{SubSec:LagN}

\paragraph{Isotropic subalgebra.} So far, we have seen the non-degenerate $\Ec$-model as a Hamiltonian field theory defined in terms of the $\df$-valued current $\Jc(x)$. Importantly, there exists a closely related theory, taking the form of a $\s$-model, that shares the same first order equations of motion and Poisson structure when written in terms of this current. In addition to the data $(\df,\psd,\Ec)$ used above, this $\s$-model depends on the choice of a maximally isotropic Lie subalgebra $\kf$ of $\df$. Concretely, this amounts to choosing a subspace $\kf \subset \df$ such that
\begin{equation}
[X,Y] \in \kf \qquad \text{ and } \qquad \ps{X}{Y} = 0\,, \qquad \forall\, X,Y\in \kf\,,
\end{equation}
and, corresponding to the maximality property,
\begin{equation}
\dim\,\kf = \frac{1}{2} \dim\,\df = d\,.
\end{equation}

\paragraph{$\bm\s$-model action.} To describe the $\s$-model corresponding to this data, we introduce the double group $D$, defined as a connected Lie group with Lie algebra $\text{Lie}(D)=\df$, and its subgroup $K \subset D$ with $\text{Lie}(K)=\kf$. As we will see, the target space of the $\s$-model will then be the left quotient $\KD$. Rather than describing the theory directly in terms of a field valued in this target space, it will in fact be more natural to use a formulation in terms of a $D$-valued field possessing a local gauge symmetry by the subgroup $K$, which effectively reduces the physical degrees of freedom of the model to the quotient $\KD$. We will denote this $D$-valued field as $\ell(t,x)$, depending on the space-time coordinates $(t,x)\in\R\times\mathbb{S}^1$. In these terms, the action of the theory then takes the form
\begin{equation}\label{Eq:ActionN}
S[\ell] = \iint_{\R\times\mathbb{S}^1} \dd t\, \dd x\; \Bigl( \ps{W^+_\ell\,\ell^{-1} \p_+ \ell}{ \ell^{-1} \p_- \ell} - \ps{\ell^{-1} \p_+ \ell}{ W^-_\ell\,\ell^{-1} \p_- \ell} \Bigr) - \Wd{\ell}\,.
\end{equation}
Let us explain the different ingredients entering this formula. The linear operators $W^\pm_\ell : \df \to \df$ are projectors on the double algebra characterised by the following kernels and images:\footnote{\label{Foot:W}We note that the subspaces $\Ad_\ell^{-1} \, \kf$ and $V_\pm$ are both of dimensions $d=\frac{1}{2}\dim\df$. The projector $W^\pm_\ell$ thus exists if and only if these subspaces have trivial intersection. In the following, we will assume that this holds, at least for generic values of $\ell$ (in the general case, we allow for the existence of specific non-generic values of $\ell\in D$ for which these subspaces intersect non-trivially: these points then correspond to singularities in the target space of the resulting $\s$-model). We finally note that this trivial intersection property always holds when the bilinear form $\pse$ is definite positive~\cite{Lacroix:2020flf}.}
\begin{equation}\label{Eq:WpmN}
\Ker\bigl( W^\pm_\ell ) = \Ad_\ell^{-1} \, \kf \qquad \text{ and } \qquad \text{Im}\bigl( W^\pm_\ell ) = V_\pm\,.
\end{equation}
In particular, the definition of $W^\pm_\ell$ depends on the choice of the operator $\Ec$ through its eigenspaces $V_\pm = \Ker(\Ec \mp \Id)$. We note for completeness that $\Tp W_\ell^\pm = 1 - W_\ell^\mp$, where $\Tp W_\ell^\pm$ is the transpose of $W_\ell^\pm$ with respect to the bilinear form $\psd$.

The last part of the action \eqref{Eq:ActionN} involves the Wess-Zumino term $\Wd{\ell}$ of the $D$-valued field $\ell$. To define it, we first introduce a 3-dimensional manifold $\mathbb{B}$ with boundary $\p\mathbb{B} = \R\times\mathbb{S}^1$ coinciding with the 2-dimensional space-time of the $\s$-model. We then consider an extension of the field $\ell : \R\times\mathbb{S}^1 \to D $ to $\mathbb{B}$, which coincides with $\ell$ on the boundary $\p\mathbb{B}$ and that we will still denote as $\ell$ for simplicity. The Wess-Zumino term is then defined as
\begin{equation}\label{Eq:WZD}
\Wd{\ell} = \frac{1}{12} \iiint_{\mathbb{B}} \; \bigl\langle \ell^{-1} \dd\ell \;  \overset{\wedge}{,} \; \bigl[ \ell^{-1} \dd\ell \;  \overset{\wedge}{,} \; \ell^{-1} \dd\ell \bigr] \bigr\rangle\,.
\end{equation}
Here $\dd$ is the de Rham differential on $\mathbb{B}$, so that $\ell^{-1} \dd\ell$ is a $\df$-valued 1-form on $\mathbb{B}$, \textit{i.e.} an element of $\df \otimes \Omega^1(\mathbb{B})$. The integrand in the above formula should then be understood as a 3-form in $\Omega^3(\mathbb{B})$, built from $\ell^{-1} \dd\ell$ using the commutator $[\cdot,\cdot]$ and bilinear form $\psd$ on the Lie algebra factor $\df$ and exterior products on the factor $\Omega^1(\mathbb{B})$. It is a standard result that this 3-form is closed, so that the expression \eqref{Eq:WZD} is locally equivalent to a 2-dimensional term involving the boundary field $\ell : \R\times\mathbb{S}^1 \to D $. It is however not exact in general, meaning that the Wess-Zumino term cannot be written as a 2-dimensional integral globally and usually depends on the choice of extension of $\ell$ to $\mathbb{B}$. This dependence is classified by the 3$^{\text{rd}}$ homology group of $D$ and crucially does not change the contribution of the Wess-Zumino term to the classical equation of motion derived from extremizing the $\s$-model action \eqref{Eq:ActionN}.

\paragraph{$\bm K$-gauge symmetry.} As mentioned above, one of the crucial property of the action \eqref{Eq:ActionN} is that it possesses a gauge symmetry with respect to the subgroup $K$. More precisely, let us consider the local transformation
\begin{equation}\label{Eq:GaugeK}
\ell(t,x) \longmapsto k(t,x)\, \ell(t,x)\,, \qquad \text{ with } \qquad k(t,x) \in K
\end{equation}
an arbitrary $K$-valued function. One then checks that the action \eqref{Eq:ActionN} is invariant under this action. The physical target space of the $\s$-model is thus the quotient $\KD$, which is of dimension $\frac{1}{2} \dim \df = d$.

If this quotient admits a section $s: \KD \to D$, which smoothly assigns a unique choice of representative in $D$ to each equivalence class in $\KD$, then one can explicitly eliminate the non-physical degrees of freedom of the theory by imposing that the field $\ell$ is valued in the image of $s$. This amounts to a gauge-fixing of the theory and allows to concretely describe it as a $\s$-model with a $(\KD)$-valued field, through the isomorphism between $\KD$ and $\text{Im}(s) \subset D$ provided by the section.\\

For completeness, let us briefly point out the key steps of the proof of the $K$-gauge invariance. We first note that, as $K$ is a subgroup of $D$, one has $\Ad_k^{-1} \kf = \kf$. Thus, using equation \eqref{Eq:WpmN}, the kernels and images of the projectors $W^\pm_{k\ell}$ exactly coincide with that of $W^\pm_\ell$, hence $W^\pm_{k\ell}=W^\pm_{\ell}$. In other words, the projectors $W^\pm_\ell$ are gauge invariant. Moreover, we note that under the gauge transformation, the currents $\ell^{-1} \p_\pm \ell$ transform as
\begin{equation}
\ell^{-1} \p_\pm \ell \longmapsto \ell^{-1} \p_\pm \ell + \Ad_\ell^{-1} \bigl( k^{-1} \p_\pm k \bigr)\,.
\end{equation}
In particular, we see that the second term in this transformed current belongs to $\Ad_\ell^{-1} \kf = \Ker(W^\pm_\ell)$ and thus is annihilated by the projectors $W^\pm_\ell$. We then find that the quantities $W^\pm_\ell\, \ell^{-1} \p_\pm \ell$ are gauge invariant. This result, together with the property $\Tp W_\ell^\pm = 1 - W_\ell^\mp$ discussed earlier, allows one to easily determine the gauge transformation of the first two terms in the action \eqref{Eq:ActionN}.

There is thus left to describe the gauge transformation of the Wess-Zumino term $\Wd{\ell}$. To do so, it is useful to recall the Polyakov-Wiegmann formula~\cite{Polyakov:1983tt}, which states that for any two $D$-valued fields $\ell$ and $\ell'$, one has
\begin{equation}\label{Eq:PW}
\Wd{\ell'\ell} = \Wd{\ell} + \Wd{\ell'} +  \iint_{\R\times\mathbb{S}^1} \dd t \, \dd x \; \Bigl( \ps{\p_+\ell\,\ell^{-1}}{\ell'\,^{-1}\p_- \ell'} - \ps{\ell'\,^{-1}\p_+ \ell'}{\p_-\ell\,\ell^{-1}} \Bigr) \,.
\end{equation}
To compute the gauge transformed Wess-Zumino term $\Wd{k\ell}$, we apply this formula and use the fact that $\Wd{k}=0$ due to the isotropy of $\kf$. Combining this with the results obtained above, one then finally finds that the action \eqref{Eq:ActionN} is gauge invariant, as claimed.

\paragraph{Canonical analysis and current $\bm\Jc$.} So far, we have not explained how the $\s$-model on $\KD$ built in this subsubsection is related to the $\Ec$-model described in the previous one as a Hamiltonian field theory with a current Poisson algebra. To do so, let us consider the canonical analysis of the $\s$-model, thus passing to the Hamiltonian formulation. We will not enter into the details of this analysis here and will simply describe the main features that we shall need later. The most important ones concern the $\df$-valued current
\begin{equation}\label{Eq:Jn}
\Jc = W^+_\ell \, \ell^{-1} \p_+ \ell - W^-_\ell \, \ell^{-1} \p_- \ell\,.
\end{equation}
Passing to the Hamiltonian formulation and computing the Poisson bracket of this quantity, one finds that it satisfies the current algebra \eqref{Eq:PbJ}. Moreover, performing the Legendre transform to derive the Hamiltonian of the theory, one observes that the latter exactly takes the form \eqref{Eq:Hn} once rewritten in terms of $\Jc$. Thus, we find that the Hamiltonian formulation of the $\s$-model considered here, when expressed in terms of the current $\Jc$, is nothing but the non-degenerate $\Ec$-model introduced in the previous subsubsection. In particular, the second order equations of motion obeyed by the $\s$-model field $\ell$ can be recast as the first order equation \eqref{Eq:EoMn} in terms of $\Jc$.

Let us also note that the current $\Jc$ is gauge invariant under the local $K$-action \eqref{Eq:GaugeK}. Indeed, we have observed in the previous paragraph that the quantities $W^\pm_\ell\,\ell^{-1}\p_\pm \ell$ are left unchanged under this transformation. In practice, it means that all the computations involving $\Jc$ can be performed independently of how we treat the $K$-gauge symmetry. For instance, we can keep $\Jc$ expressed in terms of the whole $D$-valued field $\ell$ as in equation \eqref{Eq:Jn} or in contrast impose a gauge-fixing condition as sketched in the previous paragraph and work with an expression of $\Jc$ explicitly in terms of physical fields of the model. In particular, the current algebra \eqref{Eq:PbJ} obeyed by $\Jc$ does not depend on whether we work with a gauged formulation or a gauge-fixed one.

Recall finally from equation \eqref{Eq:WpmN} that the image of the projector $W^\pm_\ell$ is $V_\pm$. Thus it is clear that the two terms in the expression \eqref{Eq:Jn} of $\Jc$ are respectively valued in $V_+$ and $V_-$. In other words, the decomposition $\Jc=\Jc_+ - \Jc_-$ of the current is simply given by
\begin{equation}\label{Eq:JpmLagN}
\Jc_+ = W^+_\ell \, \ell^{-1} \p_+ \ell \qquad \text{ and } \qquad \Jc_- = W^-_\ell \, \ell^{-1} \p_- \ell\,.
\end{equation}

\paragraph{Summary.} Let us briefly summarise the construction of this subsection. The $\s$-model that we built is characterised by the following defining data:\vspace{-3pt}
\begin{enumerate}[(i)]\setlength\itemsep{0.2pt}
\item a real connected Lie group $D$ of even dimension, with Lie algebra $\df=\text{Lie}(D)$ ;
\item a non-degenerate ad-invariant split symmetric bilinear form $\psd$ on $\df$ ;
\item a symmetric operator $\Ec : \df \to \df$ such that $\Ec^2=\Id$ and $V_\pm = \Ker(\Ec\mp \Id)$ have equal dimension ;
\item a subgroup $K\subset D$ such that $\kf=\text{Lie}(K)$ is a maximally isotropic subalgebra of $\df$. \vspace{-4pt}
\end{enumerate}
From this data, the $\s$-model is built as follows. One first considers a $D$-valued field $\ell(t,x)$ on $\R\times\mathbb{S}^1$ and constructs the projectors $W^\pm_\ell$ according to equation \eqref{Eq:WpmN}. The action of the $\s$-model is then defined by the formula \eqref{Eq:ActionN} and is invariant under the $K$-gauge symmetry $\ell \mapsto k\ell$, $k\in K$, making the physical target space of the model the quotient $\KD$. An important observable in this theory is the $\df$-valued field $\Jc$, built from $\ell$ as in \eqref{Eq:Jn}. In particular, it satisfies the current Poisson algebra \eqref{Eq:PbJ}. Moreover, the momentum, Hamiltonian and equations of motion of the model take the simple forms \eqref{Eq:P}, \eqref{Eq:Hn} and \eqref{Eq:EoMn} in terms of $\Jc$, or equivalently \eqref{Eq:EoMnJpm} and \eqref{Eq:Ppm} in light-cone coordinates.

Let us note that although the construction of the $\s$-model and the current $\Jc$ in terms of the field $\ell$ depends on the choice of the maximally isotropic Lie algebra $\kf$, the first order Hamiltonian formulation in terms of the current $\Jc$ is independent of this choice. In practice, this means that many aspects of the theory can be studied using only the data of $(\df,\psd,\Ec)$. This includes the investigation of conformal and chiral structures in these theories, which is the main subject of this article. The choice of $\kf$ will then only appear in the parts of the paper where we treat explicit examples, as it will allow us to identify concretely the $\s$-models that we are considering.

We note that, given a double algebra $(\df,\psd)$, there can be in general several choices of maximally isotropic subalgebras $\kf$. These will correspond to different $\s$-models which however share the same underlying first order formulation when written in terms of the current $\Jc$. The typical example of such a situation occurs when the double group possesses a decomposition $D = K\cdot \widetilde{K} = \widetilde{K} \cdot K$ in terms of two subgroups $K$ and $\widetilde{K}$ whose Lie algebras are complementary maximally isotropic subalgebras of $\df$. In this case, one can build two $\s$-models based on the same underlying $\Ec$-model, with respective target spaces $\KD \simeq \widetilde{K}$ and $\widetilde{K}\!\setminus\! D \simeq K$, which are then said to be Poisson-Lie T-duals to each other~\cite{Klimcik:1995ux,Klimcik:1995dy,Klimcik:1996nq}. The study of this duality, which generalises the standard notions of abelian and non-abelian T-duality, was one of the main motivation behind the introduction of $\Ec$-models.

\subsection[Conformal non-degenerate \texorpdfstring{$\Ec$}{E}-models and their chiral Poisson algebras]{Conformal non-degenerate \texorpdfstring{$\bm\Ec$}{E}-models and their chiral Poisson algebras}
\label{SubSec:ConfN}

In the reference~\cite{Klimcik:1996hp}, Klim\v{c}\'{i}k and \v{S}evera considered a simple condition on the defining data of a non-degenerate $\Ec$-model ensuring that the associated $\s$-model is conformal. In fact, it was argued in~\cite{Klimcik:1996hp} that this $\s$-model then essentially takes the form of a well-known conformal theory, the Wess-Zumino-Witten model (at least under certain technical assumptions). In this subsection, we consider a very similar configuration, which we will call the strong conformal condition, and revisit the results of~\cite{Klimcik:1996hp}, although in a slightly different approach. This will serve as a warm-up for Subsection \ref{SubSec:Conf}, where we introduce a similar condition for degenerate $\Ec$-models, leading to the construction of conformal gauged $\s$-models. Although this degenerate case will share some similarities with the present one, it will also exhibit more involved structures and will lead to a richer variety of conformal theories.

Our approach below will be the following. We will first introduce the aforementioned strong conformal condition and will show that it implies the conformal invariance of the theory at the 1-loop quantum level. Moreover, we will explain that this condition also has important structural consequences at the classical level, namely the decomposition of the degrees of freedom of the theory into left-moving and right-moving fields. Furthermore, we will prove that the sets of left-moving and right-moving fields form two closed Poisson current algebras which mutually Poisson commute, thus showing that this chiral decomposition also occurs in the Hamiltonian structure of the theory. Such a feature is one of the hallmarks of 2-dimensional conformal field theories and paves the way for the quantisation of these models using the language of Vertex Operator Algebras, which describe the quantum chiral fields. To keep the length of the paper contained, we will not discuss these quantum aspects here but hope to return to these questions in the future. Finally, in the next subsection, we will explain the relation between this conformal non-degenerate $\Ec$-model and the Wess-Zumino-Witten theory, as stated in~\cite{Klimcik:1996hp}.

\paragraph{The strong conformal condition.} Let us consider a non-degenerate $\Ec$-model associated with the data $(\df,\psd,\Ec)$, following the notations of the previous subsection. In particular, recall that the eigenspaces $V_\pm = \Ker(\Ec \mp \Id)$ of the operator $\Ec$ allows us to write an orthogonal decomposition $\df = V_+ \oplus V_-$. We will say that the model satisfies the \textit{strong conformal condition} if the following two assumptions, which are equivalent, are verified:\vspace{8pt}\\
\begin{tabular}{cp{0.87\textwidth}}
\;\;\;(C1)\!\! & the subspaces $V_\pm$ are subalgebras of $\df$ ; \vspace{7pt} \\
\;\;\;(C2)\!\! & we have $[V_+,V_-] = \lbrace 0 \rbrace$.
\end{tabular}~\vspace{8pt}\\
As mentioned earlier, the condition (C1) is quite similar to the setup considered in the reference~\cite{Klimcik:1996hp}.\footnote{\label{Foot:ConfCond}One of the main difference is that~\cite{Klimcik:1996hp} assumed only $V_+$ or $V_-$ to be a subalgebra and not necessarily both as we do here: in that case, the property (C2) in fact does not hold. As we will see later in this subsection, this more restricted setup will play an important role in the description of the chiral fields of the theory and its decomposition into a left-moving and a right-moving sectors. Another small difference is that we do not necessarily assume here that $V_\pm$ are the Lie algebras of compact groups, as done in~\cite{Klimcik:1996hp}.} Let us briefly mention that we do not expect this condition to be the most general one ensuring the 1-loop conformal invariance of the $\Ec$-model (in other words this is a sufficient but not necessary condition), which is why we used the adjective strong to characterise it. We refer to the concluding subection \ref{SubSec:GenConf} for a brief discussion of why we expect the existence of more general conformal conditions: a more thorough analysis of theses aspects is beyond the scope of this article and is a natural future perspective for further developments.\\

Before analysing the consequences of this condition, let us first prove the equivalence of the two assertions (C1) and (C2). For instance, let us suppose that (C1) is true and consider $(X_+,X_-) \in V_+ \times V_-$. We then want to prove that $[X_+,X_-]=0$. For any element $Y=Y_+ + Y_-$ in $\df$, we find
\begin{equation*}
\ps{[X_+,X_-]}{Y} = \ps{[X_+,X_-]}{Y_+} + \ps{[X_+,X_-]}{Y_-} = -\ps{X_-}{[X_+,Y_+]} + \ps{X_+}{[X_-,Y_-]} = 0\,.
\end{equation*}
Here, we have used the ad-invariance of $\psd$ to get the second equality and the assumption (C1) to obtain the last one. Indeed, (C1) implies that $[X_\pm,Y_\pm]$ belongs to $V_\pm$ and thus is orthogonal to $X_\mp \in V_\mp$. We therefore find that $\ps{[X_+,X_-]}{Y}=0$ for all $Y\in\df$. Since $\psd$ is non-degenerate, we get $[X_+,X_-]=0$, hence proving that (C1) implies (C2).

Conversely, let us assume (C2) and consider $X_+,Y_+ \in V_+$. For any $X_- \in V_-$, using the ad-invariance of $\psd$ and (C2), we get
\begin{equation*}
\ps{[X_+,Y_+]}{X_-} = \ps{X_+}{[Y_+,X_-]} = 0\,.
\end{equation*}
Since this is true for all $X_- \in V_-$, we find that $[X_+,Y_+]$ belongs to $V_-^\perp = V_+$, hence proving that $V_+$ is a subalgebra. A similar reasoning shows that $V_-$ is also a subalgebra: thus, (C2) implies (C1).

For completeness, let us mention that there are two other conditions which are equivalent to (C1) and (C2). The first one is that $\Ec$ is an intertwiner of the adjoint representation of $\df$, \textit{i.e.} it commutes with the adjoint action of $\df$ on itself. The second one is that the bilinear form $\pse$ on $\df$ is ad-invariant.

\paragraph{Vanishing of the 1-loop $\bm\beta$-functions.} One of the advantages of the $\Ec$-model formulation of a $\s$-model is that it allows a remarkably simple characterisation of the 1-loop Renormalisation Group (RG) flow of the theory~\cite{Valent:2009nv,Sfetsos:2009dj,Avramis:2009xi,Sfetsos:2009vt}. There are several equivalent ways to formulate this RG flow: here we will use a rather compact one in terms of the quadratic form $\pse$, which will be the most adapted for our purposes. In particular, it will provide us with a very simple proof of the fact that the strong conformal condition introduced above implies the vanishing of the 1-loop $\beta$-functions of the theory.

Let us then consider the non-degenerate $\Ec$-model associated with the data $(\df,\psd,\Ec)$. In particular, the Hamiltonian \eqref{Eq:Hn} of this model is characterised by the quadratic form $\pse$. After quantisation and treatment of the UV divergences that might appear, we expect the theory to depend on the energy scale $\mu$ of the RG flow. At the 1-loop level, this dependence can be expressed in terms of $\pse$ as
\begin{equation}\label{Eq:RgN}
\frac{\dd\;}{\dd \tau} \ps{X}{\Ec X} = -4\hbar\, \Tr_{\df}\bigl( S_{X} \bigr) + O(\hbar^2)\,, \qquad \forall\,X\in\df\,.
\end{equation} 
Here $\tau = \frac{1}{2\pi} \log\mu$ is the RG parameter and $S_{X}: \df \to \df$ is a linear operator defined by
\begin{equation}\label{Eq:SN}
S_{X} = \pi_-\, \ad_{\pi_- X}\,\pi_+\,\ad_{\pi_+ X} \, ,
\end{equation}
where we recall that $\pi_\pm : \df \to V_\pm$ are the projectors associated with the decomposition $\df=V_+ \oplus V_-$.

The equation \eqref{Eq:RgN} can be understood as the RG-flow of the family of all $\Ec$-models with double algebra $(\df,\psd)$, whose defining parameters are seen as the entries of the quadratic form $\pse$, subject to the constraint that this form defines a relativistic model (we note that this property is preserved by the flow \eqref{Eq:RgN}). In practice, we will often be interested in specific families of $\Ec$-models, which correspond to a particular choice of $\pse$ in terms of parameters (coupling constants) $\lambda=( \lambda_k)_{k=1,\dots,n}$. In such a setup, we find that the theory is 1-loop renormalisable if the equation \eqref{Eq:RgN} is consistent with the functional dependence of $\pse$ in terms of $\lambda$. More concretely, this happens if we have
\begin{equation}\label{Eq:Beta}
-4\, \Tr\bigl( S_{X} \bigr) = \sum_{k=1}^n \beta^{(1)}_k(\lambda)\, \frac{\p\;}{\p\lambda_k} \ps{X}{\Ec X}  \,, \qquad \forall\,X,Y\in\df\,,
\end{equation}
for some functions $(\beta^{(1)}_k)_{k=1,\dots,n}$ of the coupling constants $\lambda$. In that case, the equation \eqref{Eq:RgN} simply induces the flow
\begin{equation}
\frac{\dd\;}{\dd\tau} \lambda_k = \hbar\,\beta^{(1)}_k(\lambda)+ O(\hbar^2)\
\end{equation}
of the coupling constants and $\beta^{(1)}_k$ is then identified as the 1-loop $\beta$-function of $\lambda_k$.\\

Having at hand these general results, let us now suppose that the $\Ec$-model we are considering satisfies the strong conformal condition introduced in the previous paragraph. In particular, the formulation (C2) of this condition, namely $[V_+,V_-]=\lbrace 0 \rbrace$, implies that the operator $\ad_{\pi_- X} \pi_+$ vanishes for all $X\in\df$. We thus conclude from equation \eqref{Eq:SN} that $S_{X}=0$ for all $X\in\df$. The RG flow \eqref{Eq:RgN} is therefore trivial, showing the 1-loop conformal invariance of the model, as claimed.

\paragraph{Classical chiral fields.} One of the key features of 2-dimensional conformal field theories is the presence of chiral degrees of freedom in the dynamics, \textit{i.e.} left-moving and right-moving fields. In the case at hand, these chiral fields are easily identified at the classical level. Indeed, recall that the components $\Jc_\pm\in V_\pm$ of the current $\Jc$ satisfy the equations of motion \eqref{Eq:EoMnJpm}. If we assume the strong conformal condition $[V_+,V_-]=\lbrace 0 \rbrace$, it is then clear that
\begin{equation}
\p_- \Jc_+ = \p_+ \Jc_- = 0\,.
\end{equation}
The field $\Jc_+$ is thus left-moving, while $\Jc_-$ is right-moving. Since the equations of motion of $\Jc_+$ and $\Jc_-$ encode the dynamics of all the degrees of freedom of the $\Ec$-model, this shows that the theory completely decouples into two chiralities.

In the rest of this paper, we will indicate left-moving and right-moving quantities with the labels $\cL$ and $\cR$. In particular, in the case of a non-degenerate $\Ec$-model satisfying the strong conformal condition, we will use the notation $\Jc^\cL=\Jc_+$ and $\Jc^\cR=\Jc_-$ to designate the left- and right-moving components of the current $\Jc$.

\paragraph{Chiral Poisson algebras.} An important characteristic of 2-dimensional conformal field theories is their algebras of extended conformal symmetry. At the classical level, these take the form of two chiral Poisson algebras, formed by the left-moving and right-moving fields of the model respectively. Indeed, we recall that the Poisson bracket of two left-moving observables is itself left-moving, meaning that the set of left-moving fields is closed under Poisson bracket (and similarly for right-moving ones). The study of these chiral Poisson algebras is an important step towards the quantisation of the theory, which can be expressed in terms of Vertex Operator Algebras, describing the quantum chiral fields and their commutation relations.

For the conformal $\Ec$-model under consideration here, the chiral Poisson algebras admit a simple description. Indeed, we have shown above that the left-moving and right-moving fields of the theory are encoded respectively in $\Jc^\cL$ and $\Jc^\cR$, which are the components in $V_+$ and $V_-$ of $\Jc$. The latter satisfies the current Poisson algebra \eqref{Eq:PbJ} associated with $\df$. Yet, under the strong conformal condition, the decomposition $\df = V_+ \oplus V_-$ becomes a direct sum of quadratic Lie algebras: $V_+$ and $V_-$ are closed under the Lie bracket by (C1), commute one with another by (C2) and are orthogonal with respect to the invariant bilinear form $\psd$. One easily checks that the current Poisson algebra for $\Jc$ then splits into two decoupled Poisson subalgebras for $\Jc^\cL$ and $\Jc^\cR$. More concretely, we find
\begin{subequations}\label{Eq:PbJlr}
\begin{align}
\bigl\lbrace \Jc^\cL(x)\ti{1}, \Jc^\cL(y)\ti{2} \bigr\rbrace &= \bigl[ \mathsf{C}^+\ti{12}, \Jc^\cL(x)\ti{1} \bigr] \,\delta(x-y) - \mathsf{C}^+\ti{12}\,\p_x\delta(x-y)\,, \label{Eq:PbJl} \\
\bigl\lbrace \Jc^\cR(x)\ti{1}, \Jc^\cR(y)\ti{2} \bigr\rbrace &= \bigl[ \mathsf{C}^-\ti{12}, \Jc^\cR(x)\ti{1} \bigr] \,\delta(x-y) + \mathsf{C}^-\ti{12}\,\p_x\delta(x-y)\,, \\
\bigl\lbrace \Jc^\cL(x)\ti{1}, \Jc^\cR(y)\ti{2} \bigr\rbrace &= 0\,,
\end{align}
\end{subequations}
where $\mathsf{C}^\pm\ti{12} = \pm (\pi_\pm \otimes \pi_\pm)\,\Cd \in V_\pm \otimes V_\pm$ is the split quadratic Casimir of the Lie algebra\footnote{Recall that a split quadratic Casimir is associated with a Lie algebra equipped with a non-degenerate invariant bilinear form. Here, $\mathsf{C}^\pm\ti{12}$ is then more precisely the quadratic Casimir of the Lie algebra $V_\pm$ equipped with $\pm\psd|_{V_\pm}$. The plus/minus sign in this bilinear form has been introduced so that it is definite positive if the $\s$-model realising the conformal $\Ec$-model under investigation has a target space metric with positive signature (\textit{i.e.} if $\pse$ is definite positive on $\df$).} $V_\pm$. In other words, the left-moving (resp. right-moving) degrees of freedom of the theory form the current Poisson algebra of $V_+$ (resp. $V_-$).\\

Important elements of these chiral algebras are the left and right-moving components of the energy-momentum tensor of the theory. In the present case, these are simply given by
\begin{equation}
T^\cL = \frac{1}{2} \psb{\Jc^\cL}{\Jc^\cL} \qquad \text{ and } \qquad T^\cR = -\frac{1}{2} \psb{\Jc^\cR}{\Jc^\cR}\,.
\end{equation}
This expression for $T^\cLR$ is the classical equivalent of the Segal-Sugawara construction, which gives the energy-momentum tensor of a quantum current algebra.

As expected, the fields $T^\cL$ and $T^\cR$ satisfy the classical Virasoro algebra. More concretely, starting from the bracket \eqref{Eq:PbJl}, one finds (with a minus sign for $T^\cL$ and a plus sign for $T^\cR$)
\begin{equation}
\bigl\lbrace T^\cLR(x), T^\cLR(y) \bigr\rbrace = \mp\bigl( T^\cLR(x) + T^\cLR(y) \bigr)\, \p_x\delta(x-y)\,.
\end{equation}

\paragraph{Chiral affine symmetries.} We will end this subsection with a discussion of the chiral affine symmetries of the conformal $\Ec$-model under consideration and the relation of the present article to the recent work~\cite{Arvanitakis:2022bnr}. To do so, let us for the moment consider a generic non-degenerate $\Ec$-model, which does not necessarily satisfy the strong conformal condition. The affine symmetries of this theory are discussed in the section 7 of reference~\cite{Arvanitakis:2022bnr}. More precisely, it studies under which conditions the action\footnote{Technically, the article~\cite{Arvanitakis:2022bnr} considers the action of a model before ``gauging'' the isotropic subgroup $K$, rather than the action \eqref{Eq:ActionN} on $\KD$: however, this will not impact the discussion and results of this paragraph.} \eqref{Eq:ActionN} is invariant under a transformation $\ell(t,x) \mapsto \ell(t,x) v(t,x)$, where $v$ is valued in $D$ and can depend on the space-time coordinates $(t,x)$. We note that such a transformation should not be interpreted as a gauge symmetry of the theory: indeed, we do not require that $\ell(t,x) \mapsto \ell(t,x) v(t,x)$ leaves the model invariant for all functions $v(t,x)$ but rather search for specific conditions on the way $v$ depends on $(t,x)$ which ensure that the action is left unchanged under the transformation. These conditions were explicitly derived in~\cite[Equation (7.7)]{Arvanitakis:2022bnr}: in our conventions, they read
\begin{equation}
\Ec = \Ad_v \circ \Ec \circ \Ad_v^{-1} \qquad \text{ and } \qquad (\p_\pm v)v^{-1} \in V_\pm = \Ker(\Ec \mp \Id)\,.
\end{equation}
Let us introduce the ``commutant'' subgroup $C(\Ec) = \bigl\lbrace v_0 \in D, \, \Ad_{v_0} \circ \Ec \circ \Ad_{v_0}^{-1} = \Ec \bigr\rbrace$ in $D$ (here, $v_0$ denotes constant elements of $D$ and not functions of $(t,x)$ valued in $D$). One can then distinguish between several cases, depending on the choice of the double group $D$ and of the operator $\Ec$. For a completely generic choice, one can expect that the subgroup $C(\Ec)$ is trivial, \textit{i.e.} that the condition $\Ec = \Ad_{v_0} \circ \Ec \circ \Ad_{v_0}^{-1}$ does not have any solutions other than the identity of $D$: the model then does not possess any non-trivial symmetries of the form $\ell \mapsto \ell v$. More specific choices of $D$ and $\Ec$ can lead to a non-trivial commutant subgroup $C(\Ec)$. In that case, one can then find symmetries $\ell \mapsto \ell v$ by searching for functions $v(t,x)$ valued in $C(\Ec)$ which further satisfy the condition $(\p_\pm v)v^{-1}\in V_\pm$. For many choices of $D$ and $\Ec$, the only solutions will be the constant functions $v(t,x)=v_0$, with $v_0\in C(\Ec)$: in that case, the model is only invariant under rigid global symmetries $\ell\mapsto \ell v_0$, which do not allow for any dependence on the space-time coordinates $(t,x)$ -- these symmetries then correspond to global isometries of the $\s$-model \eqref{Eq:ActionN}. However, very specific choices of $D$ and $\Ec$ can allow for non-constant solutions where $v$ depends on specific combinations of $(t,x)$.

The typical example of such a case, which will be relevant for the study of $\Ec$-models satisfying the strong conformal condition, arises in the following setup. Suppose that the commutant $C(\Ec)$ contains a non-trivial subgroup $D^\cL$ whose Lie algebra is a subspace of $V_+$, \textit{i.e.} $\text{Lie}\,D^\cL \subset V_+$. Then any function $v(t,x)=v^\cL(t+x)$, where $v^\cL$ is valued in $D^\cL$, satisfies $(\p_+ v) v^{-1} \in V_+$ and $(\p_- v) v^{-1} = 0 \in V_-$, thus defining a symmetry $\ell\mapsto\ell v$ of the model. This symmetry is not a constant one, in the sense that it contains an arbitrary function of the light-cone coordinate $t+x$.\footnote{However, as noted above, this is not a gauge symmetry, as $v$ is a function of only $t+x$ and does not depend on $t-x$.} Obviously, one can similarly consider symmetries depending on arbitrary functions of the other light-cone coordinate $t-x$, valued in a subgroup $D^\cR$ of $C(\Ec)$ whose Lie algebra is contained in $V_-$. If they exist, these $(D^\cL\!\times\! D^\cR)$--transformations are called chiral affine symmetries of the model.

In this language, the strong conformal condition considered in the present paper exactly corresponds to the case with ``maximal chiral symmetries''. More precisely, let us suppose that the commutant subgroup $C(\Ec)$ is maximal, in the sense that it coincides with the whole double group $D$. This amounts to requiring that the operator $\Ec$ commutes with the adjoint action of $\df$: as observed at the beginning of this subsection, this is one of the equivalent formulation of the strong conformal condition. Moreover, in that case, we know that $V_+$ and $V_-$ become commuting subalgebras of $\df$: they thus correspond to subgroups $D^\cL$ and $D^\cR$ of $D$, such that\footnote{Here, we will ignore any potential global or topological obstruction in extending the Lie algebra decomposition $\df = V_+ \oplus V_-$ to a group decomposition.} $D=D^\cL\! \times \! D^\cR$, with $\text{Lie}\,D^\cL = V_+$ and $\text{Lie}\,D^\cR = V_-$. We thus obtain chiral affine $(D^\cL\!\times\! D^\cR)$--symmetries taking the form
\begin{equation}\label{Eq:ChiralSym}
\ell(t,x) \longmapsto \ell(t,x)\, v^\cL(t+x)\, v^\cR(t-x)\,, \qquad v^\cL \in D^\cL\,, \quad v^\cR \in D^\cR\,.
\end{equation}
These symmetries are maximal, in the sense that $D^\cL$ and $D^\cR$ generate the whole group $D$. They can in fact be reinterpreted as the transformations generated by the chiral currents $\Jc^\cL$ and $\Jc^\cR$ considered in the previous paragraphs. This setup was discussed in~\cite{Arvanitakis:2022bnr} for the example of the WZW model: this theory and its interpretation as a conformal $\Ec$-model will be the subject of the next subsection (in fact, we will argue there that any non-degenerate $\Ec$-model satisfying the strong conformal condition, and thus having maximal chiral symmetries, essentially takes the form of a WZW model).

\subsection[The WZW model and the \texorpdfstring{$\lambda$}{lambda}-deformation]{The WZW model and its \texorpdfstring{$\bm\lambda$}{lambda}-deformation}
\label{SubSec:WZW}

The conformal $\Ec$-models considered in the previous subsection possess two commuting chiral current algebras. This setup is reminiscent of a well-known conformal field theory called the Wess-Zumino-Witten (WZW) model~\cite{Witten:1983ar,Witten:1983tw}. As already mentioned earlier, the results of~\cite{Klimcik:1996hp} show that a non-degenerate $\Ec$-model satisfying the strong conformal condition in fact essentially takes the form of such a WZW model once realised as a $\s$-model. In this subsection, we will revisit this result and the interpretation of the WZW model in terms of $\Ec$-models.

We will in fact make a small detour, by considering not only the WZW model but also its so-called $\lambda$-deformation. The latter was introduced in~\cite{Sfetsos:2013wia} as an integrable deformation of the former: it is a non-conformal $\s$-model, which depends on a coupling constant $\lambda$ and which gives back the WZW model when $\lambda \to 0$ (corresponding to the UV limit of the theory). It was proven in~\cite{Klimcik:2015gba} that the $\lambda$-model can be reformulated as a non-degenerate $\Ec$-model. In this subsection, we will recall this formulation and will use it to illustrate the results of this section. In particular, we will show that the conformal limit $\lambda\to 0$ exactly coincides with the point in parameter space where the underlying $\Ec$-model satisfies the strong conformal condition. This detour through the non-conformal $\lambda$-model will allow us to illustrate more concretely the role played by this strong conformal condition, for instance on the chirality properties of the theory. Moreover, we include it as it represents a prototypical example of integrable deformations of conformal $\Ec$-models, which were part of the main motivations behind the present work. We will end this subsection by coming back to the results of~\cite{Klimcik:1996hp}: namely, we will argue that all the non-degenerate $\Ec$-models satisfying the strong conformal condition can in fact be brought back to the form of a WZW model.

\paragraph{The double algebra.} Let us first describe the data defining the non-degenerate $\Ec$-model that we will consider, starting with the double algebra $\df$. We will suppose that we are given a real connected simple Lie group $G$, whose Lie algebra $\g=\text{Lie}(G)$ is then equipped with a non-degenerate invariant bilinear form\footnote{\label{FootForm}We fix the normalisation of $\kappa(\cdot,\cdot)$ so that it coincides with the opposite of the ``minimal'' bilinear form on $\g$, with respect to which the longest roots of $\g^\mathbb{C}$ have length $2$. For instance, $\kappa(\cdot,\cdot)$ coincides with the opposite of the trace form in the fundamental representation for $\g=\mathfrak{su}(n)$. The minus sign was introduced so that the bilinear form is positive definite if $G$ is compact. The main property of $\kappa(\cdot,\cdot)$ that we shall need is its relation with the Killing form, which reads
\begin{equation}
\kappa(X,Y) = - \frac{1}{2 \hv} \Tr\bigl(\ad_X \ad_Y \bigr)\,, \qquad \forall\,X,Y\in\g\,,
\end{equation}
where $\hv$ is the dual Coxeter number of $\g$.} $\kappa(\cdot,\cdot)$ (the construction below can be essentially generalised to any group $G$ whose Lie algebra possesses a bilinear form with these properties, but we will focus on the case of $G$ simple for convenience). We define the double group as the direct product $D=G\times G$, with Lie algebra $\df = \g \times \g$, whose bracket reads
\begin{equation}
\bigl[ (X_1,X_2), (Y_1,Y_2) \bigr] = \bigl( [X_1,Y_1], [X_2,Y_2] \bigr)\,,
\end{equation}
for all $X_1,X_2,Y_1,Y_2\in\g$. Moreover, we equip $\df$ with the symmetric bilinear form
\begin{equation}\label{Eq:FormLambda}
\psB{(X_1,X_2)}{(Y_1,Y_2)} = \kay\,\kappa(X_1,Y_1) - \kay\,\kappa(X_2,Y_2)\, ,
\end{equation}
where $\kay$ is an external positive parameter. One easily checks that this bilinear form is invariant and of split signature, making $\df$ an appropriate choice for the double algebra underlying an $\Ec$-model.

\paragraph{The operator $\bm\Ec$.} The next ingredient needed to define the non-degenerate $\Ec$-model is the operator $\Ec:\df \to \df$. In the case at hand, since $\df=\g\times\g$, $\Ec$ can be conveniently seen as a $2\times 2$ matrix whose entries are operators on $\g$. We will choose it to be
\begin{equation}
\Ec =  \begin{pmatrix}
\dfrac{1+\lambda^2}{1-\lambda^2} \Id & -\dfrac{2\lambda}{1-\lambda^2} \Id \\
\dfrac{2\lambda}{1-\lambda^2} \Id & -\dfrac{1+\lambda^2}{1-\lambda^2} \Id
\end{pmatrix}\,,
\end{equation}
where $\lambda\in[\,0,1)$ is an external parameter and $\Id$ stands here for the identity operator on $\g$. A straightforward computation shows that $\Ec$ is an involution, \textit{i.e.} $\Ec^2=\Id$, and is symmetric with respect to the bilinear form $\psd$ introduced in equation \eqref{Eq:FormLambda}. This operator $\Ec$ thus satisfies the right conditions to define a non-degenerate $\Ec$-model.\\

Let us quickly describe some of the properties of $\Ec$. Since it is an involution, it has eigenvalues $+1$ and $-1$. The corresponding eigenspaces are given explicitly by
\begin{equation}\label{Eq:VLambda}
V_+ = \bigl\lbrace (X,\lambda\,X), \; X\in \g \bigr\rbrace \qquad \text{ and } \qquad V_- = \bigl\lbrace (\lambda\,X,X), \; X\in \g \bigr\rbrace\,
\end{equation}
and have equal dimension $\dim V_\pm = \dim \g$ as required in the previous subsections. One also checks that $V_+$ and $V_-$ are orthogonal with respect to $\psd$, as expected. Moreover, a direct computation shows that the quadratic form $\pse$ reads
\begin{equation}\label{Eq:FormELambda}
\psB{(X_1,X_2)}{\Ec(X_1,X_2)} = \frac{\kay}{1-\lambda^2} \Bigl( \kappa(X_1 - \lambda\, X_2 , X_1 - \lambda\, X_2) + \kappa(X_2 - \lambda\, X_1 , X_2 - \lambda\, X_1) \Bigr)\,.
\end{equation}
In particular, since $\kay >0$ and $0 \leq \lambda < 1$, we see that $\pse$ is positive definite if and only if the group $G$ is compact (in which case $\kappa(\cdot,\cdot)$ is itself definite positive -- see footnote \ref{FootForm}). For $G$ compact, the $\Ec$-model under consideration thus possesses a positive Hamiltonian.

\paragraph{Isotropic subalgebra and $\bm\s$-model action.} Before we discuss the conformal limit of the $\Ec$-model we are constructing, let us describe the $\s$-model it corresponds to. For that, we will follow the procedure recalled in Subsection \ref{SubSec:LagN} and will first choose a maximally isotropic subalgebra $\kf$ of $\df$. To obtain the $\lambda$-model, we will take it to be
\begin{equation}
\kf = \g_{\diag} = \bigl\lbrace (X,X), \; X\in \g \bigr\rbrace\,,
\end{equation}
which is clearly maximally isotropic with respect to the bilinear form \eqref{Eq:FormLambda}. According to Subsection \ref{SubSec:LagN}, the associated $\s$-model has then for target space the quotient $\KD = G_{\diag} \setminus (G\times G)$ and is described by a field $\ell(t,x)$ valued in $G\times G$ with a left gauge symmetry $\ell \mapsto k\ell$, $k\in G_{\diag}$. For simplicity, we will fix this gauge symmetry from the start by taking $\ell$ to be of the form
\begin{equation}
\ell(t,x) = \bigl( g(t,x), \Id \bigr), \qquad g(t,x) \in G\,.
\end{equation}
Indeed, it is clear that such a form can always be attained by a gauge transformation $\ell \mapsto k\ell$, $k\in G_{\diag}$. In practice, we are thus identifying the quotient target space $G_{\diag} \setminus (G\times G)$ with the subspace $G \times \lbrace \Id \rbrace$, therefore describing the model in terms of a $G$-valued field $g(t,x)$.\\

To derive the $\s$-model action, the next step is to compute the projectors $W^\pm_\ell$, defined by their kernels and images \eqref{Eq:WpmN}. In the present case, the images $\text{Im}(W^\pm_\ell) = V_\pm$ have been described explicitly in equation \eqref{Eq:VLambda}. Moreover, for the choice of isotropic subalgebra $\kf = \g_{\diag}$ and the gauge-fixing $\ell=(g,\Id)$, the kernel of these operators is given by
\begin{equation}
\Ker(W^\pm_\ell) = \Ad_\ell^{-1}\kf = \bigl\lbrace \bigl(\Ad_g^{-1} X, X \bigr), \, X\in \g \bigr\rbrace\,.
\end{equation}
It is then a simple algebraic exercise to determine the explicit expressions of the projectors $W^\pm_\ell$, yielding
\begin{equation}\label{Eq:WLambda}
W^+_\ell = \begin{pmatrix}
\dfrac{\Ad_g}{\Ad_g-\lambda}& -\dfrac{1}{\Ad_g-\lambda} \\
\dfrac{\lambda\,\Ad_g}{\Ad_g-\lambda}& -\dfrac{\lambda}{\Ad_g-\lambda}
\end{pmatrix} \qquad \text{ and } \qquad W^-_\ell = \begin{pmatrix}
-\dfrac{\lambda\,\Ad_g}{1-\lambda\,\Ad_g} & \dfrac{\lambda}{1-\lambda\,\Ad_g} \\
-\dfrac{\Ad_g}{1-\lambda\,\Ad_g} & \dfrac{1}{1-\lambda\,\Ad_g}
\end{pmatrix} \,.
\end{equation}

We now have all the ingredients necessary to compute the action \eqref{Eq:ActionN} of the model. We recall that $\ell=(g,\Id)$ and thus that $\ell^{-1} \dd \ell = (g^{-1} \dd g, 0)$. Taking this into account and using the expression \eqref{Eq:FormLambda} of the form $\psd$ in the case at hand, we find that the last term $- \Wd{\ell}$ in the action \eqref{Eq:ActionN} becomes $-\kay\,\Wg{g}$, where
\begin{equation}
\Wg{g} = \frac{1}{12} \iiint_{\mathbb{B}} \; \kappa\bigl( g^{-1} \dd g \;  \overset{\wedge}{,} \; \bigl[ g^{-1} \dd g \;  \overset{\wedge}{,} \; g^{-1} \dd g \bigr] \bigr)
\end{equation}
is the Wess-Zumino term of the $G$-valued field $g$, defined in terms of the bilinear form $\kappa(\cdot,\cdot)$ on $\g$. Using the expression \eqref{Eq:WLambda} of $W^\pm_\ell$, one also determines the two first terms in equation \eqref{Eq:ActionN}. Combining all these results together, we finally get the action of the theory:
\begin{equation}\label{Eq:ActionLambda}
S_{\kay,\lambda}[g] = S_{\text{WZW},\,\kay}[g] + 2\lambda\,\kay \iint_{\R\times\mathbb{S}^1} \dd t\,\dd x\; \kappa\left( g^{-1} \p_+ g, \frac{1}{1-\lambda\,\Ad_g} \p_-gg^{-1} \right)\,,
\end{equation}
where
\begin{equation}\label{Eq:WZW}
S_{\text{WZW},\,\kay}[g] = \kay \iint_{\R\times\mathbb{S}^1} \dd t\,\dd x\; \kappa\bigl( g^{-1} \p_+g, g^{-1} \p_- g \bigr) - \kay\,\Wg{g}
\end{equation}
is the Wess-Zumino-Witten action with level $\kay$~\cite{Witten:1983ar,Witten:1983tw}. One recognises in the equation \eqref{Eq:ActionLambda} the action of the $\lambda$-model first introduced in~\cite{Sfetsos:2013wia}.

\paragraph{RG-flow and conformal limit.} The 1-loop RG-flow of the $\lambda$-model can be derived using the general formula \eqref{Eq:RgN}. The operator $S_{X}$ appearing in this formula is defined in equation \eqref{Eq:SN} and in particular depends on the eigenspaces $V_\pm$ of the operator $\Ec$. In the present case, the latter are explicitly given by \eqref{Eq:VLambda}. After a few manipulations, we find that for any $X=(X_1,X_2)$ in $\df=\g\times\g$:
\begin{equation*}
\Tr\bigl( S_{X} \bigr) = \frac{\lambda^2}{(1-\lambda^2)^2(1+\lambda)^2} \Bigl( (1+\lambda^2) \Tr\bigl( \ad_{X_1} \ad_{X_2} \bigr) - \lambda \, \Tr\bigl(\ad_{X_1} \ad_{X_1} +  \ad_{X_2} \ad_{X_2} \bigr) \Bigr)\,.
\end{equation*}
Using the relation $\Tr(\ad_X \ad_Y) = -2\hv \, \kappa(X,Y)$ (see footnote \ref{FootForm}), the flow \eqref{Eq:RgN} then takes the form
\begin{equation*}
\frac{\dd\;}{\dd \tau} \ps{X}{\Ec X} = \frac{8\hbar\hv\,\lambda^2}{(1-\lambda^2)^2(1+\lambda)^2} \Bigl( (1+\lambda^2) \, \kappa(X_1, X_2) - \lambda \,  \kappa(X_1, X_1) - \lambda\, \kappa(X_2, X_2)  \Bigr)+ O(\hbar^2)\,.
\end{equation*}
From the expression \eqref{Eq:FormELambda} of $\ps{X}{\Ec X}$ for the case at hand, we find that this induces the following flow on the parameters of the model:
\begin{equation}\label{Eq:RgLambda}
\frac{\dd\lambda}{\dd\tau} = -\frac{2\hbar\hv}{\kay} \frac{\lambda^2}{(1+\lambda)^2} + O(\hbar^2) \qquad \text{ and } \qquad \frac{\dd\kay}{\dd\tau} = O(\hbar^2)\,.
\end{equation}
In particular, we find that the Wess-Zumino level $\kay$ is not renormalised (at least at 1-loop): this is to be expected due to the 3-dimensional origin of the Wess-Zumino term. Moreover, we recover in this way the already known 1-loop $\beta$-function of the $\lambda$-model~\cite{Itsios:2014x}.\\

Given this RG-flow, one can look at the high-energy / UV limit of the model, corresponding to $\tau\to +\infty$. A standard analysis of the ODE \eqref{Eq:RgLambda} shows that this limit simply gives
\begin{equation}
\lambda \xrightarrow{\tau\to+\infty} 0\,.
\end{equation}
In particular, the UV limit of the action \eqref{Eq:ActionLambda} yields the WZW model action $S_{\text{WZW},\,\kay}[g]$, which defines a well-known conformal theory (as one should expect).

Let us now reinterpret this in view of the results of Subsection \ref{SubSec:ConfN}. There, we have formulated a strong conformal condition in terms of the eigenspaces $V_\pm$ and showed that this condition ensures the 1-loop conformal invariance of non-degenerate $\Ec$-models. In the present case, these eigenspaces are given in equation \eqref{Eq:VLambda}. The commutator between generic elements $X_+ = (\lambda\,X,X)$ and $Y_- = (Y,\lambda\,Y)$ of $V_+$ and $V_-$ (with $X,Y\in\g$) then reads $[X_+,Y_-]=\lambda\bigl([X,Y],[X,Y]\bigr)$. In particular, we see that the locus in parameter space where the strong conformal condition $[V_+,V_-]=\lbrace 0 \rbrace$ holds exactly coincides with the UV fixed-point $\lambda=0$ (one easily sees that $V_+$ and $V_-$ also become subalgebras of $\df$ at this point, as expected from the analysis of Subsection \ref{SubSec:ConfN}). This illustrates concretely how this strong conformal condition leads to conformal $\Ec$-models.

\paragraph{Kac-Moody currents of the $\bm\lambda$-model.} We have explained in Subsection \ref{SubSec:EModelsN} that the Hamiltonian degrees of freedom of a non-degenerate $\Ec$-model naturally form a $\df$-valued field $\Jc$ satisfying a Poisson current algebra. For the example of the $\lambda$-model considered in this subsection, the double algebra $\df$ is given by the direct product $\g\times\g$: the corresponding current Poisson algebra then splits into two decoupled subalgebras. To describe this explicitly, let us introduce $\g$-valued fields $\Kc_1$ and $\Kc_2$ through
\begin{equation}\label{Eq:JLambda}
\Jc = \frac{1}{\kay}\bigl( \Kc_1, -\Kc_2 \bigr),
\end{equation}
where the prefactor and the minus sign have been introduced for future convenience, to cancel contributions coming from the bilinear form \eqref{Eq:FormLambda}. We then find that $\Kc_1$ and $\Kc_2$ are commuting $\g$-Kac-Moody currents with levels $\ell_1=+\kay$ and $\ell_2=-\kay$, \textit{i.e.} fields satisfying two decoupled current Poisson algebras of $\g$. Explicitly, the bracket \eqref{Eq:PbJ} with the parametrisation \eqref{Eq:JLambda} implies
\begin{equation}\label{Eq:KmLambda}
\bigl\lbrace \Kc_r(x)\ti{1}, \Kc_s(y)\ti{2} \bigr\rbrace  = \delta_{rs} \bigl( \bigl[ \Cg, \Kc_r(x)\ti{1} \bigr] \,\delta(x-y) - \ell_r\,\Cg\,\p_x\delta(x-y) \bigr)\,, \qquad r,s\in\lbrace 1,2 \rbrace\,.
\end{equation}
Here $\Cg \in \g\otimes\g$ is the split quadratic Casimir of $\g$, defined as $\Cg = \kappa^{ab}\, {\tt t}_a \otimes {\tt t}_b$ in terms of a basis $\lbrace {\tt t}_a \rbrace_{a=1,\dots,\dim\g}$ of $\g$ and the inverse $\kappa^{ab}$ of $\kappa({\tt t}_a,{\tt t}_b)$.

In general, the expression of the current $\Jc$ in terms of the field of the $\s$-model is given by the formula \eqref{Eq:Jn}. In the present case, using equation \eqref{Eq:WLambda}, this yields
\begin{subequations}
\begin{align}
\Kc_1 &= \kay \left(  \frac{1}{1-\lambda\,\Ad_g^{-1}} g^{-1} \p_+ g + \frac{\lambda}{1-\lambda\,\Ad_g} \p_- gg^{-1}  \right) \,, \\
\Kc_2 &= -\kay \left( \frac{\lambda}{1-\lambda\,\Ad_g^{-1}} g^{-1} \p_+ g + \frac{1}{1-\lambda\,\Ad_g} \p_- gg^{-1} \right) \,.
\end{align}
\end{subequations}

\paragraph{Chiral Kac-Moody currents in the conformal limit.} Rather than using the decomposition $\df=\g\times\g$ to describe the current $\Jc$ in terms of $\Kc_1$ and $\Kc_2$, as we did in the previous paragraph, one can also use the components $\Jc_\pm \in V_\pm$, following the decomposition $\df=V_+ \oplus V_-$. In particular, we know from Subsections \ref{SubSec:EModelsN} and \ref{SubSec:ConfN} that this description is the most adapted one to reflect the light-cone structure of the model and study its conformal limit. For the $\lambda$-model under consideration, we find that the currents $\Jc_\pm$ are given by
\begin{equation}
\Jc_+ = (1 , \lambda ) \otimes \frac{1}{1-\lambda\,\Ad_g^{-1}} g^{-1} \p_+ g \qquad \text{ and } \qquad \Jc_- = - (\lambda,1) \otimes \frac{1}{1-\lambda\,\Ad_g} \p_- gg^{-1}\,,
\end{equation}
where we used the tensorial notation $(a,b)\otimes X = (aX,bX)$ to describe pure products in $\df \simeq \R^2 \otimes \g$.

One of the main advantage of the currents $\Jc_\pm$ is that they become chiral when the strong conformal condition is satisfied (see Subsection \ref{SubSec:ConfN}). Here, this corresponds to taking the UV limit $\lambda=0$, in which case we get
\begin{equation}
\Jc^\cL = \Jc_+ \bigl|_{\lambda=0}\, = (g^{-1}\p_+ g,0) \qquad \text{ and } \qquad \Jc^\cR = \Jc_- \bigl|_{\lambda=0} \, = -(0,\p_- gg^{-1})\,.
\end{equation}
In the above equation, we used the notations $\cL$ and $\cR$ to label left-moving and right-moving fields at the conformal point, as in Subsection \ref{SubSec:ConfN}. We recover this way standard results on the WZW model \eqref{Eq:WZW}: indeed, it is well-known that the dynamics of this theory is such that the fields $g^{-1}\p_+ g$ and $\p_- gg^{-1}$ are respectively left-moving and right-moving. In other words, the equations of motion at the conformal point $\lambda=0$ take the form
\begin{equation}\label{Eq:EoM-WZW}
\p_-\bigl( g^{-1} \p_+ g \bigr) = 0\,, \qquad \text{ or equivalently } \qquad \p_+\bigl( \p_-gg^{-1}  \bigr) = 0\,.
\end{equation}

As explained in Subsection \ref{SubSec:ConfN}, the chiral fields $\Jc^{\cLR}$ at the conformal point also form two decoupled closed Poisson algebras. More precisely, in the present case, $\Jc^\cL$ satisfies the current bracket associated with the Lie algebra $V_+ = \g \times \lbrace 0 \rbrace$ and $\Jc^\cR$ the one associated with $V_- = \lbrace 0 \rbrace \times \g$. In fact, we see that, at the conformal point $\lambda=0$, the left-moving field $\Jc^\cL$ essentially corresponds to the Kac-Moody current $\Kc_1$ introduced above, while the right-moving field $\Jc^\cR$ corresponds to $\Kc_2$. More precisely:
\begin{equation}
\Jc^\cL = \frac{1}{\kay} \bigl( \Kc_1|_{\lambda=0}, 0  \bigr) \qquad \text{ and } \qquad \Jc^\cR = \frac{1}{\kay} \bigl( 0, \Kc_2|_{\lambda=0} \bigr)\,.
\end{equation}
The current algebra obeyed by $\Jc^\cL$ and $\Jc^\cR$ is then equivalent to the Kac-Moody bracket \eqref{Eq:KmLambda}. To summarise, we have seen that the $\lambda$-model is always described by two commuting Kac-Moody currents $\Kc_1$ and $\Kc_2$, even for the non-conformal model, and that these currents become chiral in the conformal limit, giving back the standard chiral Kac-Moody algebras underlying the WZW model. This illustrates the general ideas of this section, namely that all non-degenerate $\Ec$-models are described by a current algebra and that the strong conformal condition further ensures that this current algebra splits into two chiral components.

\paragraph{Chiral symmetries of the WZW model.} The conformal structure of the WZW model is also reflected in the presence of chiral symmetries in its dynamics, which can be recovered from the general discussion at the end of Subsection \ref{SubSec:ConfN} -- see also~\cite{Arvanitakis:2022bnr}. There, it was argued that a non-degenerate $\Ec$-model satisfying the strong conformal condition is invariant under chiral symmetries \eqref{Eq:ChiralSym}. In the present case, recall that the subgroups of $D = G\times G$ whose Lie algebras are $V_+$ and $V_-$ are respectively $D^\cL= G \times \lbrace \Id \rbrace$ and $D^\cR=\lbrace \Id \rbrace \times G$. The chiral symmetries \eqref{Eq:ChiralSym} then act on $\ell(t,x)$ by multiplications $\ell(t,x) ( v^\cL(t+x), v^\cR(t-x) )$, where $v^\cL$ and $v^\cR$ are arbitrary functions valued in $G$. Recall moreover that in this subsection, we fixed the left $G_\diag$--gauge symmetry of the model by setting $\ell(t,x) = ( g(t,x), \Id )$. This gauge is not preserved by the above chiral action on $\ell$ and one thus have to perform a left multiplication by $\bigl( v^\cR(t-x), v^\cR(t-x)  \bigr)^{-1}\in G_\diag$ to bring $\ell$ back to the right form. In the end, we thus obtain the following transformation of the physical field $g$:
\begin{equation}
g(t,x) \longmapsto v^\cR(t-x)^{-1}\,g(t,x)\,v^\cL(t+x)\,.
\end{equation}
One easily checks that this transformation preserves the equations of motion \eqref{Eq:EoM-WZW} and thus defines a chiral symmetry of the WZW model, as expected from~\cite{Arvanitakis:2022bnr} and the general discussion in Subsection \ref{SubSec:ConfN}, and as already well-established in the literature.

\paragraph{From the strong conformal condition to the WZW structure.} We end this subsection by arguing that, under mild technical assumptions, any non-degenerate $\Ec$-model satisfying the strong conformal condition in fact takes the form of a WZW model, as pointed out in~\cite{Klimcik:1996hp}.\footnote{As explained in footnote \ref{Foot:ConfCond}, the setup considered in the present section is very similar to the one of reference~\cite{Klimcik:1996hp}, although there are a few minor differences. The approach that we will follow here to establish the relation with WZW models will be a bit different from the one of~\cite{Klimcik:1996hp} but will rely on the same crucial property, namely the complementarity of $\kf$ and $V_\pm$ in $\df$.} The main idea will be to show that the double algebra $\df$ of this model always have the structure of a direct product $\g\times\g$, as considered earlier in this subsection.

Let us then start with a general non-degenerate $\Ec$-model satisfying the strong conformal condition, so that $\df=V_+ \oplus V_-$ where $V_+$ and $V_-$ are orthogonal commuting subalgebras of dimension $d=\frac{1}{2}\dim\df$ (see Subsection \ref{SubSec:ConfN}). In order to obtain a conformal $\s$-model from it, one has to follow the recipe of Subsubsection \ref{SubSec:LagN}. In particular, one has to find a maximally isotropic subalgebra $\kf$ in $\df$ and construct the projector $W^\pm_\ell$, with kernel and image as in equation \eqref{Eq:WpmN}. For this projector to exist, one has to assume that these kernel and image have trivial intersection and thus in particular that $\kf\, \cap V_\pm = \lbrace 0 \rbrace$.\footnote{As mentioned in footnote \ref{Foot:W}, one can in principle ask for the property $\Ad_{\ell}^{-1}\kf \,\cap V_\pm = \lbrace 0 \rbrace$ to hold only for generic values of $\ell\in D$. In that more general setup, the point $\ell=\Id$ could be one of the non-generic values where this property fails and $\kf$ and $V_\pm$ would thus have a non-trivial intersection. The reasoning of this paragraph then fails and one could in principle get other types of theories than WZW models. It is however not clear whether such a case exists in practice.} We are thus in a situation where we have the vector spaces decompositions $\df=V_+ \oplus V_- = \kf \oplus V_\pm$, with $\dim V_\pm=\dim\kf=d$. With a little bit of linear algebra, one shows that in this configuration, there always exist bases $\lbrace U_a \rbrace_{a=1}^d$ and $\lbrace \Ub_a \rbrace_{a=1}^d$ of $V_+$ and $V_-$ such that
\begin{equation}
K_a = U_a + \Ub_a\,, \qquad a=1,\dots,d\,,
\end{equation}
form a basis of $\kf$. Using the fact that $V_+$, $V_-$ and $\kf$ are closed under the Lie bracket and that $[V_+,V_-]=\lbrace 0 \rbrace$, one further find that
\begin{equation}
[U_a, U_b] = f_{ab}^{\;\;\,c}\,U_c\,, \qquad [\Ub_a, \Ub_b] = f_{ab}^{\;\;\,c}\,\Ub_c\,, \qquad [K_a, K_b] = f_{ab}^{\;\;\,c}\,K_c\,,
\end{equation}
for some structure constants $f_{ab}^{\;\;\,c}$, which crucially are the same for the three subalgebras. To prove this, one starts with different structure constants for $U_a$ and $\Ub_a$ and write down the commutation relations of $K_a = U_a + \Ub_a$, using the fact that $[U_a,\Ub_b]=0$: requiring that these commutators close in $\kf$, we then find that the structure constants of $U_a$, $\Ub_a$ and $K_a$ all coincide. In particular, this means that the three subalgebras $V_+$, $V_-$ and $\kf$ are all isomorphic. To make the relation with the language of this subsection, we denote by $\g$ the abstract Lie algebra defined by their common commutation relations. It is then easy to see that we have an isomorphism $\df \simeq \g \times \g$, identifying the double algebra $\df$ with the direct product $\g\times\g$, in such a way that
\begin{equation}
V_+ \simeq \g \times \lbrace 0 \rbrace\,, \qquad V_- \simeq \lbrace 0 \rbrace \times \g \qquad \text{ and } \qquad
\kf \simeq \g_{\diag} = \bigl\lbrace (X,X), \; X\in \g \bigr\rbrace\,.
\end{equation}
We recognise here the structure of the $\Ec$-model underlying the WZW model on $G$, as described earlier in this subsection. To complete this identification, we finally have to study the image of the bilinear form $\psd$ through this isomorphism. Using the fact that $V_+$ and $V_-$ are orthogonal with respect to $\psd$ while $\kf$ is isotropic, one easily checks that the restriction of $\psd$ on $V_+ \simeq \g \times \lbrace 0 \rbrace$ induces a non-degenerate invariant bilinear form on $\g$ while the restriction on $V_- \simeq \lbrace 0 \rbrace \times \g$ yields the opposite of this form. Through the identification $\df\simeq \g\times\g$, the bilinear form $\psd$ is then sent to the form \eqref{Eq:FormLambda} considered earlier in this subsection. This achieves the identification of the conformal $\Ec$-model under consideration with the WZW model on $G$. We note that, in general, this group $G$ is not necessarily semi-simple (it only needs to have a Lie algebra with a non-degenerate invariant bilinear form).

\section{Conformal degenerate \texorpdfstring{$\bm\Ec$}{E}-models}
\label{Sec:D}

This section is the main component of the article and concerns the study of conformal and chiral structures in \textit{degenerate} $\Ec$-models. The latter~\cite{Klimcik:1996np} resemble the non-degenerate $\Ec$-models considered in the previous section but admit an additional gauge symmetry, generated by a subalgebra of the current Poisson bracket underlying the theory. At the Hamiltonian level, the physical observables of the theory are thus described by the Poisson reduction of a current algebra.

At the Lagrangian level, the degenerate $\Ec$-models are naturally related to $\s$-models, as was the case for the non-degenerate ones. More precisely, recall from the previous section that a non-degenerate $\Ec$-model with double group $D$ can be realised as a $\s$-model with target space $\KD$, where $K \subset D$ is a subgroup whose Lie algebra is maximally isotropic. In the degenerate setup, the target space of the $\s$-model becomes a double quotient $K \backslash D / H$, where $H \subset D$ is a subgroup encoding the additional gauge symmetry and whose Lie algebra is isotropic (but not maximal).

The first subsection below is a review of the construction of degenerate $\Ec$-models and of the related $\s$-models. The rest of the section is devoted to the study of conformal occurrences of such theories and the description of their chiral fields and Poisson algebras, generalising the results of the previous section to the degenerate case. Explicit examples of such conformal degenerate $\Ec$-models are discussed in the next section \ref{Sec:Examples}.

\subsection[Reminder about degenerate \texorpdfstring{$\Ec$}{E}-models]{Reminder about degenerate \texorpdfstring{$\bm\Ec$}{E}-models}
\label{SubSec:EModels}

The degenerate $\Ec$-models and their associated $\s$-models were first introduced in~\cite{Klimcik:1996np} (see also~\cite{Sfetsos:1999zm,Squellari:2011dg} for related constructions). Below, we review the definition and main properties of these theories, following mostly the more recent references~\cite{Klimcik:2019kkf,Klimcik:2021bjy,Klimcik:2021bqm} (in the literature, these models are also sometimes referred to as dressing cosets: for simplicity, we will only use the denomination degenerate $\Ec$-models in this article). Since they consist of a gauged version of the non-degenerate $\Ec$-models discussed in Subsection \ref{SubSec:EModelsN}, we will use the notations and results already introduced there to avoid repetition and will mainly focus on the new aspects related to the additional gauge symmetry.

\subsubsection{Hamiltonian formulation in terms of a reduced current Poisson algebra}
\label{SubSec:Ham}

\paragraph{Double Lie algebra and gauge subalgebra.} As in the non-degenerate setup, one of the main data defining a degenerate $\Ec$-model is a double Lie algebra $\df$ of dimension $2d$, equipped with an invariant symmetric bilinear form $\psd$ with split signature $(d,d)$ (see Subsection \ref{SubSec:EModelsN} for details and conventions). The key additional ingredient in the degenerate case is the choice of an isotropic subalgebra $\hf\subset \df$, which thus satisfies $[\hf,\hf]\subset \hf$ and $\ps{\hf}{\hf}=0$. Note that $\hf$ is not required to be maximally isotropic and thus can have any dimension smaller than $d$: we let $p=d-\dim\hf$, which is then a positive integer (as we will see later, this integer will correspond to the dimension of the target space of the $\s$-models associated to this degenerate $\Ec$-model).

We denote by $\hf^\perp$ the orthogonal subspace of $\hf$ in $\df$ with respect to $\psd$. Since $\hf$ is isotropic, we have $\hf \subset \hf^\perp$. Moreover, this orthogonal subspace has dimension $2d-\dim\hf=d+p$. The ad-invariance of $\psd$ implies that $\hf^\perp$ is stable under the adjoint action of $\hf$, \textit{i.e.} that $[\hf,\hf^\perp] \subset \hf^\perp$.

\paragraph{Current algebra and gauge symmetry.} As in the non-degenerate case, the degenerate $\Ec$-model is a Hamiltonian field theory described in terms of a $\df$-valued field $\Jc(x)=\Jc_A(x)T^A$, satisfying the current Poisson algebra \eqref{Eq:PbJA}, which we recall here for the reader's convenience:
\begin{equation}\label{Eq:PbJA2}
\bigl\lbrace \Jc_A(x), \Jc_B(y) \bigr\rbrace = \F{AB}C\,\Jc_C(x)\,\delta(x-y) - \eta_{AB}\,\p_x\delta(x-y)\,.
\end{equation}

However, for degenerate models, not all degrees of freedom in $\Jc$ are physical, due to the presence of a gauge symmetry, which we now describe. We choose the basis $\lbrace T_A \rbrace_{A=1}^{2d}$ of $\df$ such that the first $d-p$ elements $\lbrace T_\alpha \rbrace_{\alpha=1}^{d-p}$ form a basis of the subalgebra $\hf$ introduced earlier. In what follows, we will use greek indices $\alpha,\beta,\dots$ to label basis elements of $\hf$, while we keep upper-case latin letters $A,B,\dots$ for basis elements of the full double algebra $\df$, as before. Let us then consider the components
\begin{equation}
\Jc_\alpha(x) = \ps{T_\alpha}{\Jc(x)}\,, \qquad \alpha\in\lbrace 1,\dots,d-p\rbrace\,,
\end{equation}
of the current $\Jc(x)$. They form a Poisson subalgebra of \eqref{Eq:PbJA2}, namely
\begin{equation}\label{Eq:PbJH}
\bigl\lbrace \Jc_\alpha(x), \Jc_\beta(y) \bigr\rbrace = \F{\alpha\beta}\gamma\,\Jc_\gamma(x)\,\delta(x-y)\,.
\end{equation}
Here, the labels $\alpha,\beta,\gamma$ run only over $\lbrace 1,\dots,d-p\rbrace$ due to the closedness of $\hf$: in particular, the coefficients $\F{\alpha\beta}\gamma$ are the structure constants of $\hf$. Moreover, we see that there is no term proportional to $\p_x\delta(x-y)$ in the bracket \eqref{Eq:PbJH}: this is due to the isotropy of $\hf$ with respect to $\psd$.

We will interpret the components $\Jc_\alpha(x)$ as the generators of a local gauge symmetry. The corresponding infinitesimal action on any observable $\Oc$ built from $\Jc$ is given by
\begin{equation}\label{Eq:TransfO}
\delta_\epsilon \Oc = \int_0^{2\pi} \epsilon^\alpha(x)\,\bigl\lbrace \Jc_\alpha(x), \Oc \bigr\rbrace\,\dd x\,,
\end{equation}
where $\epsilon(x)=\epsilon^\alpha(x)\,T_\alpha$ is the infinitesimal local parameter of the symmetry, which is an arbitrary $\hf$-valued function. Since the generators $\Jc_\alpha(x)$ satisfy the current algebra \eqref{Eq:PbJH}, one sees that this transformation defines a local action of the Lie algebra $\hf$, in the sense that $[\delta_\epsilon,\delta_{\epsilon'}] = \delta_{[\epsilon,\epsilon']}$. In particular, it is straightforward to compute the action of this symmetry on the current $\Jc$ using the bracket \eqref{Eq:PbJA2}, yielding
\begin{equation}
\delta_\epsilon \Jc (x) = \bigl[ \Jc(x), \epsilon(x) \bigr] + \p_x \epsilon(x)\,.
\end{equation}
This infinitesimal $\hf$-transformation can be easily lifted to a local action of a group $H$ with $\text{Lie}(H)=\hf$:
\begin{equation}\label{Eq:GaugeJ}
\Jc(x) \longmapsto \Jc^h(x) = h(x)^{-1}\Jc(x)\,h(x) + h(x)^{-1}\p_x h(x)\,,
\end{equation}
where $h(x)$ is an arbitrary function valued in $H$.\\

The physical observables of the degenerate $\Ec$-model are obtained by the Hamiltonian reduction of the current Poisson algebra with respect to this local $H$-action. This means that we are treating the generators of the action as constraints that we set to zero, following the approach of Dirac~\cite{dirac1964lectures} (see also~\cite{Henneaux:1992ig} for a standard textbook on the subject):
\begin{equation}
\Jc_\alpha(x) \approx 0\,, \qquad \forall\,\alpha\in\lbrace 1,\dots,d-p \rbrace\,.
\end{equation}
Here, we use the notation `$\approx$' to designate weak equalities, which are true when the constraints are imposed. In contrast, the standard equality sign `$=$' will be reserved for strong equalities, which hold even when the constraints are not imposed. The reduced algebra is then obtained from the initial current algebra in two steps: we first consider the observables up to the constraints $\Jc_\alpha \approx 0$ (which mathematically corresponds to taking a quotient of the current algebra) and then restrict to observables which are invariant under the gauge transformation \eqref{Eq:GaugeJ} generated by these constraints.

An important property ensuring the consistency of this procedure is the fact that the constraints are first-class, meaning that their Poisson brackets weakly vanish:
\begin{equation}\label{Eq:FirstClass}
\bigl\lbrace \Jc_\alpha(x), \Jc_\beta(y) \bigr\rbrace \approx 0\,, \qquad \forall\,\alpha,\beta\in\lbrace 1,\cdots,d-p \rbrace\,.
\end{equation}
This is a direct consequence of the equation \eqref{Eq:PbJH}, which ensures that the local $\hf$-transformation \eqref{Eq:TransfO} generated by $\Jc_\alpha$ preserves the constraints $\Jc_\beta \approx 0$. We note that this first-class property reflects the fact that $\hf$ is an isotropic subalgebra of $\df$. Indeed, the isotropy of $\hf$ ensures that there are no components proportional to $\p_x\delta(x-y)$ in the Poisson brackets of the generators $\lbrace \Jc_\alpha \rbrace_{\alpha=1}^{d-p}$, while the closedness of $\hf$ under the Lie bracket ensures that the component proportional to $\delta(x-y)$ is expressed in terms of the generators themselves and thus vanishes weakly.

Let us quickly discuss the number of physical degrees of freedom obtained by this reduction. The initial current algebra is described by $\dim\df=2d$ fields (namely the components of the current $\Jc$). Imposing the constraints $\Jc_\alpha \approx 0$ eliminates $\dim\hf=d-p$ of these degrees of freedom. Moreover, the local symmetry with gauge group $H$ removes $\dim\hf=d-p$ additional ones. We thus end up with $2p$ degrees of freedom. This means that the reduced algebra can be thought of as a Poisson algebra generated by $2p$ fields. In practice, these fields can be difficult to describe and manipulate explicitly. It is thus often easier to work with the unreduced current $\Jc$, keeping in mind that physical observables are subject to the constraints $\Jc_\alpha \approx 0$ and the gauge symmetry \eqref{Eq:GaugeJ}.\\

We note that imposing the constraints $\Jc_\alpha = \ps{T_\alpha}{\Jc} \approx 0$ does not correspond to putting the $\hf$-valued component of $\Jc$ to zero\footnote{In fact, defining the $\hf$-valued component of $\Jc$ would first require choosing a complement of $\hf$ in $\df$, which is arbitrary.}. Rather, it corresponds to killing the component of $\Jc$ along any subspace dual to $\hf$ with respect to the bilinear form $\psd$. In particular, the constraint can be expressed as the following orthogonality condition:
\begin{equation}\label{Eq:JPerp}
\Jc \text{ weakly belongs to } \hf^\perp\,.
\end{equation}
Recall from above that $\hf^\perp$ is stable under the adjoint action of $\hf$ and thus under the lifted adjoint action of the group $H$: in other words, $h^{-1} X \,h \in \hf^\perp$ for all $h\in H$ and $X\in\hf^\perp$. Moreover, by isotropy of $\hf$, we have $\hf \subset \hf^\perp$, showing that $h^{-1}\p_x h$ belongs to $\hf^\perp$ if $h$ is a $H$-valued function. We thus conclude that the gauge transformed current $h^{-1}\Jc\,h + h^{-1}\p_x h$ also weakly belongs to $\hf^\perp$. This is another way of stating the preservation of the constraints under the gauge transformation \eqref{Eq:GaugeJ}.

\paragraph{Hamiltonian and operator $\bm\Eh$.} Let us now define the Hamiltonian $\Hc$ of the theory. As in the non-degenerate case, we will choose the density of $\Hc$ to be quadratic in the components of the current $\Jc$. However, in the degenerate case, the presence of a constraint and a gauge symmetry requires two additional important considerations. Firstly, the only physically relevant information in the Hamiltonian is its weak definition, \textit{i.e.} its expression when the constraint is imposed. In particular, different choices for the Hamiltonian that weakly coincide will give rise to the same theory. Recall from equation \eqref{Eq:JPerp} that imposing the constraint amounts to restricting $\Jc$ to the subspace $\hf^\perp$. Defining a weak quadratic Hamiltonian then exactly becomes specifying a choice of symmetric bilinear form $E(\cdot,\cdot)$ on $\hf^\perp$ and letting
\begin{equation}\label{Eq:HWeak}
\Hc \approx \frac{1}{2} \int_0^{2\pi} E\bigl(\Jc(x),\Jc(x) \bigr)\, \dd x\,.
\end{equation}
The second additional consideration to take into account in the degenerate case is that this Hamiltonian should be invariant under the gauge transformation \eqref{Eq:GaugeJ} of $\Jc$ (which we recall weakly preserves the fact that $\Jc$ belongs to $\hf^\perp$). This condition is equivalent to the bilinear form $E(\cdot,\cdot)$ being invariant under the adjoint action of $H$ on $\hf^\perp$ and having a kernel containing $\hf$ (\textit{i.e.} $E(\hf,\cdot)=0$). In particular, this shows that $E(\cdot,\cdot)$ is degenerate as a bilinear form: this is a manifestation of the ``degenerate'' nature of the $\Ec$-model under construction here.\\

In analogy with the non-degenerate case, it is useful to represent the bilinear form $E(\cdot,\cdot)$ as $\langle\cdot,\Eh\cdot\rangle$, with $\Eh:\hf^\perp\to\hf^\perp$ a linear operator. This is always possible but in fact does not uniquely define $\Eh$. From a practical perspective, it is often useful to give a strong definition of the Hamiltonian $\Hc$ rather than just the weak one \eqref{Eq:HWeak}. This can be done in practice by extending the operator $\Eh$ from $\hf^\perp$ to the full double algebra $\df$ and defining
\begin{equation}\label{Eq:H}
\Hc = \frac{1}{2} \int_0^{2\pi} \psb{\Jc(x)}{\Eh(\Jc(x))}\,\dd x = \frac{1}{2}  \int_0^{2\pi} \Eh^{AB} \, \Jc_A(x)\,\Jc_B(x)\,\dd x\,,
\end{equation}
where $\Eh^{AB} = \psb{T^A}{\Eh (T^B)}$, in the dual basis $\lbrace T^A \rbrace_{A=1}^{2d}$ of $\df$. The non-unicity in defining such an extended operator $\Eh:\df \to \df$ encodes the freedom of adding to the strong expression of the Hamiltonian any term proportional to the constraints $\Jc_\alpha$ (indeed, such a term will not change the weak expression \eqref{Eq:HWeak} of $\Hc$). In the following, we suppose that the operator $\Eh: \df \to \df$ satisfies:\vspace{-3pt}
\begin{enumerate}[(i)]\setlength\itemsep{0.5pt}
\item $\Eh$ is symmetric with respect to $\psd$ ;
\item $\Ker\bigl(\Eh\bigl|_{\hf^\perp}\bigr) = \hf$ ;
\item $\Eh\bigl|_{\hf^\perp}$ commutes with the adjoint action of $H$ ;
\item $\Eh\bigl|_{\hf^\perp}^3=\Eh\bigl|_{\hf^\perp}$.
\end{enumerate}
The second of these properties means that the bilinear form $\langle\cdot,\Eh\,\cdot\rangle$ on $\df$ is degenerate and that the gauge subalgebra $\hf$ is contained in its kernel, extending to the full algebra $\df$ the result stated above for the restriction $\langle\cdot,\Eh\,\cdot\rangle|_{\hf^\perp}=E(\cdot,\cdot)$ on $\hf^\perp$. This can be traced back to the degenerate nature of the $\Ec$-model under consideration and thus to the presence of a gauge symmetry. The latter is also reflected in the condition (iii), which translates the invariance of $\langle\cdot,\Eh\,\cdot\rangle|_{\hf^\perp}$ under the adjoint action of $H$. The property (iv) essentially corresponds to the requirement that the field theory defined by this choice of Hamiltonian is relativistic and is the degenerate equivalent of the requirement that $\Ec^2=\Id$ in the non-degenerate case (see the next paragraph for more details).

\paragraph{Decomposing $\bm\df$.} The above conditions (ii) and (iv) obeyed by $\Eh$ imply that
\begin{equation}\label{Eq:DecoPerp}
\hf^\perp = \hf \oplus V_+ \oplus V_-\,, \qquad \text{ with } \qquad V_\pm = \Ker\bigl(\Eh\bigl|_{\hf^\perp} \mp \,\Id \bigr)\,.
\end{equation}
In particular, $\Eh$ is an involution (\textit{i.e.} it squares to the identity) when restricted to $V_+ \oplus V_-$. Similarly to the non-degenerate case, this property is the condition ensuring the relativistic invariance of the theory. The particularity of the degenerate setup is that it only holds on a subspace which is a complement of $\hf$ in $\hf^\perp$, while it held on the full double algebra $\df$ in the non-degenerate one. In the rest of this section, we will make the additional assumption that the two subspaces $V_\pm$ have the same dimension, which is then equal to $\dim V_\pm = \frac{1}{2}(\dim\hf^\perp-\dim\hf) = p$. Since $\Eh$ is symmetric, we also find that $V_+$ and $V_-$ are orthogonal with respect to $\psd$ and that $\psd$ is non-degenerate on $V_\pm$.\footnote{Let us note that the Hamiltonian $\Hc$ of the theory is positive, and thus bounded below, if and only if $\psd|_{V_+}$ and $\psd|_{V_-}$ are respectively positive definite and negative definite. At the Lagrangian level, this setup will correspond to $\s$-models whose target space is of euclidean signature. Although this condition is often taken as an additional assumption in the literature, we will not impose it here to also include $\s$-models with non-euclidean signature.}

It will be useful to complete the above decomposition of $\hf^\perp$ to one for the full double algebra. For that, we observe that there exists a unique subspace $\hf' \subset \df$ which is of dimension $\dim\hf'=\dim\hf$, isotropic with respect to $\psd$ and orthogonal to $V_+ \oplus V_-$. Since $\psd$ is non-degenerate, this subspace has to pair non-degenerately with $\hf$ and thus should be a complement of $\hf^\perp$, hence $\df = \hf \oplus \hf' \oplus V_+ \oplus V_-$. Using the symmetry of $\Eh$ and the fact $\Eh(\hf) = \lbrace 0 \rbrace$, we find that $\Eh(\hf') \subset \hf$. The condition (iii) obeyed by $\Eh$, \textit{i.e.} the fact that $\Eh\bigl|_{\hf^\perp}$ commutes with the adjoint action of $H$, also has consequences on this decomposition. Indeed, it implies that $V_+$, $V_-$ and $\hf'$ are all stable under this adjoint action. Finally, one checks that $[V_+,V_-] \subset \hf^\perp$.\\

Let us summarise the results of this paragraph. We have the decomposition
\begin{equation}\label{Eq:DecoD}
\df = \hf \oplus \hf' \oplus V_+ \oplus V_-\,,
\end{equation}
with $\dim\hf=\dim\hf'=d-p$ and $\dim V_+ = \dim V_- = p$. Moreover, we have
\begin{equation}\label{Eq:PsDeco}
\ps{\hf}{\hf} = \langle\hf',\hf'\rangle = 0\, \qquad \ps{\hf}{V_\pm} =  \langle\hf',V_\pm\rangle = 0\,, \qquad \ps{V_+}{V_-}=0\,,
\end{equation}
while $\psd$ pairs non-degenerately $\hf$ with $\hf'$, $V_+$ with itself and $V_-$ with itself. In particular, the subspace $\hf'$ is naturally identified with the dual $\hf^\ast$ of $\hf$ through this pairing. The action of the operator $\Eh$ on this decomposition satisfies
\begin{equation}\label{Eq:EDeco}
\Eh|_{\hf} = 0\,, \qquad \Eh(\hf') \subset \hf \qquad \text{ and } \qquad \Eh|_{V_\pm} = \pm \Id\, .
\end{equation}
Similarly, the adjoint action of $H$ is such that, for any $h\in H$,
\begin{equation}\label{Eq:AdHDeco}
\Ad_h \,\hf = \hf\,, \qquad \Ad_h\, \hf' = \hf' \qquad \text{ and } \qquad \Ad_h\, V_\pm = V_\pm \,.
\end{equation}
Finally, we have
\begin{equation}\label{Eq:ComPm}
[V_+,V_-] \subset \hf^\perp = \hf \oplus V_+ \oplus V_-\,.
\end{equation}
We will denote by $\pi_{\hf}$, $\pi_{\hf'}$ and $\pi_\pm$ the projections on $\hf$, $\hf'$ and $V_\pm$ along the decomposition \eqref{Eq:DecoD}.

\paragraph{Decomposing $\bm\Jc$.} Let us now consider the decomposition of the $\df$-valued current $\Jc$ along the direct sum \eqref{Eq:DecoD}:
\begin{equation}
\Jc(x) = \Jc_{\hf}(x) + \Jc'(x) + \Jc_+(x) - \Jc_-(x)\,,
\end{equation}
with $\Jc_{\hf}=\pi_{\hf}(\Jc)\in\hf$, $\Jc'=\pi_{\hf'}(\Jc)\in\hf'$ and $\Jc_\pm =\pm\pi_{\pm}(\Jc) \in V_\pm$ (the minus sign in $\Jc_-$ has been introduced for further convenience). Since $\hf$ is orthogonal to $\hf \oplus V_+ \oplus V_-$ with respect to $\psd$, we find that $\Jc_\alpha = \ps{ T_\alpha}{ \Jc} = \ps{ T_\alpha}{ \Jc' }$, where we recall that $\lbrace T_\alpha \rbrace_{\alpha=1}^{d-p}$ is a basis of $\hf$. Thus, imposing the constraints $\Jc_\alpha \approx 0$ amounts to requiring
\begin{equation}
\Jc'(x) \approx 0\,.
\end{equation}
The field $\Jc'(x)$ valued in $\hf' \simeq \hf^\ast$ can be understood as the generator/moment-map of the $\hf$-gauge symmetry, which is put to zero when we perform the corresponding Hamiltonian reduction. In particular, the gauge-transformation \eqref{Eq:TransfO} of an observable $\Oc$ with local infinitesimal parameter $\epsilon(x)\in\hf$ can be rewritten in a basis-independent way in terms of $\Jc'(x)$, as
\begin{equation}\label{Eq:TransfO2}
\delta_\epsilon \Oc = \int_0^{2\pi} \ps{ \epsilon(x)}{\bigl\lbrace \Jc'(x), \Oc \bigr\rbrace}\,\dd x\,,
\end{equation}
using the fact that $\psd$ defines a non-degenerate pairing on $\hf \times \hf'$. We note that, weakly,
\begin{equation}\label{Eq:DecoJw}
\Jc(x) \approx \Jc_{\hf}(x) + \Jc_+(x) - \Jc_-(x)\,.
\end{equation}
This expression belongs to $\hf\oplus V_+ \oplus V_- = \hf^\perp$, in agreement with equation \eqref{Eq:JPerp}.\\

Let us consider the gauge transformation \eqref{Eq:GaugeJ} of the current $\Jc(x)$ by a local $H$-valued parameter $h(x)$. Using the fact that $h^{-1}\p_x h$ is valued in $\hf$ and the property \eqref{Eq:AdHDeco} of the adjoint action of $H$, we find that the components of the current transform as
\begin{equation}\label{Eq:GaugeDeco}
\Jc_{\hf} \longmapsto \Jc_{\hf}^h = h^{-1}\Jc_{\hf}\,h + h^{-1}\p_x h\,, \qquad \Jc_\pm \longmapsto \Jc_\pm^h = h^{-1}\Jc_{\pm}\,h\,, \qquad \text{ and } \qquad \Jc' \longmapsto h^{-1}\Jc'\,h\,.
\end{equation}
In particular, this confirms that the constraint $\Jc' \approx 0$ is preserved under gauge transformations, as previously observed. Moreover, we see that the component $\Jc_{\hf}$ behaves as the spatial part of a $\hf$-valued gauge-field while the components $\Jc_\pm$ transform covariantly.

\paragraph{Dynamics and Lagrange multiplier.} We now come to the description of the dynamics of the degenerate $\Ec$-model. It is prescribed by the choice of Hamiltonian $\Hc$ made in equation \eqref{Eq:H}. If $\Oc$ is a gauge-invariant observable, its time evolution is given by the standard formula $\p_t \Oc \approx \lbrace \Hc, \Oc \rbrace$, where the equality should be considered weakly. The gauge-invariance of $\Hc$ and $\Oc$ means that they weakly Poisson-commute with the constraints $\Jc_\alpha$ and thus that $\lbrace \Hc, \Oc \rbrace$ is insensible to adding to $\Hc$ or $\Oc$ any term proportional to the constraints. This is natural, since the physical observables are defined only up to the constraints. In other words, $\lbrace \Hc, \cdot \rbrace$ defines a consistent dynamics on the reduced algebra.\\

The above discussion defines the time-evolution of gauge-invariant observables. However, it is often useful to also consider the dynamics of non-gauge-invariant quantities. For instance, in the present case, it will be quite simpler to write the time-evolution of the model in terms of the current $\Jc$, which is not gauge-invariant, since the physical degrees of freedom contained in $\Jc$ are not easily identifiable in general. It is tempting to define the time evolution of any non-gauge-invariant observable $\Oc$ by the standard formula $\lbrace \Hc, \Oc \rbrace$. This however poses two problems. Firstly, as $\Oc$ is non-gauge-invariant, it does not Poisson commute with the constraints and the bracket $\lbrace \Hc, \Oc \rbrace$ is thus sensitive to adding a term proportional to the constraints to the Hamiltonian, which is incompatible with the construction. Secondly, $\Oc$ contains non-physical degrees of freedom whose time-evolution should thus involve arbitrary functions of the time coordinate: such functions do not seem to appear in the bracket $\lbrace \Hc, \Oc \rbrace$. Following the standard approach of Dirac~\cite{dirac1964lectures,Henneaux:1992ig}, the solution to both these problems is given by introducing a new ingredient in the theory, the so-called Lagrange multiplier. The latter is a field $\mu$ valued in the gauge subalgebra $\hf$. It is a new degree of freedom of the theory, independent of the ones contained in the current $\Jc$, whose physical interpretation we will discuss below.

Given this new field, we introduce the so-called total Hamiltonian of the model as
\begin{equation}
\Hc_T = \Hc + \int_0^{2\pi} \ps{\mu}{\Jc'}\,\dd x\,,
\end{equation}
where we recall that the field $\Jc'$, valued in $\hf' \simeq \hf^\ast$, is the generator of the gauge-symmetry. Since the constraint amounts to setting $\Jc' \approx 0$, we see that $\Hc_T$ reduces to our initial choice of Hamiltonian $\Hc$ weakly. We then define the time evolution of any (potentially non-gauge-invariant) quantity $\Oc$ as
\begin{equation}\label{Eq:Dyn}
\p_t \Oc \approx \lbrace \Hc_T, \Oc \rbrace \approx \lbrace \Hc, \Oc \rbrace + \int_0^{2\pi} \ps{\mu}{ \bigl\lbrace \Jc', \Oc \bigr\rbrace}\,\dd x\,.
\end{equation}
The key point here is that the dynamics is given by the flow of the total Hamiltonian $\Hc_T$, which contains the Lagrange multiplier $\mu$, and not the initial Hamiltonian $\Hc$ built only in terms of $\Jc$. In particular, the presence of the new field $\mu$ in $\Hc_T$ encodes the freedom of adding to the generator of time-evolution an arbitrary term proportional to the constraint: such a transformation simply amounts to a redefinition of $\mu$. This solves the first problem raised above in defining the dynamics of a non-gauge-invariant quantity.

To understand the physical interpretation of the Lagrange multiplier, let us compare the last term in the time evolution \eqref{Eq:Dyn} with the equation \eqref{Eq:TransfO2}. We then see that this term exactly takes the form of a gauge transformation with infinitesimal parameter $\mu\in\hf$. We thus conclude that the presence of the Lagrange multiplier in the dynamics \eqref{Eq:Dyn} represents the freedom of performing arbitrary gauge transformations along the time evolution of the model. Once the differential equation \eqref{Eq:Dyn} with respect to $t$ is integrated, this will lead to the presence of arbitrary functions in the time dependence of non-gauge-invariant quantities, reflecting the non-physicality of these degrees of freedom. This solves the second problem raised earlier when discussing the dynamics of non-gauge-invariant quantities. If $\Oc$ is a gauge-invariant observable, which then Poisson-commutes with the constraint $\Jc'$, the last term in equation \eqref{Eq:Dyn} disappears and we recover the temporal flow $\p_t \Oc \approx \lbrace \Hc, \Oc \rbrace$ considered earlier for gauge-invariant quantities. In that case, the Lagrange multiplier drops out of the time dependence of $\Oc$, as expected since we are now dealing with purely physical degrees of freedom.\\

Having defined the dynamics of the theory, we can now see all its fields, \textit{i.e.} the current $\Jc$ and the Lagrange multiplier $\mu$, as being dependent on the time coordinate $t$ as well, making them functions of $(t,x)$. In particular, we can now check that the equation of motion \eqref{Eq:Dyn} of the model is gauge-invariant, as one should expect. Although so far we were working on a fixed time-slice and thus considered gauge transformations with a local parameter $h(x)\in H$ depending only on the space variable $x$, we now want to include gauge transformations which are also time-dependent. For instance, we now consider the following transformation of the current $\Jc$:
\begin{equation}
\Jc(t,x) \longmapsto h(t,x)^{-1} \Jc(t,x) h(t,x) + h(t,x)^{-1} \p_x h(t,x)\,,
\end{equation}
where $h(t,x)$ is an arbitrary function on space-time valued in the group $H$. The equation of motion \eqref{Eq:Dyn} is invariant under this gauge symmetry if we also take into account the fact that the Lagrange multiplier $\mu(t,x)$ transforms as well under the symmetry, in the following way~\cite{dirac1964lectures,Henneaux:1992ig}:
\begin{equation}\label{Eq:GaugeMu}
\mu(t,x) \longmapsto h(t,x)^{-1} \mu(t,x) h(t,x) + h(t,x)^{-1} \p_t h(t,x)\,.
\end{equation}
In particular, we see that the presence of $\mu$ in the dynamics is necessary to recover the gauge-invariance under local transformations which are explicitly time-dependent. This is consistent with the above interpretation of $\mu$ as encoding the arbitrary time-dependence of non-gauge-invariant fields.

\paragraph{Equation of motion of the current.} Now that we have defined the dynamics of the degenerate $\Ec$-model, let us write down the equation of motion of the current $\Jc(t,x)$. Applying equation \eqref{Eq:Dyn} and using the current Poisson bracket, we find
\begin{equation}\label{Eq:dtJ}
\p_t \Jc \approx \p_x \bigl( \Ec\Jc + \mu \bigr) + [ \Jc, \Ec\Jc + \mu ]\,.
\end{equation}
We note that this takes the form of a zero curvature equation
\begin{equation}\label{Eq:Zce}
\bigl[ \p_x + \Jc , \p_t + \Ec\Jc + \mu \bigr] \approx 0\,.
\end{equation}
It will be useful to decompose the $\df$-valued equation \eqref{Eq:dtJ} along the direct sum $\df = \hf' \oplus \hf \oplus V_+ \oplus V_-$. The operator $\Eh$ acts on this decomposition according to equation \eqref{Eq:EDeco}. Moreover recall that imposing the constraint amounts to setting $\Jc' \approx 0$ in $\Jc=\Jc'+\Jc_\hf+\Jc_+-\Jc_-$. Using these properties, we can rewrite the commutator in equation \eqref{Eq:dtJ} as
\begin{equation*}
[ \Jc, \Ec\Jc + \mu ] \approx [ \Jc_\hf, \mu ] - [\mu - \Jc_\hf, \Jc_+] + [\mu + \Jc_\hf, \Jc_-] + 2 [\Jc_+, \Jc_-]\,.
\end{equation*}
From the equations \eqref{Eq:AdHDeco} and \eqref{Eq:ComPm}, we deduce that the four commutators in the above equation are respectively valued in $\hf$, $V_+$, $V_-$ and $\hf^\perp=\hf \oplus V_+ \oplus V_-$. We can now easily decompose the equation \eqref{Eq:dtJ}. Let us start with its component along $\hf'$: it is clear that it simply reads
\begin{equation}
\p_t \Jc' \approx 0\,.
\end{equation}
This shows that the definition of the dynamics \eqref{Eq:Dyn} is compatible with the constraint $\Jc' \approx 0$, as one should expect. Let us now turn to the $\hf$-valued component: we find
\begin{equation}
\p_t \Jc_\hf \approx \p_x \mu + [\Jc_\hf, \mu] + 2\pi_{\hf} [\Jc_+,\Jc_-]\,,
\end{equation}
where we recall that $\pi_\hf$ is the projector along $\hf$.  Finally, for the components in $V_\pm$, we get
\begin{equation}
\p_t \Jc_\pm = \pm \p_x \Jc_\pm - [\mu \mp \Jc_\hf, \Jc_\pm] \pm 2 \pi_{\pm} [\Jc_+,\Jc_-]\,.
\end{equation}

It is instructive to rewrite the above equations  using light-cone coordinates $x^\pm = t \pm x$ and their derivatives $\p_\pm = \frac{1}{2} (\p_t \pm \p_x)$, as well as the fields
\begin{equation}\label{Eq:Apm}
\Ac_\pm = \frac{1}{2} (\mu \pm \Jc_\hf)\,.
\end{equation}
Indeed, we then get
\begin{subequations}\label{Eq:EoMpm}
\begin{eqnarray}
&\p_+ \Ac_- - \p_- \Ac_+ + [\Ac_+, \Ac_-] = -\pi_\hf [\Jc_+,\Jc_-]\,, \label{Eq:EoMA} \\
&\p_- \Jc_+ + [\Ac_-, \Jc_+] = + \pi_+ [\Jc_+,\Jc_-]\,, \label{Eq:EomJp} \\
&\p_+ \Jc_- + [\Ac_+, \Jc_-] = - \pi_- [\Jc_+,\Jc_-]\,.\label{Eq:EomJm} 
\end{eqnarray}
\end{subequations}
We note that the index $\pm$ in $\Ac_\pm$ does not mean that these fields are valued in $V_\pm$: in fact, they belong to the gauge subalgebra $\hf$. Rather, we interpret them as the light-cone components of a $\hf$-valued connection $(\p_+ + \Ac_+,\p_- + \Ac_-)$. The equation of motion \eqref{Eq:EoMA} then concerns the curvature of this connection, while the equations \eqref{Eq:EomJp} and \eqref{Eq:EomJm} describe the covariant light-cone derivatives of $\Jc_+$ and $\Jc_-$ with respect to this connection. We observe moreover that the zero-curvature equation \eqref{Eq:Zce} can be rephrased in light-cone terms as the flatness of the connection $
\p_\pm + \Ac_\pm + \Jc_\pm$.\\

Let us finally discuss the gauge-invariance of the light-cone field equations \eqref{Eq:EoMpm}. Recall the form of the gauge transformation \eqref{Eq:GaugeDeco} of the currents $\Jc_{\hf}$ and $\Jc_\pm$ as well as the one \eqref{Eq:GaugeMu} of the Lagrange multiplier $\mu$. The gauge symmetry then acts on $(\Ac_\pm,\Jc_\pm)$ as
\begin{equation}\label{Eq:GaugePm}
\Ac_{\pm} \longmapsto \Ac_{\pm}^h = h^{-1}\Ac_{\pm}\,h + h^{-1}\p_\pm h \qquad \text{ and } \qquad \Jc_\pm \longmapsto \Jc_\pm^h = h^{-1}\Jc_{\pm}\,h\,.
\end{equation}
The $\hf$-valued current $\Ac_\pm$ can thus be interpreted as a gauge field while the currents $\Jc_\pm$ are covariant fields valued in the representations $V_\pm$ of $H$. The left-hand sides of the light-cone field equations \eqref{Eq:EoMpm} involve the curvature of the gauge field $\Ac_\pm$ and the covariant derivatives of $\Jc_\pm$: they thus transform covariantly, \textit{i.e.} by conjugation by $h$. The right-hand sides of these equations share the same transformation rule since the adjoint action of $h$ commutes with the projectors $\pi_{\hf}$ and $\pi_\pm$ by equation \eqref{Eq:AdHDeco}. This shows that the equations of motion in light-cone form \eqref{Eq:EoMpm} are manifestly gauge-invariant, as expected.\footnote{For completeness, let us compare this discussion to the one around~\cite[Equation (3.21)]{Klimcik:2021bqm}. The fields $(A_\pm,j_\pm)$ in this reference essentially correspond to the ones $(\Ac_\pm,\Jc_\pm)$ in the present paper, up to signs and factors of 2 due to differences in conventions. However, we note that in~\cite{Klimcik:2021bqm}, the time component of the gauge field $\Ac_t=\Ac_++\Ac_-$ vanishes, while here it coincides with the Lagrange multiplier $\mu$. This is due to the fact that such a Lagrange multiplier was not included in the definition of the dynamics in~\cite{Klimcik:2021bqm}. As observed there, this corresponds to working in a gauge where $\Ac_t=0$: this is always possible but in fact only fixes the gauge partially. Indeed, it is preserved by gauge transformations where the local parameter $h\in H$ is time-independent, \textit{i.e.} depends on $x$ but not on $t$. The field equations in~\cite{Klimcik:2021bqm} are then only invariant under these restricted transformations. This is consistent with the interpretation of the Lagrange multiplier as encoding the time-dependence in gauge transformations, as explained below equation \eqref{Eq:Dyn}.}

\subsubsection[Lagrangian formulation and gauged \texorpdfstring{$\s$}{sigma}-models]{Lagrangian formulation and gauged \texorpdfstring{$\bm \s$}{sigma}-models}\label{SubSec:Lag}

\paragraph{Setup.} Let us now describe the $\s$-models associated with the degenerate $\Ec$-model constructed above. The starting point will be quite similar to the one discussed in Subsubsection \ref{SubSec:LagN} for the non-degenerate case: there, we considered $\s$-models on quotients $\KD$, where $D$ is the double group with Lie algebra $\text{Lie}(D)=\df$ and $K$ is a subgroup of $D$ whose Lie algebra $\text{Lie}(K)=\kf$ is maximally isotropic with respect to $\psd$. In the degenerate case, we also start with a similar setup: the main difference will be that the resulting $\s$-model will have an additional local symmetry with gauge group $H$ and thus that the physical target space of the theory will be the double quotient $\KDH$.

Below, we will use the definitions, notations and conventions of Subsubsection \ref{SubSec:LagN}: in particular, $\kf = \text{Lie}(K)$ is a maximally isotropic subalgebra of $\df$, which is then of dimension $\frac{1}{2}\dim\df=d$. Moreover, we will describe the $\s$-model in terms of a field $\ell(t,x)$ on $\R\times\mathbb{S}^1$, valued in the double Lie group $D$. As in the previous subsubsection, we will not recall in details the points which are common to the degenerate and non-degenerate cases and will instead focus on the additional considerations that arise in the degenerate case, in particular the presence of the $H$-gauge symmetry.

\paragraph{Projectors $\bm{W_\ell^\pm}$.} Let us recall from the previous subsection that $\hf$ is a subalgebra of $\df$ of dimension $\dim\hf=d-p$ and that $V_\pm$ is a subspace of dimension $\dim V_\pm = p$. We introduce projectors $W^\pm_\ell$ on $\df$ characterised by their kernels and images
\begin{equation}\label{Eq:Wpm}
\Ker\bigl( W^\pm_\ell ) = \Ad_\ell^{-1} \, \kf \qquad \text{ and } \qquad \text{Im}\bigl( W^\pm_\ell ) = \hf \oplus V_\pm\,.
\end{equation}
In particular, we note that $\dim ( \Ad_\ell^{-1} \, \kf ) = \dim \kf = d$ and $\dim (\hf \oplus V_\pm) = \dim \hf + \dim V_\pm = d$, so that these dimensions sum to $\dim\df=2d$, as required for the projectors $W^\pm_\ell$ to exist\footnote{This is a necessary but not sufficient condition for the existence of $W^\pm_\ell$. In addition, we have to assume that the subspaces $\Ad_\ell^{-1} \, \kf$ and $\hf \oplus V_\pm$ have trivial intersection. Here, we will suppose that this holds, at least for generic values of the field $\ell\in D$. The points in $D$ where this fails (if they exist) will correspond to singularities in the resulting $\s$-model's target space.}. These projectors are the analogues of the ones defined in equation \eqref{Eq:WpmN} for non-degenerate $\Ec$-models: the main difference in the present case is that the image of $W^\pm_\ell$ is not composed only by the subspace $V_\pm$ but also contains the gauge subalgebra $\hf$. This image is contained in $\hf^\perp$, so that we have the property
\begin{equation}\label{Eq:PsWh}
\ps{W^\pm_\ell X}{Y} = 0\,, \qquad \forall\,X\in\df, \; Y \in \hf\,.
\end{equation}

One can determine the transpose of $W_\ell^\pm$ with respect to $\psd$ using the isotropy of $\kf$ and the orthogonality properties \eqref{Eq:PsDeco} of $\hf$ and $V_\pm$. We find $
\Tp W_\ell^\pm = 1 - W_\ell^\mp$, as in the non-degenerate case.

\paragraph{The $\bm\s$-model action.} In terms of the $D$-valued field $\ell(t,x)$, the action of the $\s$-model associated with the degenerate $\Ec$-model is given by
\begin{equation}\label{Eq:Action}
S[\ell] = \iint_{\R\times\mathbb{S}^1} \dd t\, \dd x\; \Bigl( \ps{W^+_\ell\,\ell^{-1} \p_+ \ell}{ \ell^{-1} \p_- \ell} - \ps{\ell^{-1} \p_+ \ell}{ W^-_\ell\,\ell^{-1} \p_- \ell} \Bigr) - \Wd{\ell}\,,
\end{equation}
where we recall that $\Wd{\ell}$ is the Wess-Zumino term \eqref{Eq:WZD} of the field $\ell$. This action takes the exact same form as in the non-degenerate case \eqref{Eq:ActionN}. The only difference here is in the construction of the projectors $W^\pm_\ell$: as explained above, the latter indeed possess slightly different properties in the degenerate case. This will in the end be responsible for the presence of the additional $H$-gauge symmetry in the theory (see next paragraph).

\paragraph{$\bm K$- and $\bm H$-gauge symmetries.} Let us now discuss the gauge symmetries of the $\s$-model \eqref{Eq:Action}. It is invariant under the local transformations
\begin{equation}\label{Eq:GaugeKH}
\ell(t,x) \longmapsto k(t,x)\, \ell(t,x)\,h(t,x)\,, \qquad \text{ with } \qquad k(t,x) \in K \quad \text{and} \quad h(t,x) \in H\,.
\end{equation}
It thus possesses a left $K$-gauge symmetry and a right $H$-gauge symmetry. The first one is exactly the same as in the non-degenerate case, see equation \eqref{Eq:GaugeK}. In particular, it can be proven in a very analogous way: we thus refer to the discussion below equation \eqref{Eq:GaugeK} for the proof. In contrast, the $H$-gauge symmetry is a new feature of the degenerate case, which reflects, at the Lagrangian level, the Hamiltonian reduction of the current Poisson algebra. Let us quickly prove it: we thus consider the transformation $\ell \mapsto \ell h$ with $h$ a $H$-valued field. We first note that
\begin{equation}
(\ell h)^{-1}\p_\pm (\ell h) = \Ad_h^{-1} \bigl( \ell^{-1}\p_+ \ell) + h^{-1} \p_\pm h\,.
\end{equation}
Using the fact that $\Ad_h$ preserves $\hf$ and $V_\pm$ -- see equation \eqref{Eq:AdHDeco}, one sees that $W^\pm_{\ell h}$ is a projector with kernel $\Ad_h^{-1} \Ker(W^\pm_\ell)$ and image $\Ad_h^{-1} \text{Im}(W^\pm_\ell)$, thus proving that $W^\pm_{\ell h} = \Ad_h^{-1} \circ W^\pm_\ell \circ \Ad_h$. Since $\hf$ is contained in the image of $W^\pm_\ell$ and $\Ad_h( h^{-1}\p_\pm h)$ belongs to $\hf$, we thus get
\begin{equation}\label{Eq:GaugeWl}
W^\pm_{\ell h} (\ell h)^{-1}\p_\pm (\ell h) = \Ad_h^{-1} \bigl( W^\pm_{\ell } \ell^{-1} \p_\pm \ell \bigr) + h^{-1} \p_\pm h\,.
\end{equation}
Combining these results with the property \eqref{Eq:PsWh}, the invariance of $\psd$ under $\Ad_h$, the Polyakov-Wiegmann formula \eqref{Eq:PW} and the isotropy of $\hf$, a straightforward computation shows that the action \eqref{Eq:Action} is indeed invariant under the transformation $\ell\mapsto\ell h$.\\

As a result of the gauge symmetries \eqref{Eq:GaugeKH}, the $\s$-model \eqref{Eq:Action} has for physical target space the double quotient $\KDH$. We note that this space has dimension
\begin{equation}
\dim \KDH = \dim \df - \dim \kf - \dim \hf = p\,.
\end{equation}
This echoes the discussion below equation \eqref{Eq:FirstClass}, concerning the reduced current algebra: indeed, we observed that the latter is described by $2p$ physical degrees of freedom, which matches the number of canonical fields in the Hamiltonian formulation of a $\s$-model with a target space of dimension $p$.

Rather than treating this target space as an abstract quotient, it is often practical to realise it concretely as a $p$-dimensional submanifold $s(\KDH)$ of the double group $D$, by choosing a unique representative in $D$ for each equivalence class in the quotient, supposing that we can do so smoothly (mathematically, this corresponds to building a section $s: \KDH \to D$ of the quotient). In that case, the functional \eqref{Eq:Action} can be made into the action of a $\s$-model on $\KDH$ by imposing a gauge-fixing condition, namely that the field $\ell(t,x)$ belongs to the subspace $s(\KDH)$: by definition, this condition is always attainable and leaves no residual gauge symmetry. We are thus left with a theory where all the degrees of freedom, valued in $s(\KDH) \simeq \KDH$, are physical. Such a procedure is purely geometric, in the sense that it involves a gauge-fixing written directly in terms of the $D$-valued field $\ell(t,x)$ and not its derivatives. In particular, this ensures that many of the natural observables of the model, such as the currents $\ell^{-1} \p_\pm \ell$ and $W^\pm_\ell \ell^{-1} \p_\pm \ell$, are expressed in a purely local way in terms of the $s(\KDH)$-valued physical degrees of freedom, once gauge-fixed.

Sometimes, it is also useful to consider only a partial gauge-fixing. For instance, one can work with a formulation of the theory where we gauge-fixed the $K$-symmetry while keeping the other one untouched. This means that the model is now described by a field valued in the left quotient $\KD$ (or an embedding thereof in $D$) and possesses a residual $H$-gauge symmetry.

\paragraph{Canonical analysis and current $\bm\Jc$.} As in the non-degenerate case, the $\s$-model \eqref{Eq:Action} is a Lagrangian realisation of the degenerate $\Ec$-model considered in the previous subsubsection. The key step to make this relation explicit is the identification of the $\Ec$-model current $\Jc$ in terms of the $\s$-model field $\ell$. This is given by
\begin{equation}\label{Eq:J}
\Jc \approx W^+_\ell \, \ell^{-1} \p_+ \ell - W^-_\ell \, \ell^{-1} \p_- \ell\,.
\end{equation}
We note that this relation only holds weakly, \textit{i.e.} when the constraints are imposed. When we start from the Lagrangian point of view, these constraints arise when we perform the canonical analysis of the action. They take the form of relations among the momentum fields of the theory, defined as the canonical conjugate of the coordinate fields contained in $\ell$. These relations simply follow from the expression of the conjugate momenta as derivatives of the Lagrangian density and reflect the local invariance of the action under the $H$-gauge symmetry. In the Lagrangian formulation, the constraints then always hold and should be seen as identities relating together some of the conjugate momenta associated with the action. It is only when we go to the Hamiltonian formulation of the model that it is useful to consider these momenta as purely independent fields: in that case, we then see the constraints as additional relations to be imposed by hand on these fields.

The main reason why we have to consider this extended ``strong'' formulation, where all the momentum fields are treated as independent, is to make sense of the canonical Poisson bracket. In particular, this allows us to compute the Poisson bracket of the current $\Jc$ defined above, once extended strongly in an appropriate way. Doing so, one finds that $\Jc$ satisfies the current Poisson algebra \eqref{Eq:PbJA2}, making the link with the $\Ec$-model construction described earlier. Moreover, one checks that the constraints obtained above from the canonical analysis of the action coincide with the constraints $\Jc_\alpha = \ps{T_\alpha}{\Jc} \approx 0$ of the degenerate $\Ec$-model. In particular, the Poisson bracket of these constraints with the field $\ell$ takes the form of an infinitesimal version of the gauge symmetry $\ell \mapsto \ell h$, confirming the interpretation of the constraints as the Hamiltonian generators of the gauge symmetry. Finally, one can determine the Hamiltonian of the model by performing a Legendre transform of the Lagrangian and write it in terms of the current $\Jc$: one then finds that it agrees with the Hamiltonian \eqref{Eq:H} of the degenerate $\Ec$-model. This achieves the relation between the $\s$-model constructed above and the degenerate $\Ec$-model built in terms of the current $\Jc$ in the previous subsubsection.\\

The $H$-gauge symmetry and its associated constraints played a somehow important role in the discussion above. In contrast, we did not mention the $K$-gauge symmetry: let us briefly comment on this. As already observed in the non-degenerate case, the current $\Jc$, defined in terms of $\ell$ by equation \eqref{Eq:J}, is in fact gauge-invariant under the transformations $\ell \mapsto k\ell$, $k\in K$. Since $\Jc$ is the main ingredient used to relate the $\s$-model \eqref{Eq:Action} to the degenerate $\Ec$-model, the $K$-gauge symmetry then plays a quite minimal role in this relation. For instance, the current Poisson algebra of $\Jc$ and the expression of the Hamiltonian in terms of $\Jc$ are not affected by $K$-gauge transformations. In particular, this means that we could have started the discussion of this paragraph with the model in which we gauge-fixed this $K$-gauge symmetry, \textit{i.e.} with a theory on $\KD$ with a residual $H$-gauge symmetry. The above relation with the current algebra and the degenerate $\Ec$-model in the Hamiltonian formulation will then stay the same. Instead, if one considers the Hamiltonian formulation of the fully gauged model on $D$, then additional constraints will arise due to the $K$-gauge symmetry: however, the presence of this symmetry and these constraints will not affect the construction of the current $\Jc$ and the relation to the degenerate $\Ec$-model formulation. For this reason, we will mainly ignore the details of how we treat this symmetry in the rest of this section.

In contrast, the $H$-gauge symmetry has a non-trivial interplay with $\Jc$. In particular, $\Jc$ is not invariant under the transformations $\ell \mapsto \ell h$, $h\in H$. In fact, using the identity \eqref{Eq:GaugeWl}, one finds that the Lagrangian expression \eqref{Eq:J} for $\Jc$ is mapped to $h^{-1}\Jc\,h+h^{-1}\p_x h$. This is consistent with the equation \eqref{Eq:GaugeJ} derived earlier in the Hamiltonian formulation. In particular, since $\Jc$ is not $H$-gauge-invariant, fixing this gauge would modify the current Poisson bracket of $\Jc$, potentially in a rather complicated way. The relation with the degenerate $\Ec$-model is then technically made through the unreduced formulation of the theory and not the completely gauge-fixed $\s$-model on $\KDH$.

\paragraph{Lagrange multiplier.} Recall from the equation \eqref{Eq:Dyn} and the discussion below that to completely describe the dynamics of the degenerate $\Ec$-model in the Hamiltonian setup, we had to introduce a Lagrange multiplier $\mu$ to encode the time evolution of non-gauge-invariant degrees of freedom. This Lagrange multiplier also has an interpretation when we start with the $\s$-model action \eqref{Eq:Action} and perform a canonical analysis, as considered above. Indeed, recall that the $H$-gauge symmetry of the model implied the presence of constraints, in the form of relations between the canonical conjugate momenta derived from the Lagrangian density. In particular, this has for consequence that the relation between the momentum fields and the time derivatives of coordinate fields is not invertible: schematically, the momentum fields fail to capture the time derivative of the degrees of freedom in $\ell$ which are non-physical due to the presence of the $H$-gauge symmetry. In this context, the Lagrange multiplier $\mu\in\hf$ can then be seen as encoding these missing time derivatives. In the present case, one finds that the Lagrangian expression of $\mu$ is given by
\begin{equation}
\mu = \pi_{\hf} \bigl( W^+_\ell \, \ell^{-1} \p_+ \ell + W^-_\ell \, \ell^{-1} \p_- \ell \bigr) = \frac{1}{2} \pi_\hf (W_\ell^+ + W_\ell^-) \ell^{-1} \p_t \ell + \frac{1}{2} \pi_\hf (W_\ell^+ - W_\ell^-) \ell^{-1} \p_x \ell \,.
\end{equation}
Using the identity \eqref{Eq:GaugeWl}, one checks that under an $H$-gauge transformation $\ell \mapsto \ell h$, this Lagrangian expression for $\mu$ is mapped to $h^{-1}\mu\,h+h^{-1}\p_t h$, which is consistent with the transformation rule \eqref{Eq:GaugeMu} discussed earlier in the Hamiltonian formulation.

\paragraph{Lagrangian expression of $\bm{\Ac_\pm}$ and $\bm{\Jc_\pm}$.} Since we have determined the realisation of $\Jc$ and $\mu$ in terms of the $\s$-model field $\ell$, we can derive the Lagrangian expression of the currents $\Ac_\pm$ and $\Jc_\pm$, using the relations \eqref{Eq:DecoJw} and \eqref{Eq:Apm} and the fact that $W^\pm_\ell$ are projectors with image $\hf \oplus V_\pm$. We get
\begin{equation}\label{Eq:AJ}
\Ac_\pm = \pi_{\hf} \bigl( W^\pm_\ell \ell^{-1}\p_\pm \ell \bigr) \qquad \text{ and } \qquad \Jc_\pm = \pi_{\pm}\bigl( W^\pm_\ell \ell^{-1}\p_\pm \ell\bigr)\,.
\end{equation}
Alternatively, we can write this equation as
\begin{equation}
W^+_\ell \ell^{-1}\p_+ \ell = \Ac_+ + \Jc_+  \qquad \text{ and } \qquad W^-_\ell \ell^{-1}\p_- \ell = \Ac_- + \Jc_- \,.
\end{equation}
This is to be compared with equation \eqref{Eq:JpmLagN} in the non-degenerate case: the main difference here is that $W^\pm_\ell \ell^{-1}\p_\pm \ell$ does not entirely coincides with the $V_\pm$-valued current $\Jc_\pm$ but also contains the $\hf$-valued gauge field $\Ac_\pm$.

\paragraph{Summary.} We end this subsection with a brief summary, gathering the main results that we will need in the rest of the paper. The gauged $\s$-model that we built above is characterised by the following defining data:\vspace{-3pt}
\begin{enumerate}[(i)]\setlength\itemsep{0.2pt}
\item a real connected Lie group $D$ of even dimension, with Lie algebra $\df=\text{Lie}(D)$ ;
\item a non-degenerate ad-invariant split symmetric bilinear form $\psd$ on $\df$ ;
\item a subgroup $H \subset D$ such that $\hf = \text{Lie}(H)$ is an isotropic subalgebra of $\df$ ;
\item a symmetric operator $\Eh : \df \to \df$ such that $\Eh|_{\hf^\perp}^{\,3}=\Eh|_{\hf^\perp}$, $\Ker(\Eh|_{\hf^\perp})=\hf$, $V_\pm = \Ker(\Eh|_{\hf^\perp} \mp \Id)$ have equal dimension and $\Eh|_{\hf^\perp}$ commutes with the adjoint action of $H$ ;
\item a subgroup $K\subset D$ such that $\kf=\text{Lie}(K)$ is a maximally isotropic subalgebra of $\df$. \vspace{-4pt}
\end{enumerate}
From this data, the $\s$-model is built as follows. One first considers a $D$-valued field $\ell(t,x)$ on $\R\times\mathbb{S}^1$ and constructs the projectors $W^\pm_\ell$ according to equation \eqref{Eq:Wpm}. We then define the action of the $\s$-model as the functional \eqref{Eq:Action} of $\ell(t,x)$. This action is invariant under the gauge symmetries $\ell \mapsto k\ell h$, with $k\in K$ and $h \in H$, making the physical target space of the model the double quotient $\KDH$. Alternatively, it is often useful to see the theory as a gauged $\s$-model on $\KD$, with an $H$-gauge symmetry. A key role in this model is played by the fields $\Ac_\pm$ and $\Jc_\pm$, defined in terms of $\ell$ by equation \eqref{Eq:AJ}, and valued respectively in $\hf$ and $V_\pm$. In particular, in terms of these, the equations of motion take the simple first-order form \eqref{Eq:EoMpm}. In this formulation, $\Ac_\pm$ plays the role of a gauge field, while $\Jc_\pm$ are covariant currents.

This gauged $\s$-model is the Lagrangian realisation of a degenerate $\Ec$-model, obtained by passing to the Hamiltonian formulation. The presence of the $H$-gauge symmetry implies the apparition of constraints when we perform the canonical analysis of the model: in the Hamiltonian formulation, these are interpreted as the generators of the symmetry. In this setup, one can construct a $\df$-valued field $\Jc$ which satisfies the current algebra \eqref{Eq:PbJA2}, given by $\Jc \approx \Ac_+ - \Ac_- + \Jc_+ - \Jc_-$ weakly (\textit{i.e.} up to the constraints). Moreover, the Hamiltonian of the theory is given by the formula \eqref{Eq:H}, in terms of $\Jc$ and the operator $\Eh$, making the relation with the degenerate $\Ec$-model.

As in the non-degenerate case, the first-order equations of motion and the degenerate $\Ec$-model formulation of the theory are independent of the choice of $K$, and are thus characterised solely by the data (i) to (iv) above. The datum (v) of the subgroup $K$ intervenes only when we look for a concrete $\s$-model realisation of this theory. Different $\s$-models can correspond to the same underlying degenerate $\Ec$-model, but with different choices of $K$, leading in particular to the notion of Poisson-Lie T-duality (see the non-degenerate discussion at the end of Subsubsection \ref{SubSec:LagN} for more details). The rest of this section, which concerns the study of conformal and chiral structures in degenerate $\Ec$-models will rely only on the data (i) to (iv), independently of the choice of $K$. We will come back to the latter in the next section, when studying explicit examples in their $\s$-model form.

\subsection[Conformal degenerate \texorpdfstring{$\Ec$}{E}-models and their chiral Poisson algebras]{Conformal degenerate \texorpdfstring{$\bm\Ec$}{E}-models and their chiral Poisson algebras}
\label{SubSec:Conf}

We now come to the main part of the paper, namely the discussion of conformal and chiral structures in degenerate $\Ec$-models. We will discuss a simple condition on the data defining such a model which ensures its 1-loop conformal invariance. Moreover, we will show that this condition is also reflected at the classical level through the apparition of chiral observables in the dynamics of the theory, \textit{i.e.} left- and right-moving fields. Finally, we will study the Poisson algebras formed by these chiral fields: we will find that left-moving observables commute with right-moving ones and that the fields of a given chirality satisfy a gauged Poisson current algebra. This paves the way for an exact quantisation of the model as a 2-dimensional gauged CFT (a more thorough discussion of these aspects is out of the scope of the present paper but constitutes a natural perspective for further developments).

There are two main differences with the case of conformal non-degenerate $\Ec$-models treated earlier in Subsection \ref{SubSec:ConfN}. The first one is that we obtain gauged current algebras rather than ungauged ones: this is a direct consequence of the presence of a gauge symmetry in degenerate $\Ec$-models. The second difference is that the chiral fields of the theory are in general non-local objects when written in terms of the $\Ec$-model current (or the $\s$-model fields). This non-locality is also visible in the chiral Poisson algebra formed by these fields, which generally (although not necessarily) contains $\epsilon$-functions: more precisely, we argue in this subsection that these non-local chiral fields satisfy a parafermionic-like Poisson algebra, which arises as a natural gauge-fixing of the gauged current algebra.

A natural question in this situation is whether there exist well-chosen combinations of the non-local chiral fields which are local. The answer to this question is positive. A typical example of such a local chiral field is given by the left- or right-moving component of the energy-momentum tensor of the model, which can be seen as a particular quadratic combination of the parafermionic fields that turns out to be local. This is the first element in a whole family of local chiral fields which we will describe in this subsection. This family is closed under Poisson bracket and thus forms a local Poisson subalgebra of the parafermionic chiral algebra. More precisely, it takes the form of a classical higher-spin $\Wc$-algebra. In this subsection, we will develop a systematic procedure to build elements of this $\Wc$-algebra. As we shall see in Section \ref{Sec:Examples}, this gives a concrete and efficient way to find local chiral fields in various explicit examples of conformal $\s$-models.

\subsubsection[Strong conformal condition and vanishing of the 1-loop \texorpdfstring{$\beta$}{beta}-functions]{Strong conformal condition and vanishing of the 1-loop \texorpdfstring{$\bm\beta$}{beta}-functions}\label{SubSec:Beta}

\paragraph{Strong conformal condition.} Consider the degenerate $\Ec$-model defined in the previous subsection and, in particular, recall the decomposition $\df = \hf \oplus \hf' \oplus V_+ \oplus V_-$ of the double algebra. We will say that this model satisfies the strong conformal condition if the following equivalent properties hold:\vspace{8pt}\\
\begin{tabular}{cp{0.87\textwidth}}
\;\;\;(C1)\!\! & the subspaces $\df_\pm = \hf \oplus \hf' \oplus V_\pm$ are subalgebras of $\df$ ; \vspace{7pt} \\
\;\;\;(C2)\!\! & we have $[V_+,V_-] = \lbrace 0 \rbrace$ and $\hf \oplus \hf'$ is a subalgebra of $\df$.
\end{tabular}~\vspace{8pt}\\
As in the non-degenerate case treated in Subsection \ref{SubSec:ConfN}, we use the terminology strong conformal condition as we expect that it is not the most general one ensuring the 1-loop conformal invariance of the model, \textit{i.e.} it is sufficient but not necessary. We restrict to this more constraining case in this paper as it allows a simple characterisation of the chiral Poisson algebras of the model, as we shall see below. We refer to the concluding subsection \ref{SubSec:GenConf} for additional details and perspectives on more general conformal conditions.

The claimed equivalence of the above conditions (C1) and (C2) can be proven in an elementary way using the ad-invariance of $\psd$ and the property \eqref{Eq:PsDeco} of the decomposition $\df = \hf \oplus \hf' \oplus V_+ \oplus V_-$. We will not enter into the details of this proof, as it is similar to the one for non-degenerate models detailed in Subsection \ref{SubSec:ConfN}. The main difference in the present degenerate case is the additional treatment of the subspace $\hf\oplus\hf'$. When proving the statement, one actually derives more precise results on the structure of the subalgebras $\df_\pm = \hf \oplus \hf' \oplus V_\pm$, namely
\begin{equation}\label{Eq:ComHVpm}
[\hf,\hf] \subset \hf\,,  \qquad [\hf,\hf'] \subset \hf'\,,  \qquad [\hf',\hf'] \subset \hf  \qquad \text{ and } \qquad [\hf \oplus \hf', V_\pm ] \subset V_\pm\,.
\end{equation}

\paragraph{Vanishing of the 1-loop $\bm\beta$-functions.} Let us now prove that the strong conformal condition introduced above implies the vanishing of the 1-loop $\beta$-functions of the degenerate $\Ec$-model, as we did in the non-degenerate case in Subsection \ref{SubSec:ConfN}. For that, we first need to describe the 1-loop RG flow of a generic degenerate $\Ec$-model, which was initially determined in~\cite{Severa:2018pag}. In this reference, it was written in the form of a flow of the subspace $V_+$ (see also~\cite[Subsection 6.2]{Klimcik:2019kkf}). Here, we will use an alternative but equivalent formulation, in terms of the quadratic form $E(\cdot,\cdot)=\langle\cdot,\Eh\,\cdot\rangle|_{\hf^\perp}$ which characterises the Hamiltonian of the theory (see previous subsection). Namely, we have
\begin{equation}\label{Eq:Rg}
\frac{\dd\;}{\dd \tau} \langle X, \Eh\, X \rangle = -4\hbar\, \Tr_{\hf^\perp}\bigl( S_{X} \bigr) + O(\hbar^2)\,, \qquad \forall\,X\in\hf^\perp\,,
\end{equation} 
where $\tau = \frac{1}{2\pi} \log\mu$ is the RG parameter and $S_{X}: \hf^\perp \to \hf^\perp$ is a linear operator defined by
\begin{equation}\label{Eq:S}
S_{X} = \pi_-\, \ad_{\pi_- X}\,\pi_+\,\ad_{\pi_+ X}   +  \ad_{\pi_- X}  \, \ad_{\pi_+ X} \,\pi_{\hf} \,.
\end{equation}
(We recall that the operators $\pi_\hf$ and $\pi_\pm$ are the projectors along $\hf$ and $V_\pm$ in the decomposition $\hf^\perp = \hf \oplus V_+ \oplus V_-$.) The above formulae for the RG-flow and the operator $S_{X}$ are the degenerate equivalents of the equations \eqref{Eq:RgN} and \eqref{Eq:SN} discussed in Subsection \ref{SubSec:ConfN} for non-degenerate $\Ec$-models. The main difference here is the second term in the definition of $S_{X}$, involving $\pi_{\hf}$, and the fact that we consider the quadratic form $\langle\cdot,\Eh\,\cdot\rangle|_{\hf^\perp}$ and the operator $S_{X}$ on the subspace $\hf^\perp$ rather than the full double algebra $\df$. One easily checks that the flow \eqref{Eq:Rg} preserves the fact that $\langle\cdot,\Eh\,\cdot\rangle|_{\hf^\perp}$ has kernel $\hf$ and is invariant under the adjoint action of $\hf$ on $\hf^\perp$: this ensures that the gauge symmetry of the degenerate $\Ec$-model stays the same along the RG-flow, as one should expect.

When treating concrete examples, one typically considers a specific degenerate $\Ec$-model with a given choice of Hamiltonian, characterised by a quadratic form $\langle\cdot,\Eh\,\cdot\rangle|_{\hf^\perp}$ defined in terms of certain parameters (coupling constants) $\lambda=(\lambda_k)_{k=1,\dots,n}$. In that case, we say that this model is 1-loop renormalisable if the flow \eqref{Eq:Rg} is compatible with the functional dependence of $\langle\cdot,\Eh\,\cdot\rangle|_{\hf^\perp}$ on $\lambda$, \textit{i.e.} if it can be completely reabsorbed into a running of the coupling constants $\lambda$ (for more details, see equation \eqref{Eq:Beta} and the associated discussion in the non-degenerate case). \\ 

We now consider the case of a degenerate $\Ec$-model satisfying the strong conformal condition introduced in the previous paragraph and show that this implies that the model is a fixed-point of the 1-loop RG flow \eqref{Eq:Rg}. Let $X\in\hf^\perp$. Recall that $[V_+,V_-]=\lbrace 0 \rbrace$ (by the formulation (C2) of the strong conformal condition): thus, the operator $\ad_{\pi_- X}\,\pi_+$ vanishes. Similarly, since $[V_+,\hf] \subset V_+$, the operator $\ad_{\pi_- X}  \, \ad_{\pi_+ X}\,\pi_{\hf}$ maps to $[V_-,V_+]$ and hence vanishes by (C2). We thus have $S_{X}=0$ for all $X\in\hf^\perp$. This ensures that the right-hand side of the RG-flow \eqref{Eq:Rg} vanishes, proving our claim that the model possesses 1-loop conformal invariance.

\subsubsection{Non-local chiral fields and their parafermionic Poisson algebras}
\label{SubSec:Para}

\paragraph{Non-local chiral fields.} We now come to the description of the chiral fields of the theory. For that it will be useful to recall the light-cone formulation \eqref{Eq:EoMpm} of the equations of motion, in terms of the $\hf$-valued gauge field $\Ac_\pm$ and the covariant currents $\Jc_\pm$ valued in $V_\pm$. As we suppose in this subsection that the model satisfies the strong conformal condition and thus that $[V_+,V_-]=\lbrace 0 \rbrace$, it is clear that these equations of motion now take the form
\begin{subequations}\label{Eq:CEoMpm}
\begin{eqnarray}
&\p_+ \Ac_- - \p_- \Ac_+ + [\Ac_+, \Ac_-] = 0\,, \label{Eq:CEoMA} \\
&\p_- \Jc_+ + [\Ac_-, \Jc_+] = 0\,, \label{Eq:CEomJp} \\
&\p_+ \Jc_- + [\Ac_+, \Jc_-] = 0\,.\label{Eq:CEomJm} 
\end{eqnarray}
\end{subequations}
In particular, the first of these equations ensures that $\Ac_\pm$ is on-shell flat. Therefore, there exists a $H$-valued field $\eta(t,x)$ such that
\begin{equation}\label{Eq:EtaA}
\Ac_\pm = -\p_\pm \eta\,\eta^{-1} \,.
\end{equation}
Recall from the previous subsection that the spatial part of this gauge field is $\Jc_{\hf} = \Ac_+ - \Ac_-$, \textit{i.e.} the $\hf$-valued component of the current $\Jc$. The field $\eta$ is thus related to a path-ordered exponential of $\Jc_{\hf}$ and more precisely takes the form
\begin{equation}\label{Eq:EtaPexp}
\eta(t,x) = \Pexp \left( - \int_{0}^x \Jc_{\hf}(t,y)\,\dd y \right)\,\eta_0(t)\,,
\end{equation}
where $\eta_0(t)=\eta(t,0)$ is a $H$-valued observable independent of the space coordinate $x$ but which can however still be a dynamical quantity.\\

Recall that the fields $\Ac_\pm$ and $\Jc_\pm$ transform under $H$-gauge transformations according to the formula \eqref{Eq:GaugePm}. In particular, this means that the quantity $\eta(t,x)$ can alternatively be characterised as the $H$-valued field such that the gauge transformation with local parameter $\eta(t,x)$ brings the gauge field $\Ac_\pm$ to zero, \textit{i.e.}
\begin{equation}
\Ac_\pm^\eta = \eta^{-1}\Ac_{\pm}\,\eta + \eta^{-1}\p_\pm \eta = 0\,.
\end{equation}
We now introduce fields $\Psi^\cL$ and $\Psi^\cR$, defined as the image of the currents $\Jc_+$ and $\Jc_-$ after a gauge transformation by $\eta$. More concretely, we then have
\begin{equation}\label{Eq:Psi}
\Psi^\cL = \Jc_+^\eta = \eta^{-1}\Jc_+\,\eta \qquad \text{ and } \qquad \Psi^\cR = \Jc_-^\eta = \eta^{-1}\Jc_-\,\eta\,,
\end{equation}
which are respectively valued in $V_+$ and $V_-$. Since the equations of motion \eqref{Eq:CEoMpm} of the model are gauge-invariant, they hold with $(\Ac_\pm,\Jc_+,\Jc_-)$ replaced by $(\Ac_\pm^\eta,\Jc_+^\eta,\Jc_-^\eta)=(0,\Psi^\cL,\Psi^\cR)$. We thus conclude that
\begin{equation}
\p_- \Psi^\cL = 0 \qquad \text{ and } \qquad \p_+ \Psi^\cR = 0 \,,
\end{equation}
\textit{i.e.} $\Psi^\cL$ is a left-moving field and $\Psi^\cR$ is a right-moving field.

Since $\eta$ is expressed in terms of a path-ordered exponential of $\Jc_\hf$ -- see equation \eqref{Eq:EtaPexp}, the chiral fields $\Psi^\cL$ and $\Psi^\cR$ are non-local observables of the theory. This non-locality can be traced back to the degenerate nature of the $\Ec$-model under consideration: indeed, contrarily to the non-degenerate case discussed in Subsection \ref{SubSec:ConfN}, the components $\Jc_\pm$ of the current are not chiral here but are rather chiral up to a gauge transformation by a non-local field. Since $\Ac_\pm=-\p_\pm \eta\,\eta^{-1}$, we note that under a gauge transformation $\Ac_\pm \mapsto h^{-1}\Ac_\pm h + h^{-1}\p_\pm h$, $\eta$ is sent to $h^{-1}\eta$. In particular, this means that the fields $\Psi^\cL$ and $\Psi^\cR$ defined in equation \eqref{Eq:Psi} are gauge-invariant, as the transformation of $\eta$ cancels the one of $\Jc_\pm$ by conjugation. These chiral fields are thus physical observables of the model.

Let us quickly comment on the number of these chiral fields. Recall from the previous subsection that the subspaces $V_+$ and $V_-$, in which $\Psi^\cL$ and $\Psi^\cR$ are valued, are both of dimension $p$. We thus extract $p$ left-moving fields from $\Psi^\cL$ and $p$ right-moving fields from $\Psi^\cR$. Combining these together, we then obtain $2p$ degrees of freedom. This coincides with the number of physical fields of the degenerate $\Ec$-model (see previous subsection). We thus conclude that the dynamics of this theory is completely decoupled into a left-moving sector and a right-moving sector.

\paragraph{Boundary conditions.} At this point, a comment is in order regarding boundary conditions. Recall that we are considering our 2d field theory on the cylinder $\R\times \mathbb{S}^1$, assuming that the fields $\Ac_\pm$ and $\Jc_\pm$ are periodic with respect to the space variable, \textit{i.e.} that $\Ac_\pm(t,x+2\pi)=\Ac_\pm(t,x)$ and $\Jc_\pm(t,x+2\pi)=\Jc_\pm(t,x)$. In particular, when performing gauge transformations $(\Ac_\pm,\Jc_\pm) \mapsto (\Ac_\pm^h,\Jc_\pm^h)$, we should restrict to parameters $h(t,x)$ which are themselves periodic, in order to preserve the space of physically allowed field configurations. However, the field $\eta(t,x)$ introduced by equation \eqref{Eq:EtaA} is not periodic: in fact, the zero curvature equation \eqref{Eq:CEoMA} of $\Ac_\pm$ does not ensure the existence of a field $\eta$ satisfying \eqref{Eq:EtaA} on the cylinder $\R\times \mathbb{S}^1$, since the latter is not simply connected, but rather on its cover $\R^2$, where the spatial coordinate $x$ has been ``decompactified'' and is now seen as a variable in $\R$. As fields on $\R^2$, $\eta(t,x)$ and $\eta(t,x+2\pi)$ satisfy the same partial differential equation \eqref{Eq:EtaA}, using the periodicity of $\Ac_\pm$: this means that $\eta(t,x+2\pi)=\eta(t,x)\,\gamma$, for some matrix $\gamma$ which is independent of $t$ and $x$. This matrix is in general non-trivial, showing that $\eta$ is not a periodic field.

The transformation from $(\Ac_\pm,\Jc_+,\Jc_-)$ to $(\Ac_\pm^\eta,\Jc_+^\eta,\Jc_-^\eta)=(0,\Psi^\cL,\Psi^\cR)$ is thus not a physically allowed gauge transformation. In particular, one finds that the chiral fields $\Psi^\cLR$ are not periodic but rather quasi-periodic, in the sense that $\Psi^\cLR(t,x+2\pi) = \gamma^{-1}\, \Psi^\cLR(t,x)\, \gamma$. In other words, they are not single-valued fields on the cylinder: they have a non-trivial monodromy along the spatial circle and are thus defined only on the cover $\R^2$.\footnote{This is a common feature in 2-dimensional coset CFTs such as gauged WZW models, which possess multivalued parafermionic fields. At the quantum level, this property is reflected in the presence of non-integer powers in the correlation functions and operator product expansions of these fields -- see for instance~\cite{Lepowsky:1984,Fateev:1985mm,Ninomiya:1986dp,Gepner:1987sm}.} Technically, this means that the configuration $(0,\Psi^\cLR)$ should not be understood as a physically allowed gauge choice for the fields $(\Ac_\pm,\Jc_\pm)$. However, we stress that the way we build $\Psi^\cLR$ from $\Jc_\pm$ still takes the same symbolic form as a gauge transformation: this ensures that the reasoning proposed above works. In particular, one can formally replace the fields $(\Ac_\pm,\Jc_\pm)$ by $(\Ac_\pm^\eta,\Jc_\pm^\eta)=(0,\Psi^\cLR)$ in the equations of motion to find $\p_-\Psi^\cL=\p_+\Psi^\cR=0$, even though $\eta$ is not a physical gauge parameter: indeed, one easily checks that the equations of motion \eqref{Eq:CEoMpm} are also invariant under transformations $(\Ac_\pm,\Jc_\pm) \mapsto (\Ac_\pm^h,\Jc_\pm^h)$ with $h$ non-periodic. Similarly, the other properties of $\Psi^\cLR$ that we will derive in the rest of this section will rely on the idea that many statements about gauge transformations in fact still hold for non-periodic gauge parameters.\footnote{This is clearly the case for the results of Subsubsection \ref{SubSec:W}, where we will use gauge-invariant local fields to build the $\Wc$-algebra of the theory: indeed, these fields are also invariant under non-periodic transformations. That it holds for the computations of the next paragraph, which concern the Dirac bracket of $\Psi^\cLR$, is a bit more subtle: however, we do expect the reasoning to go through. In particular, we note that the bracket \eqref{Eq:PbPara} that we will obtain there should be understood only as a result for $0 < x,y < 2\pi$. Extending it to $x,y$ in the whole cover $\R$ would require defining carefully the extended distributions $\delta(x-y)$, $\p_x\delta(x-y)$ and $\epsilon(x-y)$ and how they act on non-periodic test functions, in a way that makes the bracket consistent with the quasi-periodicity of $\Psi^\cLR$ (taking into account that the matrix $\gamma$	can itself have non-trivial Poisson brackets). A thorough analysis of these aspects is however beyond the scope of this paper.} For this reason and by a slight abuse of terminology, we will still refer to $(\Ac_\pm,\Jc_\pm) \mapsto (0,\Psi^\cLR)$ as a gauge transformation in the rest of the article.
 
\paragraph{Parafermionic chiral Poisson algebras.} We now want to determine the Poisson brackets of the chiral fields $\Psi^\cL$ and $\Psi^\cR$. This can in principle seems like a difficult task since they are defined in terms of the complicated non-local field $\eta$, given by equation \eqref{Eq:EtaPexp}. In particular, the quantity $\eta_0$ appearing in this equation is in general a dynamical field and can have non-trivial Poisson bracket with other observables, making the computation rather involved. However, there is a simple and powerful way to overcome these difficulties, using the notion of Dirac bracket. Recall that $\Psi^\cL$ and $\Psi^\cR$ are obtained as the image of $\Jc_+$ and $\Jc_-$ under the gauge transformation with local parameter $\eta$. As explained above, this gauge transformation is the one that brings the gauge field $\Ac_\pm$ to zero. Going back to the temporal and spatial components of this field -- see equation \eqref{Eq:Apm}, we see that this transformation thus sends the Lagrange multiplier $\mu$ and the field $\Jc_{\hf}$ to zero. In terms of the current Poisson algebra underlying the degenerate $\Ec$-model, the chiral observables $\Psi^\cL$ and $\Psi^\cR$ thus coincide with the fields $\Jc_+$ and $\Jc_-$ under the gauge-fixing condition setting $\Jc_{\hf}$ to zero.

The Poisson brackets of gauge-fixed observables can be conveniently computed using the so-called Dirac bracket: this formalism can thus be used here to determine the Poisson algebra formed by the chiral fields $\Psi^\cL$ and $\Psi^\cR$. We describe the main steps of this computation in the appendix \ref{App:Dirac} and summarise the outcome here. The first result that one obtains is that the left-moving fields Poisson commute with the right-moving ones, \textit{i.e.}
\begin{equation}\label{Eq:PbPsiLR}
\bigl\lbrace \Psi^\cL(x)\ti{1}, \Psi^\cR(y)\ti{2} \bigr\rbrace = 0\,.
\end{equation}
Here, we used the tensorial notations $\bm{\underline{i}}$ to conveniently write down the Poisson brackets between the different components of $\Psi^\cL$ and $\Psi^\cR$ (see the discussion around equation \eqref{Eq:PbJ} for conventions and more details on these notations). The above equation shows that the decoupling of the theory into the two chiral sectors is also visible at the level of the Poisson structure.\\

There is thus left to describe the Poisson brackets of $\Psi^\cL$ with itself and $\Psi^\cR$ with itself. To do so, it will be useful to introduce a few notations. We define
\begin{equation}
D\hf = \hf \oplus \hf'\,,
\end{equation}
which we recall is a subalgebra of $\df$ by the strong conformal condition (see beginning of this subsection). The projector on this subalgebra, according to the orthogonal decomposition $\df = D\hf \oplus V_+ \oplus V_-$, is then given by $\pi_{D\hf}=\pi_{\hf}+\pi_{\hf'}$. Recall also that $\Cd$ is the $(\df\otimes\df)$-valued split quadratic Casimir of the double Lie algebra $\df$. We introduce its projections
\begin{equation}\label{Eq:CasDeco}
\Cdh  = (\pi_{D\hf}\otimes\pi_{D\hf}) \Cd \qquad \text{ and } \qquad \Cpm = \pm (\pi_{\pm}\otimes\pi_\pm) \Cd\,,
\end{equation}
respectively valued in $\hf\otimes\hf' + \hf' \otimes \hf \subset D\hf \otimes D\hf$ and $V_\pm \otimes V_\pm$ (the minus sign in $\Cm$ has been introduced for future convenience). We now have enough notations to write down the Poisson algebra of the non-local chiral fields $\Psi^\cL$ and $\Psi^\cR$, as obtained from the Dirac bracket computation detailed in appendix \ref{App:Dirac}. Namely, we have
\begin{subequations}\label{Eq:PbPara}
\begin{align}
\bigl\lbrace \Psi^\cL(x)\ti{1}, \Psi^\cL(y)\ti{2} \bigr\rbrace &= \left( \bigl[ \Cp , \Psi^\cL(x)\ti{1}  \bigr] + \bigl[ \Cdh, \Psi^\cL(x)\ti{2}  \bigr] \right)\,\delta(x-y) \\
& \hspace{30pt} - \Cp \,\p_x\delta(x-y) - \frac{1}{2} \Bigl[ \bigl[ \Cdh , \Psi^\cL(x)\ti{1}  \bigr]  , \Psi^\cL(y)\ti{2}  \Bigr] \, \epsilon(x-y)\,,  \notag \\[5pt]
\bigl\lbrace \Psi^\cR(x)\ti{1}, \Psi^\cR(y)\ti{2} \bigr\rbrace &= \left( \bigl[ \Cm, \Psi^\cR(x)\ti{1}  \bigr] - \bigl[ \Cdh, \Psi^\cR(x)\ti{2}  \bigr] \right)\,\delta(x-y) \\
& \hspace{30pt} + \Cm \,\p_x\delta(x-y) - \frac{1}{2} \Bigl[ \bigl[ \Cdh , \Psi^\cR(x)\ti{1}  \bigr]  , \Psi^\cR(y)\ti{2}  \Bigr] \, \epsilon(x-y)\,,  \notag
\end{align}
\end{subequations}
where $\epsilon(x-y)$ is the sign function, defined by $\epsilon(x)=-1$ if $-2\pi < x < 0$ and $\epsilon(x)=+1$ if $0 < x < 2\pi$, such that $\p_x \epsilon(x) = 2\delta(x)$. One checks that these brackets are valued in $V_+ \otimes V_+$ and $V_- \otimes V_-$ respectively, as expected since $\Psi^\cL\in V_+$ and $\Psi^\cR\in V_-$. The non-locality of the fields $\Psi^\cL$ and $\Psi^\cR$ is reflected in their Poisson algebra through the presence of the sign function: indeed, the Poisson bracket of $\Psi^\cLR(x)$ with $\Psi^\cLR(y)$ is not a distribution localised at $x=y$ but also contains non-local terms which have support on the regions $x<y$ and $x>y$. \\

Recall that $\df_\pm = D\hf \oplus V_\pm$ is a subalgebra of $\df$ by the strong conformal condition. The restricted bilinear form $\pm\psd|_{\df_\pm}$ on this subalgebra is invariant and non-degenerate. We can thus consider the current algebras associated with $\df_+$ and $\df_-$: the Poisson brackets \eqref{Eq:PbPara} of the chiral fields $\Psi^\cL$ and $\Psi^\cR$ can then be seen as the Hamiltonian reduction of these current algebras with respect to $\hf$. Thus, the left- and right-moving sectors of the theory are described by reduced / gauged current algebras. Such algebras, their description in terms of non-local fields with brackets of the form \eqref{Eq:PbPara} and their quantisation have been the subjects of many works in the literature, at least for a number of important cases\footnote{For instance when $\df_\pm$ and $\hf$ are simple Lie algebras, or more generally reductive ones. Many of these works were motivated by the study of gauged Wess-Zumino-Witten models, which are specific conformal $\s$-models described by chiral reduced current algebras: as we shall see in Subsection \ref{SubSec:gWZW}, these theories belong to the class of conformal degenerate $\Ec$-models studied in this paper.}: see for instance~\cite{Lepowsky:1984,Fateev:1985mm,Ninomiya:1986dp,Gepner:1987sm,Karabali:1989dk,Bardakci:1990lbc,Bardakci:1990ad}. In this context, $\Psi^\cL$ and $\Psi^\cR$ are generally called parafermionic fields, which is the terminology that we will use in the rest of this paper.\footnote{Let us also note that parafermionic-like Poisson algebras already appeared in the study of degenerate $\Ec$-models in the work~\cite{Sfetsos:1999zm}. The main difference with the case studied here is that the $\Ec$-model considered in this reference was in general not conformal and the parafermionic fields not chiral.}

\paragraph{Decomposition in a basis.} For concreteness, it can be helpful to write down the parafermionic Poisson algebras using an explicit decomposition of $\Psi^\cL$ and $\Psi^\cR$ along bases of $V_+$ and $V_-$. To do so, let us introduce a few notations. Recall that $\lbrace T_\alpha \rbrace_{\alpha=1}^{d-p}$ is a basis of $\hf$. Since $\hf'$ is paired non-degenerately with $\hf$, there exists a basis $\lbrace T^\alpha \rbrace_{\alpha=1}^{d-p}$ of $\hf'$, which is dual to $\lbrace T_\alpha \rbrace_{\alpha=1}^{d-p}$. Moreover, we let $\lbrace U_a \rbrace_{a=1}^p$ and $\lbrace \Ub_{\ab} \rbrace_{\ab=1}^p$ be bases of $V_+$ and $V_-$ respectively (recall from subsection \ref{SubSec:Conf} that these are both of dimension $p$). Here and in what follows, we use greek indices $\alpha,\beta,\dots\null\in\lbrace1,\dots,d-p\rbrace$ to label the basis of $\hf$ and its dual $\hf'$, lower-case latin indices $a,b,\dots\null\in\lbrace1,\dots,p\rbrace$ to label the basis of $V_+$ and lower-case latin barred indices $\ab,\bb,\dots\null\in\lbrace1,\dots,p\rbrace$ to label the basis of $V_-$. The family $\lbrace T_\alpha, T^\alpha \rbrace_{\alpha=1}^{d-p} \sqcup \lbrace U_a \rbrace_{a=1}^p \sqcup \lbrace \Ub_{\ab} \rbrace_{\ab=1}^p$ then forms a basis of the full double algebra $\df$. The non-zero entries of the bilinear form $\psd$ in this basis read
\begin{equation}
\ps{T^\alpha}{T_\beta}=\delta^{\alpha}_{\;\,\beta}, \qquad \ps{U_a}{U_b} = \rho_{ab} \qquad \text{ and } \qquad \ps{\Ub_{\ab}}{{\Ub}_{\bb}} = -\rhob_{\,\ab\bb}\,,
\end{equation}
for some $p\times p$ symmetric matrices $\rho$ and $\rhob$ (the sign in the third equation has been introduced so that the matrices $\rho$ and $\rhob$ are both positive definite if the Hamiltonian of the $\Ec$-model is positive). We will denote by $\rho^{\,ab}$ and $\rhob^{\,\ab\bb}$ the entries of the inverse of these matrices. The projections of the quadratic Casimir introduced in equation \eqref{Eq:CasDeco} can then be written as
\begin{equation}
\Cdh = T_\alpha \otimes T^\alpha + T^\alpha \otimes T_\alpha\,, \qquad \Cp = \rho^{\,ab}\; U_a \otimes U_b\,, \qquad \Cm = \rhob^{\,\ab\bb} \; \Ub_{\ab} \otimes \Ub_{\bb}\,,
\end{equation}
where we sum over repeated indices.

Recall that $[V_\pm,V_\pm]\subset V_\pm \oplus \hf \oplus \hf'$ and $[V_\pm , \hf \oplus \hf'] \subset V_\pm$. We will need the following commutation relations written in the basis introduced above:
\begin{equation}
[ U_a, U_b ] = \F{ab}{\!c} \, U_c + \F{ab}{\!\alpha} \, T_\alpha + \F{ab\alpha}{\null} \, T^\alpha\,, \qquad [ \Ub_{\ab}, \Ub_{\bb} ] = \F{\ab\bb}{\cb} \, \Ub_{\cb} + \F{\ab\bb}{\!\alpha} \, T_\alpha + \F{\ab\bb\alpha}{\null} \, T^\alpha\,,
\end{equation}
\begin{equation*}
[ U_a, T_\alpha ] = \F{a\alpha}{\!b} \, U_b\,, \qquad [ U_a, T^\alpha ] = \F{a}{\alpha b}\,U_b\, \qquad [ \Ub_{\ab}, T_\alpha ] = \F{\ab\alpha}{\bb} \, \Ub_{\bb}\qquad \text{and} \qquad [ \Ub_{\ab}, T^\alpha ] = \F{\ab}{\alpha\bb}\,\Ub_{\bb}\, .
\end{equation*}
The structures constants appearing in these commutation relations and the coefficients $\rho_{ab}$ and $\rhob_{\,\ab\bb}$ obey certain identities due to the ad-invariance of $\psd$.\\

Let us finally introduce the parametrisation of the parafermionic fields $\Psi^\cL$ and $\Psi^\cR$ along these bases. We define
\begin{equation}
\psi^\cL_a(x) = \psb{U_a}{\Psi^\cL(x)} \qquad \text{ and } \qquad \psi^\cR_{\ab}(x) = -\psb{\Ub_{\ab}}{\Psi^\cR(x)}\,,
\end{equation}
such that $\Psi^\cL(x) = \rho^{ab}\, \psi_a^\cL(x)\, U_b$ and $\Psi^\cR(x) = \rhob^{\,\ab\bb}\, \psi^\cR_{\ab}(x)\, \Ub_{\bb}$. Using the various results above, we can then extract the Poisson brackets of the fields $\psi^\cL_a$ and $\psi^\cR_a$ from the parafermionic algebra \eqref{Eq:PbPara}. This gives
\begin{subequations}\label{Eq:PbParaBasis}
\begin{align}
\bigl\lbrace \psi_a^\cL(x), \psi_b^\cL(y) \bigr\rbrace &= \F{ab}{\!c}\,\psi_c^\cL(x)\,\delta(x-y) - \rho_{ab}\,\p_x\delta(x-y) \label{Eq:PbParaBasisL} \\
& \hspace{40pt} - \frac{1}{2} \bigl( \F{a\alpha}{\!c} \, \F{b}{\alpha d} + \F{a}{\alpha c} \, \F{b\alpha}{\!d} \bigr)\; \psi^\cL_c(x)\, \psi^\cL_d(y) \, \epsilon(x-y)\,, \notag \\[5pt]
\bigl\lbrace \psi_{\ab}^\cR(x), \psi_{\bb}^\cR(y) \bigr\rbrace &= \F{\ab\bb}{\cb}\,\psi_{\cb}^\cR(x)\,\delta(x-y) + \rhob_{\,\ab\bb}\,\p_x\delta(x-y) \label{Eq:PbParaBasisR} \\
&\hspace{40pt} - \frac{1}{2} \bigl( \F{\ab\alpha}{\!\cb} \, \F{\bb}{\alpha \bar{d}} + \F{\ab}{\alpha\cb} \, \F{\bb\alpha}{\!\bar{d}} \bigr)\, \psi^\cR_{\cb}(x)\, \psi^\cR_{\bar d}(y) \, \epsilon(x-y)\,. \notag
\end{align}
\end{subequations}

\subsubsection[Local chiral fields and \texorpdfstring{$\Wc$}{W}-algebra]{Local chiral fields and \texorpdfstring{$\bm\Wc$}{W}-algebra}
\label{SubSec:W}

In the previous subsubsection, we have studied the parafermionic chiral fields of the model, which in particular are non-local observables. This raises the natural question of whether there also exist local chiral fields, which we answer positively in this subsubsection. These fields can be seen as specific combinations of the parafermions which turn out to be local and form a closed Poisson algebra that is called the $\Wc$-algebra. Below, we will characterise the latter in terms of the initial current algebra and will describe an efficient way of building its elements. These results were inspired by the analysis of~\cite[Subsection 6.3]{Kotousov:2022azm}, which can be seen as a particular case of the present construction.

\paragraph{Energy-momentum tensor.} We start by describing the simplest examples of local chiral fields, namely the components of the energy-momentum tensor. In the present case, they are given in terms of the current $\Jc$ by the expression
\begin{equation}\label{Eq:T}
T^\cL(x) = \frac{1}{2} \ps{\Jc_+(x)}{\Jc_+(x)} \qquad \text{ and } \qquad T^\cR(x) = -\frac{1}{2} \ps{\Jc_-(x)}{\Jc_-(x)}\,.
\end{equation}
One easily checks that $T^\cL$ is left-moving and $T^\cR$ right-moving (in fact, this is always the case classically, even for a $\Ec$-model which does not satisfy the strong conformal condition). Recall from equation \eqref{Eq:GaugeDeco} that $\Jc_\pm$ transforms by conjugation under the $H$-gauge symmetry: this ensures that the fields $T^\cL$ and $T^\cR$ are gauge-invariant observables.

We can compute the Poisson brackets of $T^\cL$ and $T^\cR$ starting from the current bracket for $\Jc$. We then find that they satisfy two decoupled classical Virasoro algebras:
\begin{subequations}
\begin{align}
\bigl\lbrace T^\cL(x), T^\cL(y) \bigr\rbrace &= -\bigl( T^\cL(x) + T^\cL(y) \bigr)\, \p_x\delta(x-y)\,, \label{Eq:VirL} \\
\bigl\lbrace T^\cR(x), T^\cR(y) \bigr\rbrace &= +\bigl( T^\cR(x) + T^\cR(y) \bigr)\, \p_x\delta(x-y)\,, \\
\bigl\lbrace T^\cL(x), T^\cR(y) \bigr\rbrace &= 0\,.
\end{align}
\end{subequations}
Let us also note that, with respect to the relativistic symmetry of the model, the fields $T^\cL$ and $T^\cR$ have spin $+2$ and $-2$ respectively.\\ 

Since $T^\cL$ and $T^\cR$ are gauge-invariant, we can evaluate their expression \eqref{Eq:T} in any gauge. In particular, recall from the previous subsubsection that in the gauge where we set $\Jc_{\hf}$ to zero, the fields $\Jc_+$ and $\Jc_-$ become the parafermionic fields $\Psi^\cL$ and $\Psi^\cR$. We can thus rewrite the chiral components of the energy-momentum tensor as
\begin{equation}\label{Eq:TPara}
T^\cL(x) = \frac{1}{2} \psb{\Psi^\cL(x)}{\Psi^\cL(x)} \qquad \text{ and } \qquad T^\cR(x) = - \frac{1}{2} \psb{\Psi^\cR(x)}{\Psi^\cR(x)}\,.
\end{equation}
This expression makes the chirality of $T^\cL$ and $T^\cR$ manifest, since $\Psi^\cL$ and $\Psi^\cR$ are themselves left- and right-moving. In contrast, it makes the locality of $T^\cLR$ less apparent, as the parafermions are non-local fields: we can thus see $T^\cLR$ as a particular combination of these non-local fields which turn out to be local\footnote{This can be seen directly from the expression \eqref{Eq:TPara}. Indeed, the parafermions $\Psi^\cLR$ are non-local due to the conjugacy by the field $\eta$ in their definition \eqref{Eq:Psi}. Yet the bilinear form $\psd$ is conjugacy invariant, ensuring that this non-local contribution cancels in equation \eqref{Eq:TPara}.}.

\paragraph{The $\bm{\Wc}$-algebra from the gauged current algebra.} Our goal now is to generalise the construction of the previous paragraph to find more general local chiral fields than $T^\cL$ and $T^\cR$. For instance, let $F$ be a polynomial on $V_+$ which is invariant under the adjoint action of $H$ and consider the field $F(\Jc_+)$. This is obviously a local field of the degenerate $\Ec$-model. Moreover, since $\Jc_+$ is transformed by conjugation under the $H$-gauge symmetry and $F$ is assumed to be invariant under such transformations, the field $F(\Jc_+)$ is by construction gauge-invariant. In particular, we can evaluate it after a gauge transformation by $\eta$, \textit{i.e.} in the gauge where $\Jc_{\hf}$ is set to zero and $\Jc_+$ becomes the parafermionic field $\Psi^\cL$: we then find $F(\Jc_+)=F(\Psi^\cL)$. Therefore, $F(\Jc_+)$ is a left-moving local field, as we were aiming to build. This construction is a clear generalisation of the one of $T^\cL$ in the previous paragraph, which corresponds to the case where $F$ is the second degree invariant polynomial on $V_+$ defined by the invariant bilinear form $\psd|_{V_+}$.\\

The above construction can be further generalised. Let us consider a field $W^\cL(x)$ which is built as a differential polynomial in the components of $\Jc_{\hf}(x)$ and $\Jc_+(x)$. To be more explicit, let us introduce the decompositions $\Jc_{\hf}(x)=\Jc_\hf^\alpha(x)\,T_\alpha$ and $\Jc_+(x) = \Jc_+^a(x)\,U_a$ along the bases $\lbrace T_\alpha \rbrace_{\alpha=1}^{d-p}$ and $\lbrace U_a \rbrace_{a=1}^p$ of $\hf$ and $V_+$ introduced earlier. The differential polynomial $W^\cL(x)$ is then defined as a linear combination of monomials of the form
\begin{equation}
\p_x^{p_1} \Jc_\hf^{\alpha_1}(x) \cdots \p_x^{p_m} \Jc_\hf^{\alpha_m}(x) \; \p_x^{q_1} \Jc_+^{a_1}(x) \cdots \p_x^{q_n} \Jc_+^{a_n}(x)\,,
\end{equation}
where $p_1,\dots,p_m,q_1,\dots, q_n \in \Z_{\geq 0}$ are non-negative integers, $\alpha_1,\dots,\alpha_m$ are labels in $\lbrace 1,\dots,d-p \rbrace$ and $a_1,\dots,a_n$ are labels in $\lbrace 1,\dots,p \rbrace$. We further assume that $W^\cL(x)$ is invariant under the $H$-gauge transformation \eqref{Eq:GaugeDeco}. By construction, this is a local field of the degenerate $\Ec$-model, directly expressed from the current and its derivatives. Moreover, since it is gauge-invariant, we can evaluate it after a gauge transformation by $\eta$, which sends $\Jc_{\hf}(x)$ to zero and $\Jc_+(x)$ to $\Psi^\cL(x)$. Since the latter is left-moving, so is $W^\cL(x)$. We thus conclude that $W^\cL(x)$ is a gauge-invariant left-moving local field, as wanted.  We will denote by $\Wc^\cL$ the space of such fields. The construction of the previous paragraph can be seen as a special case of this one, where $W^\cL(x)$ is built as a polynomial of only $\Jc_+(x)$ and not its derivatives nor $\Jc_{\hf}(x)$.

There is of course a similar construction of a set $\Wc^\cR$ of right-moving local fields, built as gauge-invariant differential polynomials in the components of $\Jc_{\hf}(x)$ and $\Jc_-(x)$. Since the fields in $\Wc^\cL$ and $\Wc^\cR$ can be expressed in terms of $\Psi^\cL$ and $\Psi^\cR$ respectively and the latter Poisson-commute by equation \eqref{Eq:PbPsiLR}, we see that the sets $\Wc^\cL$ and $\Wc^\cR$ are mutually Poisson-commuting. Moreover, the Poisson bracket of two fields in $\Wc^\cL$ (resp. $\Wc^\cR$) can be computed starting from the bracket \eqref{Eq:PbPara} of the parafermions or the initial current bracket of $\Jc$ (using the different expressions of these fields in different gauges): one checks that this bracket takes the form of a linear combination of derivatives of the Dirac distribution whose coefficients are themselves elements of $\Wc^\cL$ (resp. $\Wc^\cR$). The sets $\Wc^\cL$ and $\Wc^\cR$ thus form closed local Poisson algebras, which we call the $\Wc$-algebras or algebras of extended conformal symmetry of the model (they always contain the Virasoro subalgebra formed by the conformal generators $T^\cL(x)$ and $T^\cR(x)$, but are in general bigger, hence the name).\\

The definition of the $\Wc$-algebra $\Wc^\cLR$ can also be interpreted in terms of the Hamiltonian reduction of the current algebra associated with $\df_\pm$. Recall that $\df_\pm = \hf \oplus \hf' \oplus V_\pm$ is a subalgebra of $\df$ equipped with the invariant non-degenerate restricted bilinear form $\pm\psd|_{\df_\pm}$: we can thus consider the associated current algebra, which is realised here as the Poisson subalgebra of the initial one for $\Jc$ generated by the component $\Jc_\pm \pm \Jc_{\hf} \pm \Jc'$ valued in $\df_\pm$. Under the Hamiltonian reduction by the $\hf$-gauge symmetry, $\Jc'$ is set to zero as a first-class constraint and the reduced $\df_\pm$-current algebra is further obtained by considering the gauge-invariant observables built from the remaining components $\Jc_\pm$ and $\Jc_{\hf}$. In this context, the $\Wc$-algebra $\Wc^\cLR$ then corresponds to restricting ourselves to the gauge-invariant observables which are built only as local differential polynomials in $\Jc_\pm$ and $\Jc_{\hf}$. In contrast, we can obtain a bigger reduced algebra by also allowing non-local gauge-invariant combinations: this way, we find the parafermionic fields $\Psi^{\cLR}$ and the corresponding brackets \eqref{Eq:PbPara}.\\

Our goal in the rest of this subsection is to give a concrete procedure to find the gauge-invariant differential polynomials forming the $\Wc$-algebras $\Wc^\cLR$. In order to describe this general construction, it will be useful to first introduce a few additional ingredients, such as the covariant derivative.

\paragraph{Covariant derivative.} We say that a $\df$-valued field $\mathcal{K}(x)$ is covariant if it transforms by conjugation $\Kc \mapsto h^{-1} \Kc h$ under a gauge transformation with parameter $h\in H$. For instance, the components $\Jc_+$ and $\Jc_-$ of $\Jc$ are covariant fields. We define the covariant derivative of $\Kc(x)$ as
\begin{equation}
\nabla_x \Kc(x) = \p_x\Kc(x) + \bigl[ \Jc_{\hf}(x), \Kc(x) \bigr]\,.
\end{equation}
It takes its name from the property that $\nabla_x\Kc$ is covariant if the field $\Kc$ we start with is itself covariant. This can be checked explicitly from the transformation rules $\Kc \mapsto h^{-1} \Kc h$ and $\Jc_{\hf} \mapsto h^{-1} \Jc_{\hf} h + h^{-1}\p_x h$.

We note that the covariant derivative can be rephrased using the $H$-valued field $\eta$ given by equation \eqref{Eq:EtaPexp}. Indeed, it satisfies $\p_x \eta\,\eta^{-1} = -\Jc_{\hf}$, from which we find
\begin{equation}
\nabla_x \Kc = \Ad_{\eta}\, \p_x \bigl( \Ad_{\eta}^{-1} \Kc \bigr)\,.
\end{equation}
The covariance property of $\nabla_x\Kc$ can then be alternatively derived from the transformation rules $\Kc \mapsto h^{-1} \Kc h$ and $\eta \mapsto h^{-1} \eta$.\\

As observed above, the $V_\pm$-valued current $\Jc_\pm(x)$ is covariant. By construction, $\nabla_x^p\,\Jc_\pm(x)$ is thus also covariant for any non-negative integer $p\in\Z_{\geq 0}$. Since $\Jc_{\hf}(x)$ is valued in $\hf$, whose adjoint action preserves $V_\pm$, the $p^{\text{th}}$-covariant derivative $\nabla_x^p\,\Jc_\pm$ is also valued in $V_\pm$. This object will play a crucial role in the construction of the $\Wc$-algebra below.

\paragraph{Building fields in the $\bm\Wc$-algebra.} Let $m\in\Z_{>0}$ be a positive integer and $F : V_+^m \rightarrow \R$ be a multilinear $m$-form on $V_+$ (not necessarily symmetric), which is invariant under the adjoint action of $H$ on $V_+$. Concretely, the last condition means that
\begin{equation}
F\bigl( h^{-1} X_1 h, \dots, h^{-1} X_m h \bigr) = F(X_1,\dots,X_m)\qquad \forall \, X_1,\dots,X_m \in V_+ \,, \quad \forall\, h\in H\,.
\end{equation}
Choosing in addition a collection of $m$ non-negative integers $p_1,\dots,p_m \in \Z_{\geq 0}$, we define the field
\begin{equation}\label{Eq:WField}
W^\cL_{F\,;\;p_1,\dots,p_m}(x) = F\bigl( \nabla_x^{p_1} \Jc_+(x)\, , \, \dots \, , \, \nabla_x^{p_m} \Jc_+(x)\,\bigr)\,.
\end{equation}
By construction, this is a differential polynomial in the components of $\Jc_{\hf}(x)$ and $\Jc_+(x)$. Since all the fields $\nabla_x^m\Jc_+$ are covariant (see previous paragraph) and the form $F$ is invariant under the adjoint action of $H$, we see that the field $W^\cL_{F\,;\;p_1,\dots,p_m}(x)$ is gauge-invariant. It is thus an element of the $\Wc$-algebra $\Wc^\cL$ and in particular is a local left-moving field of the theory. With respect to the relativistic symmetry of the theory, this field has Lorentz spin
\begin{equation}
m + \sum_{k=1}^m p_k
\end{equation}
(intuitively, each current $\Jc_+$ and each covariant derivative $\nabla_x$ appearing in the definition \eqref{Eq:WField} of the field contributes to a unit of spin).\\ 

Since it is gauge-invariant, we can express the field $W^\cL_{F\,;\;p_1,\dots,p_m}$ as in equation \eqref{Eq:WField} but replacing $(\Jc_{\hf},\Jc_+)$ by $(\Jc_{\hf}^\eta,\Jc_+^\eta)=(0,\Psi^\cL)$. By construction, in this gauge, the $p^{\text{th}}$-covariant derivative $\nabla_x^p\Jc_+$ simply becomes the standard $p^{\text{th}}$-derivative $\p_x^p \Psi^\cL$ of the parafermionic field. Reinstating the time dependence of the fields and recalling that $\p_x = \p_+ - \p_-$ acts as $\p_+$ on the left-moving field $\Psi^\cL$, we can rewrite $W^\cL_{F\,;\;p_1,\dots,p_m}$ as\footnote{Let us make a quick comment on boundary conditions. It was explained in Subsubsection \ref{SubSec:Para} that the parafermionic field $\Psi^\cL(t,x)$ is not periodic but rather quasi-periodic, in the sense that $\Psi^\cL(t,x+2\pi)=\gamma^{-1}\,\Psi^\cL(t,x)\,\gamma$ for some matrix $\gamma$ independent of $t$ and $x$. Despite this, we note that the local chiral field \eqref{Eq:WFpara} is properly periodic, since the conjugations by $\gamma$ are ``cancelled'' by the ad-invariance of $F$.}
\begin{equation}\label{Eq:WFpara}
W^\cL_{F\,;\;p_1,\dots,p_m}(t,x) = F\bigl( \p_+^{p_1} \Psi^\cL(x^+) \, , \, \dots \, , \, \p_+^{p_m} \Psi^\cL(x^+) \bigr)\,.
\end{equation}
This makes manifest the left-moving dynamics of this field, but hides its locality (the latter being apparent in the initial gauge).\\

The procedure described in this paragraph produces many higher-spin fields in the $\Wc$-algebra $\Wc^\cL$. We conjecture that they in fact form the whole $\Wc$-algebra, \textit{i.e.} that any gauge-invariant differential polynomial in $\Jc_{\hf}$ and $\Jc_+$ is of the form $W^\cL_{F\,;\;p_1,\dots,p_m}$ for some invariant $m$-form $F$ and some integers $p_1,\dots,p_m$. This gives a quite efficient way to build the local chiral fields of the theory, translating the problem to a purely algebraic one, namely finding multilinear forms on $V_+$ invariant under the adjoint action of $H$. There is of course a similar construction for the right-moving $\Wc$-algebra $\Wc^\cR$, using invariant multilinear forms on $V_-$.

We will see direct applications of this procedure in explicit examples of conformal $\s$-models in Section \ref{Sec:Examples}. We note however that, although we expect it to produce all the chiral fields which are local when written in terms of the components of the current $\Jc$, this might in general not give all the chiral fields which are local when written in terms of the $\s$-model fields. This will in particular depend on the specific way we realise the degenerate $\Ec$-model as a $\s$-model, \textit{i.e.} on the choice of the maximally isotropic subalgebra $\kf$ of $\df$ (see Subsection \ref{SubSec:EModels}). For instance, we expect that two conformal $\s$-models which share the same underlying degenerate $\Ec$-model do not necessarily have equivalent sets of chiral fields local in the $\s$-model fields: more precisely, they will both have at least the ones coming from the above construction but might also admit other ones which are not visible from the $\Ec$-model perspective. The study of such fields then requires a case by case analysis.

\subsubsection{Coset CFTs and the Goddard-Kent-Olive construction}
\label{SubSec:GKO}

In the previous subsubsections, we have described the chiral fields of the conformal degenerate $\Ec$-model in terms of the Hamiltonian reduction of the current Poisson algebra associated with $\df_\pm$. This is reminiscent of a class of conformal models called coset CFTs, described by the so-called Goddard-Kent-Olive (GKO) construction~\cite{Goddard:1984vk,Goddard:1986ee}. In this subsubsection, we explain how the two constructions are related, at least under certain technical assumptions.

\paragraph{The coset setup.} In order to relate our construction to the GKO one, we will have to restrict to a particular configuration for the structure of the double Lie algebra $\df$ and the gauge subalgebra $\hf$ underlying the degenerate $\Ec$-model. More precisely, we will suppose that\vspace{-3pt}
\begin{enumerate}[(i)]\setlength\itemsep{0.2pt}
\item $\df = \g_+ \oplus \g_-$, where $\g_\pm$ are orthogonal mutually-commuting subalgebras ;
\item there exist subalgebras $\hf_\pm \subset \g_\pm$ such that $\psd|_{\hf_\pm}$ is non-degenerate ;
\item there exists a Lie algebra isomorphism $\lambda: \hf_+ \to \hf_-$ which sends $\psd|_{\hf_+}$ to $-\psd|_{\hf_-}$. \vspace{-4pt}
\end{enumerate}
In this case, we then define
\begin{equation}\label{Eq:hGKO}
\hf = \lbrace X+\lambda(X)\,,\, X\in\hf_+ \rbrace \qquad \text{ and } \qquad \hf' = \lbrace X-\lambda(X)\,,\, X\in\hf_+ \rbrace\,.
\end{equation}
One easily checks that $\hf$ is an isotropic subalgebra of $\df$, which can then serve as the gauge subalgebra of a degenerate $\Ec$-model (note that, as Lie algebras, $\hf$, $\hf_+$ and $\hf_-$ are all isomorphic). Moreover, we find that $\hf'$ is an isotropic subspace, pairing non-degenerately with $\hf$ through the bilinear form $\psd$. We finally let $V_\pm$ be the orthogonal complement of $\hf_\pm$ in $\g_\pm$ (which has trivial intersection with $\hf_\pm$ since $\psd|_{\hf_\pm}$ is non-degenerate). We then have the decomposition
\begin{equation}\label{Eq:DecoGKO}
\df = \hf_+ \oplus V_+ \oplus \hf_- \oplus V_- = \hf \oplus \hf' \oplus V_+ \oplus V_-\,,
\end{equation}
which satisfies the strong conformal condition introduced at the beginning of this subsection. This setup thus defines a conformal degenerate $\Ec$-model. The corresponding operator $\Eh:\df\to\df$ reads
\begin{equation}\label{Eq:ECoset}
\Eh|_{V_\pm} = \pm\,\Id_{V_\pm}\,, \qquad \Eh|_{\hf_+} = \frac{1}{2}\bigl( \Id_{\hf_+} + \lambda \bigr) \qquad \text{ and } \qquad \Eh|_{\hf_-} = -\frac{1}{2}\bigl( \Id_{\hf_-} + \lambda^{-1} \bigr)\,.
\end{equation}
In the notations of this section, the subalgebras $\df_\pm$ are then given by
\begin{equation}
\df_\pm = V_\pm \oplus \hf_+ \oplus \hf_- = \g_\pm \oplus \hf_\mp\,.
\end{equation}

\paragraph{Decomposition of the current and constraint.} Consider the current $\Jc$ of the degenerate $\Ec$-model. In the rest of this section, we have used its decomposition $\Jc=\Jc_+-\Jc_-+\Jc_\hf+\Jc'$, along the direct sum $\df = V_+ \oplus V_- \oplus \hf \oplus \hf'$. The setup described in the previous paragraph suggests to consider another decomposition, reading
\begin{equation}
\Jc = \Jc_+ + \Jc_{\hf_+} - \Jc_- - \Jc_{\hf_-}\,, \qquad \text{ with } \qquad \Jc_\pm \in V_\pm\,, \quad \Jc_{\hf_\pm} \in \hf_\pm\,,
\end{equation}
along the direct sum $\df=V_+\oplus\hf_+\oplus V_-\oplus\hf_-$. The $V_\pm$-valued components $\Jc_\pm$ are the same in the two decompositions, while the others are related by
\begin{equation}\label{Eq:DecoCoset}
\Jc_{\hf} = \frac{\Id_{V_+}+\lambda}{2} \left( \Jc_{\hf_+} - \lambda^{-1}(\Jc_{\hf_-}) \right) \qquad \text{ and } \qquad \Jc' = \frac{\Id_{V_+} - \lambda}{2} \left( \Jc_{\hf_+} + \lambda^{-1}(\Jc_{\hf_-}) \right) \,.
\end{equation}
The advantage of the coset setup is that we can identify many closed subalgebras of the current Poisson algebra of $\Jc$. For instance, the fields
\begin{equation}
\Jc_{\g_\pm} = \Jc_\pm + \Jc_{\hf_\pm}
\end{equation}
satisfy two Poisson-commuting current algebras associated with $\g_\pm=V_\pm \oplus \hf_\pm$. Moreover, the fields $\Jc_{\hf_\pm}$ alone form current subalgebras\footnote{Written in a basis, these two current subalgebras for $\hf_+$ and $\hf_-$ share the same $\delta(x-y)$ component (expressed in terms of the structure constants of $\hf\simeq\hf_\pm$) but have opposite $\p_x\delta(x-y)$ components.} associated with $\hf_\pm$.

Recall that the physical observables of the degenerate $\Ec$-model are obtained by a Hamiltonian reduction. The latter is performed by imposing the first-class constraint $\Jc'\approx 0$ and further restricting to observables which are invariant under the gauge symmetry it generates. In the language of this subsubsection, we see from equation \eqref{Eq:DecoCoset} that in terms of the fields $\Jc_{\hf_\pm}$, the constraint now reads
\begin{equation}\label{Eq:ConstCoset}
\Jc_{\hf_+} + \lambda^{-1}(\Jc_{\hf_-}) \approx 0\,.
\end{equation}

\paragraph{Chiral Poisson algebras as classical GKO cosets.} Let us now consider the left-moving Poisson algebra of the theory. As explained in the previous subsubsections, it can be seen as a subsector of the aforementioned Hamiltonian reduction where we start with the current Poisson subalgebra corresponding to $\df_+ \subset \df$ (more precisely, we obtain the left-moving $\Wc$-algebra $\Wc^\cL$ if we restrict to gauge-invariant local fields or the left-moving parafermionic algebra of $\Psi^\cL$ if we allow for non-local fields). In the present setup, this $\df_+$-subalgebra is generated by the $\g_+$-valued field $\Jc_{\g_+} = \Jc_+ + \Jc_{\hf_+}$ and the $\hf_-$-valued field $\Jc_{\hf_-}$. Imposing the constraint \eqref{Eq:ConstCoset}, we see that we can eliminate the latter by imposing $\Jc_{\hf_-} \approx -\lambda(\Jc_{\hf_+})$, so that we are left with observables built only from the $\g_+$-current $\Jc_{\g_+}$. To obtain the reduced algebra, we further need to restrict to gauge-invariant quantities, which Poisson-commute with the generator $\Jc_{\hf_+} + \lambda^{-1}(\Jc_{\hf_-})$: however, since we have used the constraint to only get observables built from $\Jc_{\g_+}$, their Poisson bracket with $\Jc_{\hf_+} + \lambda^{-1}(\Jc_{\hf_-})$ is the same as their bracket with $\Jc_{\hf_+}$ alone. To summarise, the reduced left-moving algebra can thus be seen as the set of observables built from the $\g_+$-current $\Jc_{\g_+}$ and which Poisson-commute with the $\hf_+$-component $\Jc_{\hf_+}$. More precisely, we obtain the $\Wc$-algebra $\Wc^\cL$ if we focus on local observables built as differential polynomials in $\Jc_{\g_+}$ or the parafermionic algebra if we also allow for non-local observables.

This is a classical equivalent of the GKO construction~\cite{Goddard:1984vk,Goddard:1986ee} for the $\g_+/\hf_+$ coset algebra. The advantage compared with the previously used formulation is that we now start with the $\g_+$-current algebra and simply look at observables commuting with the $\hf_+$-current subalgebra: we do not have to pass through the bigger algebra $\df_+$, the imposition of the constraint and the Hamiltonian reduction. There is of course an analogue construction for the right-moving chiral Poisson algebra, which starts with the algebra $\df_-$. In that case, we end up with a $\g_-/\hf_-$ GKO coset algebra.

\paragraph{The GKO energy-momentum tensor.} The GKO coset construction also gives a natural expression of the chiral components of the energy-momentum tensor. Focusing on the left-moving sector, which corresponds to a $\g_+/\hf_+$ coset, the GKO prescription is
\begin{equation}
T^\cL = T_{\g_+} - T_{\hf_+}\,,
\end{equation}
where $T_{\g_+}$ is the energy-momentum tensor of the ``quotiented'' $\g_+$-current algebra while $T_{\hf_+}$ is the energy-momentum tensor of the ``quotient'' $\hf_+$-current algebra. These two quantities are given by the (classical) Segal-Sugawara construction and we thus find
\begin{equation}
T^\cL = \frac{1}{2} \ps{\Jc_{\g_+}}{\Jc_{\g_+}} - \frac{1}{2} \ps{\Jc_{\hf_+}}{\Jc_{\hf_+}}\,.
\end{equation}
This is a local field built from the $\g_+$-current and one easily checks that it Poisson-commutes with the component $\Jc_{\hf_+}$: it thus belongs to the $\g_+/\hf_+$ coset $\Wc$-algebra introduced in the previous paragraph. Recalling the decomposition $\Jc_{\g_+} = \Jc_{\hf_+} + \Jc_+$ along the orthogonal direct sum $\g_+ = \hf_+ \oplus V_+$, we see that the above expression reduces to $\frac{1}{2} \ps{\Jc_+}{\Jc_+}$: we thus recover the definition of $T^\cL$ considered previously in equation \eqref{Eq:T}.

\paragraph{Comments.} Let us end this subsubsection with two brief comments. The GKO construction and the coset CFTs have been extensively studied in the literature, including at the quantum level. The results described above thus pave the way for the quantisation of a large class of conformal degenerate $\Ec$-models using these approaches. A natural question at this point is whether all the degenerate $\Ec$-models which satisfy the strong conformal condition fall into the GKO scheme. This is in fact not the case: indeed, we will study in Subsection \ref{SubSec:Diam} an explicit example of such a model, where the underlying double algebra does not have the ``coset'' structure described at the beginning of this subsubsection.\\

For our final comment, let us consider the linear map $\Ec: \df\to\df$ defined by $\Ec|_{\g_\pm}=\pm\,\Id_{\g_\pm}$. This is a symmetric operator satisfying $\Ec^2=\Id$: it can thus be taken as the $\Ec$-operator of a non-degenerate $\Ec$-model, as considered in Section \ref{Sec:ND}. Moreover, the eigenspaces $\Ker(\Ec\mp\Id)$ of $\Ec$ are identified with $\g_\pm$, which are orthogonal mutually-commuting subalgebras of $\df$. This means that the non-degenerate $\Ec$-model built from $\Ec$ satisfies the strong conformal condition discussed in Subsection \ref{SubSec:ConfN} and is thus conformal. The conformal degenerate $\Ec$-model that we considered in this subsubsection is obtained from this non-degenerate one by a procedure introduced in~\cite{Klimcik:2019kkf}, which ``gauges away'' the subalgebra $\hf$. As explained in~\cite{Klimcik:2019kkf}, this procedure amounts to defining a reduced operator $\Eh$ from $\Ec$, which satisfies the right properties to be the $\Eh$-operator of a degenerate $\Ec$-model. More precisely, the reduced operator $\Eh$ is defined as $\Ec$ itself on $V_\pm\oplus\hf'$ but acts trivially on $\hf$. In the present case, this procedure gives back the operator $\Eh$ considered in equation \eqref{Eq:ECoset}. We thus see that the conformal degenerate $\Ec$-models which fall into the GKO coset construction correspond to the gauging of conformal non-degenerate $\Ec$-models. This is reminiscent of the construction of gauged Wess-Zumino-Witten models from ungauged ones: as we shall see in Subsection \ref{SubSec:gWZW}, these theories are specific examples among the class of conformal degenerate $\Ec$-models studied here.

\section{Examples of conformal degenerate \texorpdfstring{$\bm\Ec$}{E}-models}
\label{Sec:Examples}

We now come to the description of explicit examples of degenerate $\Ec$-models satisfying the strong conformal condition, thus illustrating the general formalism described in the previous section.

\subsection{Gauged Wess-Zumino-Witten models}
\label{SubSec:gWZW}

Let us start with the $G/H_0$ gauged Wess-Zumino-Witten (WZW) model~\cite{Witten:1983ar,Witten:1983tw}, whose interpretation as a degenerate $\Ec$-model was first established in~\cite{Klimcik:1999ax}. It is the degenerate equivalent of the WZW model on $G$, whose interpretation as a conformal non-degenerate $\Ec$-model was recalled in Subsection \ref{SubSec:WZW}. There, we saw the WZW model as the UV fixed-point of an integrable but non-conformal $\Ec$-model called the $\lambda$-deformation. Such integrable deformations exist in the gauged case as well, at least for appropriate choice of gauged subgroup $H_0$~\cite{Sfetsos:2013wia,Hollowood:2014rla,Hollowood:2014qma,Hoare:2021dix}. However, for conciseness, we will restrict the analysis below to the conformal point only. For simplicity, we also restrict to the case of a vectorially gauged $G/H_0$ WZW model with $G$ a simple Lie group and $H_0$ a subgroup of $G$ whose Lie algebra is non-degenerate with respect to the Killing form of $\text{Lie}(G)$.

\paragraph{Double Lie algebra and gauged subalgebra.} Let us now describe the degenerate $\Ec$-model underlying the $G/H_0$ gauged WZW model. We let $G$ be a real simple connected Lie group with Lie algebra $\g=\text{Lie}(G)$. We denote by $\kappa(\cdot,\cdot)$ the ``minimal'' invariant bilinear form on $\g$, which is equal to the Killing form multiplied by an appropriate normalisation prefactor (see Subsection \ref{SubSec:WZW} and the footnote \ref{FootForm} for details). In particular, this form is definite positive if $G$ is compact. Similarly to the non-degenerate case discussed in Subsection \ref{SubSec:WZW}, we take the double group of the $\Ec$-model to be $D=G\times G$ and we equip its Lie algebra $\df=\g \times \g$ with the bilinear form \eqref{Eq:FormLambda}, which we recall here for convenience:
\begin{equation}\label{Eq:FormWZW}
\psb{(X_1,X_2)}{(Y_1,Y_2)} = \kay\,\kappa(X_1,Y_1) - \kay\,\kappa(X_2,Y_2)\, .
\end{equation}
Here, $\kay$ is a real positive parameter. This bilinear form $\psd$ is then invariant, non-degenerate and of split signature, as wanted.

The additional ingredient in the degenerate case is the choice of the gauge subalgebra $\hf\subset\df$. To define it, we will suppose that we are given a subgroup $H_0 \subset G$, with Lie algebra $\hf_0 \subset \g$, such that $\kappa(\cdot,\cdot)$ is non-degenerate on $\hf_0$. We then define
\begin{equation}
\hf = \hf_{0,\diag} = \bigl\lbrace (X,X), \, X\in\hf_0 \bigr\rbrace\,,
\end{equation}
\textit{i.e.} the diagonal embedding of $\hf_0$ in $\df=\g\times\g$. It is clear that this is a subalgebra of $\df$, isotropic with respect to the bilinear form \eqref{Eq:FormWZW}. It thus possesses the right properties to be the gauge subalgebra of a degenerate $\Ec$-model.

We also introduce the orthogonal subspace $W$ of $\hf_0$ in $\g$. Since we supposed that $\kappa(\cdot,\cdot)$ is non-degenerate on $\hf_0$, this subspace has trivial intersection with $\hf_0$ and we thus have the decomposition $\g = \hf_0 \oplus W$. We will denote by $\pi_0$ and $\pi_W$ the corresponding projectors. From the ad-invariance of $\kappa(\cdot,\cdot)$, one finds that $W$ is stable under the adjoint action of $\hf_0$. We thus have
\begin{equation}\label{Eq:ComH0}
[ \hf_0, \hf_0 ] \subset \hf_0 \qquad \text{ and } \qquad [ \hf_0, W ] \subset W \,.
\end{equation}
We also note that, in the double algebra $\df=\g\times\g$, the subspace orthogonal to $\hf$ is given by
\begin{equation}
\hf^\perp = \hf \oplus ( W\! \times\! W)\,.
\end{equation}

\paragraph{The $\bm\Eh$-operator.} Let us now define the operator $\Eh:\df\to\df$ characterising the degenerate $\Ec$-model. Since $\df=\g\times\g$, we can conveniently see the operators on this space as $2\times 2$ matrices whose entries are operators on $\g$. Using this convention, we will define $\Eh$ as:
\begin{equation}
\Eh =  \begin{pmatrix}
\frac{1}{2} \pi_0 + \pi_W & -\frac{1}{2} \pi_0 \\
\frac{1}{2} \pi_0 & -\frac{1}{2}\pi_0 - \pi_W
\end{pmatrix}\,.
\end{equation}
One easily checks that $\Eh$ is symmetric, $\Eh|_{\hf^\perp}^3=\Eh|_{\hf^\perp}$, $\Ker(\Eh)=\hf$ and $\Eh$ commutes with the adjoint action of $\hf$ (using equation \eqref{Eq:ComH0}). It thus possesses the right properties to be the $\Eh$-operator of a degenerate $\Ec$-model with gauge subalgebra $\hf$.\\

With respect to this choice of operator $\Eh$, the eigenspaces  $V_\pm=\Ker(\Eh|_{\hf^\perp} \mp \Id)$ appearing in the decomposition $\hf^\perp=\hf\oplus V_+ \oplus V_-$ are simply given by
\begin{equation}
V_+ = W \times \lbrace 0 \rbrace \qquad \text{ and } \qquad V_- = \lbrace 0 \rbrace \times W\,.
\end{equation}
Recall from Subsection \ref{SubSec:EModels} that $\hf'$ is defined as the unique subspace of $\df$ which is isotropic, orthogonal to $V_\pm$ and pairing non-degenerately with $\hf$. In the present case, we find
\begin{equation}
\hf' = \bigl\lbrace (X,-X), \, X\in\hf_0 \bigr\rbrace\,.
\end{equation}
It is clear that
\begin{equation}
\hf \oplus \hf' = \hf_0 \times \hf_0
\end{equation}
is a subalgebra of $\df=\g\times\g$ and that $[V_+,V_-]=\lbrace 0 \rbrace$. The degenerate $\Ec$-model under consideration thus satisfies the strong conformal condition introduced at the beginning of Subsection \ref{SubSec:Conf}: it should therefore define a conformal theory.

\paragraph{Maximally isotropic subalgebra and gauge symmetries.} In order to find a Lagrangian realisation of this conformal theory as a $\s$-model, we follow the construction described in Subsubsection \ref{SubSec:Lag}. The starting point is to choose a maximally isotropic subalgebra $\kf$ of the double $\df$. Here, we will take the same choice that gave us the non-gauged WZW model in the non-degenerate Subsection \ref{SubSec:WZW}, namely
\begin{equation}
\kf = \g_{\diag} = \bigl\lbrace (X,X), \; X\in \g \bigr\rbrace\,.
\end{equation}
According to Subsection \ref{SubSec:Lag}, the associated $\s$-model has then for target space the double quotient $\KDH = G_{\diag} \setminus (G\times G) / H_{0,\diag}$ and is described by a field $\ell(t,x)$ valued in $G\times G$, with a left gauge symmetry $\ell \mapsto k\ell$, $k\in G_{\diag}$, and a right one $\ell \mapsto \ell h$, $h\in H_{0,\diag}$. As in the non-degenerate subsection \ref{SubSec:WZW}, we will fix the left $G_{\diag}$-gauge symmetry from the start. More precisely, we use this symmetry to bring $\ell$ to the form
\begin{equation}
\ell(t,x) = \bigl( g(t,x), \Id \bigr), \qquad g(t,x) \in G\,,
\end{equation}
hence describing the theory in terms of a $G$-valued field $g(t,x)$.

This formulation will possess a residual $H_0$-gauge symmetry. On the initial field $\ell$, it acts by the right multiplication $\ell \mapsto \ell(h,h)$, where $h$ is valued in $H_0$, so that $(h,h)\in H_{0,\diag}$. However, this transformation does not preserve the gauge-fixing $\ell=(g,\Id)$ considered above, as it sends it to $(gh,h)$. This issue is solved by performing an appropriate left $G_{\diag}$-gauge transformation to bring back the field in the right gauge: more precisely, multiplying on the left by $(h^{-1},h^{-1})\in G_\diag$, we obtain $(h^{-1}gh,\Id)$, which respects the chosen gauge-fixing condition. To summarise, the residual $H_0$-gauge symmetry then acts on the $G$-valued field $g(t,x)$ by the local conjugation
\begin{equation}\label{Eq:GaugeWZW}
g(t,x) \longmapsto h(t,x)^{-1}g(t,x)h(t,x)\,, \qquad \text{ with } \qquad h(t,x)\in H_0\,.
\end{equation}

\paragraph{$\bm\s$-model action.} In order to derive the action of the $\s$-model, the next step is to determine the expression of the projectors $W_\ell^\pm$, whose kernel and image are defined by equation \eqref{Eq:Wpm}. In the present case, this means that
\begin{equation}
\Ker(W_\ell^\pm) = \Ad_\ell^{-1} \kf = \Ad_{(g,\Id)}^{-1}\, \g_\diag = \bigl\lbrace (g^{-1}Xg, X), \, X\in\g \bigr\rbrace\,,
\end{equation}
together with
\begin{subequations}
\begin{align}
\text{Im}(W_\ell^+) &= \hf \oplus V_+ = \hf_{0,\diag} \oplus ( W \times \lbrace 0 \rbrace ) = \bigl\lbrace (X,\pi_0 X), \, X\in\g \bigr\rbrace\,, \\
\text{Im}(W_\ell^-) &= \hf \oplus V_- = \hf_{0,\diag} \oplus ( \lbrace 0 \rbrace \times W ) = \bigl\lbrace (\pi_0X, X), \, X\in\g \bigr\rbrace\,.
\end{align}
\end{subequations}
With a little bit of algebra, this leads us to
\begin{equation}\label{Eq:WgWZW}
W^+_\ell = \begin{pmatrix}
\dfrac{1}{1-\Ad_g^{-1}\pi_0}\! & -\dfrac{1}{1-\Ad_g^{-1}\pi_0} \Ad_g^{-1} \vspace{5pt} \\
\pi_0 \dfrac{1}{1-\Ad_g^{-1}\pi_0}\! & -\pi_0\dfrac{1}{1-\Ad_g^{-1}\pi_0} \Ad_g^{-1}
\end{pmatrix}\,, \qquad W^-_\ell = \begin{pmatrix}
-\pi_0 \dfrac{1}{1-\Ad_g\pi_0}\Ad_g\! & \pi_0 \dfrac{1}{1-\Ad_g\pi_0} \vspace{5pt} \\
-\dfrac{1}{1-\Ad_g\pi_0}\Ad_g\! &  \dfrac{1}{1-\Ad_g\pi_0}
\end{pmatrix} \,.
\end{equation}
Using this and $\ell^{-1}\p_\pm \ell = (g^{-1}\p_\pm g,0)$, we can compute the $\s$-model action \eqref{Eq:Action}. We then find
\begin{equation}\label{Eq:ActionGWZW}
S_{\text{gWZW},\kay}[g] = S_{\text{WZW},\,\kay}[g] + 2\,\kay \iint_{\R\times\mathbb{S}^1} \dd t\,\dd x\; \kappa\left( g^{-1} \p_+ g, \pi_0\frac{1}{1-\Ad_g\pi_0} \p_-gg^{-1} \right)\,,
\end{equation}
where we recall that $S_{\text{WZW},\,\kay}[g]$ is the non-gauged  WZW action with level $\kay$, as defined in equation \eqref{Eq:WZW}. We recognize in the above formula the $G/H_0$ vectorially gauged WZW model~\cite{Witten:1983ar,Witten:1983tw}, which is a well-known conformal $\s$-model. One can check that this action is invariant under the gauge transformation \eqref{Eq:GaugeWZW}, as expected from the construction. As a consequence, the target space of this $\s$-model can be seen as the group $G$ quotiented by the adjoint action of the subgroup $H_0$.

\paragraph{Currents $\bm\Ac_\pm$ and $\bm\Jc_\pm$.} Using the expression \eqref{Eq:WgWZW} of the projectors $W^\pm_\ell$ and the fact that $\ell^{-1}\p_\pm \ell = (g^{-1}\p_\pm g,0)$, we can compute the currents $\Ac_\pm$ and $\Jc_\pm$, following the equation \eqref{Eq:AJ}. For the gauge field $\Ac_\pm$, we find
\begin{equation}\label{Eq:GaugeFieldWzw}
\Ac_+ = (1,1)\otimes \pi_0 \dfrac{1}{1-\Ad_g^{-1}\pi_0} g^{-1}\p_+ g\qquad \text{ and }  \qquad \Ac_- = -(1,1)\otimes \pi_0 \dfrac{1}{1-\Ad_g \pi_0} \p_-g g^{-1}\,,
\end{equation}
where we used the notation $(a,b)\otimes X=(aX,bX)$ to denote pure factors in $\df=\g\times\g \simeq \R^2\otimes\g$. The above fields are valued in $\hf=\hf_{0,\diag}=(1,1)\otimes\hf_0$, as expected. Similarly, the currents $\Jc_\pm$ read
\begin{equation}\label{Eq:JgWZW}
\Jc_+ = (1,0)\otimes \pi_W \dfrac{1}{1-\Ad_g^{-1}\pi_0} g^{-1}\p_+ g\qquad \text{ and }  \qquad \Jc_- = -(0,1)\otimes \pi_W \dfrac{1}{1-\Ad_g \pi_0} \p_-g g^{-1}\,.
\end{equation}
As wanted, these fields are valued in $V_+=W \times \lbrace 0 \rbrace$ and $V_-= \lbrace 0 \rbrace \times W$.

\paragraph{Chiral algebras.} The currents $\Ac_\pm$ and $\Jc_\pm$ computed above satisfy the equations of motion \eqref{Eq:CEoMpm}. As explained in Subsection \ref{SubSec:Conf}, this ensures the existence of chiral fields in the theory, either non-local (forming the parafermionic Poisson algebra) or local (forming the $\Wc$-algebra). We recover this way various well established results in the literature. For instance, the presence of parafermionic fields in the gauged WZW model was discussed in the references~\cite{Bardakci:1990lbc,Bardakci:1990ad,Karabali:1989dk}. We note moreover that the conformal degenerate $\Ec$-model considered here falls into the GKO coset scheme discussed in Subsubsection \ref{SubSec:GKO}. The left- and right-moving sectors of the theory can thus be seen as cosets of a $\g$-current algebra by its $\hf_0$-current subalgebra. This is compatible with the results of~\cite{Karabali:1988au,Gawedzki:1988hq}, which showed (at the quantum level) that gauged WZW models form a Lagrangian realisation of the GKO coset construction.

For simplicity, we will not discuss these chiral algebras in more detail for the general $G/H_0$ gauged WZW model. Instead, we will illustrate how this construction works concretely for a simple example, namely the $SU(2)/U(1)$ case. This will be the subject of the following subsection.

\subsection[The \texorpdfstring{$SU(2)/U(1)$}{SU(2)/U(1)} gauged WZW model]{The \texorpdfstring{$\bm{SU(2)/U(1)}$}{SU(2)/U(1)} gauged WZW model}
\label{SubSec:Su2u1}

We now specialise the setup of the previous subsection to the case where $G=SU(2)$ and $H_0=U(1)$. As we will see, in that case, it will be possible to give a very concrete description of the general objects introduced in Subsection \ref{SubSec:Conf}, such as the parafermionic fields and the $\Wc$-algebra. For completeness, we note that there exists a whole panorama of interesting variants of this setup, obtained by changing the group $G$ to $SL(2,\R)$, the subgroup $H_0$ to $\R$, the gauging from vectorial to axial or various combinations thereof. Of particular interest are the axially gauged WZW models on $SL(2,\R)/U(1)$ and $SL(2,\R)/\R$, which describe bosonic strings propagating in the 2d Euclidean and Lorentzian black holes geometry respectively~\cite{Witten:1991yr} (in the Euclidean case, the target space of the model then takes the form of the Hamilton cigar~\cite{Hamilton}). These theories have been studied in great detail in the literature and many things are known about their structure, including their parafermions~\cite{Karabali:1989dk,Bardakci:1990lbc} and their $\Wc$-algebra~\cite{Bakas:1991fs}.\footnote{For a recent and detailed discussion of the parafermionic fields and the $\Wc$-algebra of the cigar $\s$-model, we refer to the section 6 of the lecture notes of Lukyanov and Zamolodchikov, published in~\cite{Lukyanov:2019asr}.} Here, we will recover these results classically -- in the compact $SU(2)/U(1)$ case -- from the $\Ec$-model construction. For future reference, we will also briefly discuss the case of the Lorentzian black hole $\s$-model, \textit{i.e.} the axially gauged $SL(2,\R)/\R$ WZW theory, at the end of this subsection.  

\paragraph{Lie algebra conventions.} Let us consider the Lie algebra $\g=\su(2)$ with basis
\begin{equation}\label{Eq:BasisSu2}
\td_1 = \begin{pmatrix}
\ri & 0 \\ 0 & -\ri
\end{pmatrix}\,, \qquad \td_2 = \begin{pmatrix}
0 & 1 \\ -1 & 0
\end{pmatrix}\,, \qquad \td_3 = \begin{pmatrix}
0 & \ri \\ \ri & 0
\end{pmatrix}\,,
\end{equation}
written in the fundamental representation. With respect to this basis, the commutation relations and the bilinear form\footnote{For $\g=\su(2)$, this form is given by $\kappa(X,Y)=$, with the trace taken in the fundamental representation.} $\kappa(\cdot,\cdot)$ read
\begin{equation}
[\td_a, \td_b] = 2\sum_{c=1}^3 \epsilon_{abc}\,\td_c \qquad \text{ and } \qquad \kappa(\td_a,\td_b) = 2\,\delta_{ab}\,,
\end{equation}
where $\epsilon_{abc}$ is the totally antisymmetric Levi-Civita tensor. In the notations of the the previous subsection, we choose the subalgebra $\hf_0 \subset \g$ and its orthogonal subspace $W$ to be
\begin{equation}
\hf_0 = \R\,\td_1 \simeq \mathfrak{u}(1) \qquad \text{ and } \qquad W = \R\,\td_2 \oplus \R\,\td_3\,.
\end{equation}

We will visualise elements of the double group $D=SU(2)\times SU(2)$ and its Lie algebra $\df$ as block-diagonal $4\times 4$ matrices, with $2\times 2$ entries in $SU(2)$ and $\su(2)$ respectively. For instance, the gauged subalgebra $\hf=\hf_{0,\diag}$ in $\df$ can be seen as
\begin{equation}\label{Eq:u1}
\hf = \R\,T_1 \qquad \text{ with } \qquad T_1 = \text{\footnotesize $\begin{pmatrix} 
\ri & 0 & 0 & 0 \\
0 & -\ri  & 0 & 0 \\
0 & 0 & \ri & 0 \\
0 & 0 & 0 & -\ri \\
\end{pmatrix}$}\,,
\end{equation}
while the subspaces $V_+$ and $V_-$ are given by
\begin{equation}\label{Eq:VpmSu2}
V_+ =  \left\lbrace \text{\footnotesize $\begin{pmatrix}
0 & a & 0 & 0 \\
-\overline{a} & 0  & 0 & 0 \\
0 & 0 & 0 & 0 \\
0 & 0 & 0 & 0 \\
\end{pmatrix}$},\; a\in\mathbb{C} \right\rbrace \qquad \text{ and } \qquad V_- =  \left\lbrace \text{\footnotesize $\begin{pmatrix}
0 & 0 & 0 & 0 \\
0 & 0  & 0 & 0 \\
0 & 0 & 0 & a  \\
0 & 0 & -\overline{a} & 0  \\
\end{pmatrix}$},\; a\in\mathbb{C} \right\rbrace\,.
\end{equation}

\paragraph{Parametrisation of the field.} We will parametrise the $SU(2)$-valued field $g$ as
\begin{equation}\label{Eq:Euler}
g = \begin{pmatrix}
e^{\ri\alpha} \sin(\theta) & e^{-\ri\beta} \cos(\theta) \\
-e^{\ri\beta} \cos(\theta) & e^{-\ri\alpha} \sin(\theta)
\end{pmatrix} = \exp\left( \frac{\alpha-\beta}{2}\, \td_1 \right)\exp\left( \left(\frac{\pi}{2}-\theta \right)\, \td_2 \right)\exp\left( \frac{\alpha+\beta}{2}\, \td_1 \right) \,,
\end{equation}
in terms of three coordinate fields $\theta$, $\alpha$ and $\beta$, which are essentially Euler angles on $SU(2)$. The use of linear combinations $\alpha\pm\beta$ on the left and right exponentials is related to the $H_0\simeq U(1)$ gauge symmetry of the theory. Indeed, recall from the general formula \eqref{Eq:GaugeWZW} that this local symmetry acts on $g$ by conjugation by a $U(1)$-valued field $h$. The latter can be parametrised as $h=\exp(\frac{\eta}{2}\,\td_1)$. One easily checks that the transformation $g\mapsto h^{-1}gh$ then simply acts on the coordinate fields $(\theta,\alpha,\beta)$ by
\begin{equation}
\theta \longmapsto \theta\,, \qquad \alpha \longmapsto \alpha\,, \qquad \beta \longmapsto \beta + \eta\,.
\end{equation}
The fields $(\theta,\alpha)$ are thus physical degrees of freedom of the model, while $\beta$ can be gauged away.

\paragraph{Action and target space geometry.} Inserting the parametrisation \eqref{Eq:Euler} of $g$ into the action \eqref{Eq:ActionGWZW}, we can express the latter in terms of the coordinate fields. As expected from the discussion above, the field $\beta$ completely disappears throughout the computation and the action depends in the end only on the physical fields $(\theta,\alpha)$. More precisely, we find
\begin{equation}\label{Eq:Su2u1}
S[\theta,\alpha] = 2\kay\,\iint_{\R\times \mathbb{S}^1} \dd t\,\dd x \; \bigl( \p_+ \theta \, \p_- \theta + \tan^2(\theta)\,\p_+\alpha\,\p_-\alpha \bigr)\,.
\end{equation}

Let us briefly describe the geometry of the target space of this $\s$-model. In terms of the coordinates $(\theta,\alpha)$, the metric reads
\begin{equation}\label{Eq:MetSu2u1}
\dd s^2 = \kay\,\bigl(\dd \theta^2 + \tan^2(\theta)\,\dd \alpha^2 \bigr)\,.
\end{equation}
One can check explicitly that the metric \eqref{Eq:MetSu2u1} defines a conformal $\s$-model, by computing the corresponding Ricci tensor $R_{\mu\nu}$, where $\mu,\nu$ are target space indices labelling the coordinates $(\theta,\alpha)$. Doing so, we find
\begin{equation}\label{Eq:Ricci}
R_{\mu\nu} + 2\nabla_\mu \nabla_\nu \Phi = 0\,,
\end{equation}
where $\nabla_\mu$ denotes the covariant derivative and
\begin{equation}
\Phi = - \frac{1}{2} \log\bigl(\cos^2(\theta)\bigr) + \text{constant}
\end{equation}
is the dilaton. This property of $R_{\mu\nu}$ ensures that the 1-loop RG-flow~\cite{Friedan:1980jf,Fradkin:1985ys,Callan:1985ia} of the $\s$-model \eqref{Eq:Su2u1} is trivial, as expected (in mathematical terms, we say that the metric \eqref{Eq:MetSu2u1} is a steady Ricci soliton).

\paragraph{Isometry and dual field.} The metric \eqref{Eq:MetSu2u1} possesses a $U(1)$ isometry, which acts on the angle $\alpha$ as shifts $\alpha\mapsto\alpha+\xi$. In terms of the $\s$-model, this means that the action \eqref{Eq:Su2u1} is invariant under the global symmetry $\alpha(t,x)\mapsto\alpha(t,x)+\xi$ (where $\xi$ here is a constant parameter, as the symmetry is only global). This symmetry is associated with a conservation equation
\begin{equation}\label{Eq:ConsPi}
\p_+ \Pi_- + \p_- \Pi_+ = 0\,,
\end{equation}
with conserved current
\begin{equation}\label{Eq:Pi}
\Pi_\pm = \tan^2(\theta)\,\p_\pm\alpha\,.
\end{equation}
This conservation equation has an important consequence, which will be useful later. Namely, it implies the (on-shell) existence of a function $\alpha^\cD(t,x)$, which we will call the dual field, such that
\begin{equation}
\p_\pm \alpha^\cD = \pm\, \Pi_\pm\,.
\end{equation}
Integrating the spatial component of the above equation, we can realise $\alpha^\cD(t,x)$ as a non-local field
\begin{equation}\label{Eq:Dual}
\alpha^\cD(t,x) = \int^x \tan^2\bigl( \theta(t,y) \bigr) \, \p_t \alpha(t,y)\,\dd y\,.
\end{equation}

\paragraph{Gauge field and $\bm\eta$.} Recall the general expression \eqref{Eq:GaugeFieldWzw} of the gauge field $\Ac_\pm$. For the $SU(2)/U(1)$ case considered here, a direct computation shows that
\begin{equation}\label{Eq:Au1}
\Ac_\pm = \frac{1}{2} \bigl( \p_\pm\beta \pm \Pi_\pm \bigr) \,T_1\,,
\end{equation}
where $T_1$ is the generator \eqref{Eq:u1} of $\hf$ and $\Pi_\pm$ is the conserved current \eqref{Eq:Pi} associated with the $\alpha$-isometry of the model. From the general considerations of Subsection \ref{SubSec:Conf}, we expect this gauge field $\Ac_\pm$ to satisfy the flatness condition \eqref{Eq:CEoMA} on-shell. Here, the gauged subalgebra $\hf\simeq\mathfrak{u}(1)$ is abelian and this equation thus simply becomes
\begin{equation}
\p_+ \Ac_- - \p_- \Ac_+ = 0\,.
\end{equation}
It is clear that the field \eqref{Eq:Au1} satisfies this property, due to the conservation equation \eqref{Eq:ConsPi} of $\Pi_\pm$.\\

An important role in the construction of chiral observables in the general Subsection \ref{SubSec:Conf} was played by the $H$-valued field $\eta$, which is introduced by $\Ac_\pm = - (\p_\pm \eta)\eta^{-1}$. In the present case, we easily find
\begin{equation}
\eta = \exp\left( -\frac{\alpha^\cD+\beta}{2}\,T_1 \right)\,,
\end{equation}
where $\alpha^\cD$ is the dual field \eqref{Eq:Dual}, characterised by the property $\p_\pm\alpha^\cD=\pm\Pi_\pm$. In equation \eqref{Eq:EtaPexp}, we explained that $\eta$ is in general expressed in terms of a path-ordered exponential: in the present case, $H$ is abelian, so the ordering does not matter, and $\eta$ can thus be written as a simple exponential. The non-locality of the field $\eta$ is now encoded in its relation to the dual field $\alpha^\cD$, which is itself defined as a non-local integral \eqref{Eq:Dual}. All the non-locality that will arise in the rest of this subsection will always come from the presence of this dual field only. This is the main advantage of the $SU(2)/U(1)$ case, which is simple enough so that we can express quite explicitly all the relevant observables in terms of the coordinate fields of the $\s$-model and this dual field.

\paragraph{Parafermions.} Using the equation \eqref{Eq:JgWZW} and the parametrisation \eqref{Eq:Euler} of the field $g$, we can express the covariant currents $\Jc_\pm$ in terms of the coordinate fields $(\theta,\alpha,\beta)$. An additional straightforward computation allows one to determine the parafermionic fields $\Psi^\cL$ and $\Psi^\cR$, defined from $\Jc_\pm$ and $\eta$ by the equation \eqref{Eq:Psi}. More precisely, we find
\begin{subequations}\label{Eq:ParaSu2u1}
\begin{equation}
\Psi^\cL = \begin{pmatrix}
0 & -\psib^\cL & 0 & 0 \\
\psi^\cL & 0 & 0 & 0 \\
0 & 0 & 0 & 0 \\
0 & 0 & 0 & 0 
\end{pmatrix} \qquad \text{ and } \qquad \Psi^\cR = \begin{pmatrix}
0 & 0 & 0 & 0 \\
0 & 0 & 0 & 0 \\
0 & 0 & 0 & \psi^\cR \\
0 & 0 & -\psib^\cR & 0  
\end{pmatrix}\,,
\end{equation}
where
\begin{equation}
\psi^\cL = e^{\ri(\alpha-\alpha^\cD)} \bigl( \p_+ \theta + \ri\,\tan(\theta) \,\p_+\alpha)\,, \;\;\qquad \psib^\cL = e^{-\ri(\alpha-\alpha^\cD)} \bigl( \p_+ \theta - \ri\,\tan(\theta) \,\p_+\alpha)\,,
\end{equation}
\begin{equation}
\psi^\cR = e^{\ri(\alpha+\alpha^\cD)} \bigl( \p_- \theta + \ri\,\tan(\theta) \,\p_-\alpha)\,, \;\;\qquad \psib^\cL = e^{-\ri(\alpha+\alpha^\cD)} \bigl( \p_- \theta - \ri\,\tan(\theta) \,\p_-\alpha)\,.
\end{equation}
\end{subequations}
Here, rather than parametrising the $V_\pm$-valued parafermions $\Psi^\cL$ and $\Psi^\cR$ in terms of four real fields, we used the complex fields $\psi^\cL$ and $\psi^\cR$ and their complex conjugates $\psib^\cL$ and $\psib^\cR$. These quantities are non-local, due to the presence of the dual field $\alpha^\cD$. We also see that although the initial currents $\Jc_\pm$ contain the field $\beta$, the latter does not appear in the expression of the parafermionic fields: this is expected since these fields are gauge-invariant and should thus be expressed purely in terms of the physical degrees of freedom $(\theta,\alpha)$. The expression \eqref{Eq:ParaSu2u1} of the parafermionic fields, obtained here from the $\Ec$-model construction, coincides with the one initially derived in~\cite{Bardakci:1990lbc}. \\

One checks that $\psi^\cLR$ are chiral under the equations of motion derived from the action \eqref{Eq:Su2u1}, \textit{i.e.} that $\p_-\psi^\cL=\p_+\psi^\cR=0$, as expected. Performing the canonical analysis of the model, one can also compute the Poisson algebra formed by these chiral fields: we will not give the details of this analysis here but will summarise the main results. In agreement with the general expectations of Subsubsection \ref{SubSec:Para}, we find that the left-moving fields $\psi^\cL$ and $\psib^\cL$ Poisson-commute with the right-moving ones $\psi^\cR$ and $\psib^\cR$. Moreover, the left-moving Poisson algebra reads
\begin{equation*}
\bigl\lbrace \psi^\cL(x), \psi^\cL(y) \bigr\rbrace = \!\frac{1}{\kay} \, \psi^\cL(x)\, \psi^\cL(y)\,\epsilon(x-y)\,,\! \qquad \bigl\lbrace \psib^\cL(x), \psib^\cL(y) \bigr\rbrace = \!\frac{1}{\kay} \, \psib^\cL(x)\, \psib^\cL(y)\,\epsilon(x-y)\,,
\end{equation*}
\begin{equation}\label{Eq:PbParaSu2u1}
\bigl\lbrace \psi^\cL(x), \psib^\cL(y) \bigr\rbrace = -\frac{1}{\kay}\,\p_x\delta(x-y) - \frac{1}{\kay}\, \psi^\cL(x)\, \psib^\cL(y)\,\epsilon(x-y)\,.
\end{equation}
We recover here the parafermionic Poisson algebra initially derived in~\cite{Bardakci:1990lbc}. The right-moving one is similar, up to a few changes in signs. The presence of the non-local distribution $\epsilon(x-y)$ in the above brackets is due to the appearance of the dual field $\alpha^\cD$ in the expression \eqref{Eq:ParaSu2u1} of the parafermions. One can check that these Poisson brackets agree with the parafermionic Poisson algebra \eqref{Eq:PbParaBasis} derived in the general case.

\paragraph{The $\bm\Wc$-algebra.} We now turn to the study of the $\Wc$-algebras of the theory\footnote{This is the $SU(2)/U(1)$ analogue of the $SL(2,\R)/U(1)$ coset $\Wc$-algebra underlying the cigar $\s$-model~\cite{Bakas:1991fs} (see also the lecture notes~\cite{Lukyanov:2019asr} for a detailed discussion of this $\Wc$-algebra).}, which are composed by left- or right-moving local fields and which were discussed in Subsubsection \ref{SubSec:W} for a general conformal degenerate $\Ec$-model. As explained there, the simplest elements in these $\Wc$-algebras are the chiral components of the energy-momentum tensor, which are related to the parafermionic fields $\Psi^\cL$ and $\Psi^\cR$ by equation \eqref{Eq:TPara}. For the $SU(2)/U(1)$ gauged WZW model, the latter are explicitly given by \eqref{Eq:ParaSu2u1}. We then find
\begin{subequations}
\begin{align}
T^\cL = \kay\, \psi^\cL \, \psib^\cL &= \kay\bigl( (\p_+\theta)^2 + \tan^2(\theta)\,(\p_+\alpha)^2 \bigr)\,, \\
T^\cR = \kay\, \psi^\cR \, \psib^\cR &= \kay\bigl( (\p_-\theta)^2 + \tan^2(\theta)\,(\p_-\alpha)^2 \bigr)\,.
\end{align}
\end{subequations}
These fields satisfy $\p_- T^\cL=\p_+ T^\cR=0$, as expected. Moreover, they are clearly local when written in terms of the $\s$-model fields $(\theta,\alpha)$. In contrast, recall that the parafermionic fields are non-local, due to the terms $e^{\pm\ri\alpha^{\cD}}$ in the equation \eqref{Eq:ParaSu2u1}. The quantity $T^\cL=  \kay\, \psi^\cL \, \psib^\cL$ is a particular combination of these non-local fields which turns out to be local, since these exponential terms cancel. This illustrates on a simple example the general principles discussed in Subsubsection \ref{SubSec:W}.\\

In that subsubsection, we presented a general and systematic procedure to build more complicated elements of the $\Wc$-algebras than the energy-momentum tensor. Let us now illustrate how this procedure works in practice for the example of the $SU(2)/U(1)$ gauged WZW model. We will focus on the left-moving $\Wc$-algebra $\Wc^\cL$, the study of the right-moving one being very similar. As explained in Subsubsection \ref{SubSec:W}, the key ingredients of the construction are the multilinear forms on $V_+$ which are invariant under the adjoint action of $H$. In the present case, the subspaces $\hf=\text{Lie}(H)$ and $V_+$ are given explicitly by equations \eqref{Eq:u1} and \eqref{Eq:VpmSu2}. We can thus search for such invariant forms by a brute-force approach, at least for the first degrees. For the degree 1, we find that there are in fact no invariant linear forms on $V_+$. For the degree 2, we obtain one symmetric invariant bilinear form, namely the restriction $\psd|_{V_+}$, and one skew-symmetric one, defined by
\begin{equation}
\begin{array}{rccc}
\omega: & V_+ \times V_+ & \longrightarrow & \R \\
        &     (X,Y)      &   \longmapsto   & \ps{T_1}{[X,Y]}
\end{array}\,,
\end{equation}
where we recall that $T_1$ is the generator of $\hf$ defined in equation \eqref{Eq:u1}. It is clear that $\omega$ is a skew-symmetric bilinear form on $V_+$. Its invariance under the adjoint action of $H$ is proven as follows: for all $X,Y\in V_+$ and all $h\in H$, we have
\begin{equation}
\omega(\Ad_h X, \Ad_h Y) = \ps{T_1}{\Ad_h[X,Y]} = \ps{\Ad_h^{-1}T_1}{[X,Y]} = \ps{T_1}{[X,Y]} = \omega(X,Y)\,.
\end{equation}
Here, we have used successively the property that $\Ad_h$ is an automorphism, the ad-invariance of $\psd$ and the fact that $H$ is abelian (we note in passing that a similar construction will hold more generally for cases where $\hf$ possesses a non-trivial center).

We now follow the equation \eqref{Eq:WFpara} to build elements of the $\Wc$-algebra $\Wc^\cL$ from these invariant mutlinear forms and the parafermionic field $\Psi^\cL$, using the expression \eqref{Eq:ParaSu2u1} of the latter. We find that there is only one spin-2 field in $\Wc^\cL$, namely the energy-momentum tensor $W_2^\cL=T^\cL$ considered above. We further build two independent spin-3 fields. The first one is given by
\begin{equation}
\psb{\Psi^\cL}{\p_+\Psi^\cL} = \frac{1}{2}\, \p_+ \psb{\Psi^\cL}{\Psi^\cL} = \p_+ T^\cL
\end{equation}
and its thus a descendant of the energy-momentum tensor $T^\cL$. The second one is not obtained from $T^\cL$ and its derivatives and is built using the skew-symmetric form $\omega$. It reads
\begin{equation}
W^\cL_3 = \frac{1}{4}\, \omega\bigl( \Psi^\cL, \p_+ \Psi^\cL) = \frac{\kay}{2\ri}\bigl( \psi^\cL \,\p_+ \psib^\cL - \psib^\cL\,\p_+ \psi^\cL\bigr) = \kay\,\text{Im}\bigl( \psi^\cL \,\p_+ \psib^\cL \bigr) \,.
\end{equation}
It is clear that this is a real combination of the parafermionic fields $\psi^\cL$ and $\psib^\cL$ and one can moreover check that the non-local contributions $e^{\pm\ri\alpha^\cD}$ in these fields cancel in this particular combination, making $W^\cL_3$ local, as expected. When written in terms of the $\s$-model fields $(\theta,\alpha)$, we find
\begin{equation}
\kay^{-1} \, W_3^\cL = \tan^2(\theta) \bigl( \tan^2(\theta) - 1 \bigr) (\p_+\alpha)^3 + \tan(\theta)\, \bigl( \p_+^2\theta\,\p_+\alpha - \p_+\theta\, \p_+^2 \alpha \bigr) - 2 \, (\p_+\theta)^2\;\p_+\alpha\,.
\end{equation}
As a consistency check, one can verify that $\p_- W^\cL_3=0$ using the equations of motion of the fields $(\theta,\alpha)$ derived from the action \eqref{Eq:Su2u1}.

Similarly, one finds four independent spin-4 fields. Three of them are descendants of lower-spin ones, namely $T^\cL\,^2$, $\p_+^2 T^\cL$ and $\p_+ W^\cL_3$. The last one cannot be expressed as a polynomial of $T^\cL$, $W^\cL_3$ and their derivatives and can be chosen to be
\begin{equation}
W^\cL_4 =  \psb{\Psi^\cL}{\p_+^2\Psi^\cL} = \kay\bigl( \psi^\cL \,\p_+^{\hspace{1pt} 2} \psib^\cL + \psib^\cL\,\p_+^{\hspace{1pt}2} \psi^\cL  \bigr)\,.
\end{equation}
The explicit local expression of $W^\cL_4$ in terms of the $\s$-model fields $(\theta,\alpha)$ can be easily derived, but for conciseness, we will not give it here. 

Let us point out an important fact. As we said above, $W^\cL_4$ cannot be expressed as a differential polynomial of $T^\cL$ and $W^\cL_3$. However, it is not functionally independent of these two fields, as it satisfies the very simple relation
\begin{equation}
2 \, T^\cL \, W^\cL_4 - 2\,  T^\cL \, \p_+^2 T^\cL + \bigl(\p_+ T^\cL\bigr)^2 + 4W_3^\cL\,^2 = 0\,.
\end{equation}
This is in fact to be expected. Indeed, since all the densities in the $\Wc$-algebra are expressed in terms of the parafermion $\psi^\cL$ and its conjugate $\psib^\cL$, they depend in the end only on two independent real-valued fields. Thus, we expect any three of these densities to be functionally related. However, this relation can in general be rather complicated, as it will involve product, derivatives and powers of these densities. We thus make the choice of working with fields like $W_4^\cL$, which allows us to always  deal with differential polynomials only (for instance, replacing $W_4^\cL$ by its expression in terms of $T^\cL$ and $W_3^\cL$ would introduce quotients by $T^\cL$ in the construction). \\

One can easily go on with the procedure initiated above, at least for the first few spins. We expect that for any $s\geq2$, it produces one spin-$s$ local field $W^\cL_s$ which is not expressed as a differential polynomial in the lower-spin ones, generalising the construction of $W^\cL_2=T^\cL$, $W^\cL_3$ and $W^\cL_4$ above. We stress that this scheme is somehow quite efficient to find the chiral local fields of the theory, as it relates this problem to a quite simpler purely algebraic question, namely the determination of invariant multilinear forms on $V_+$. As a matter of comparison, a brute force search for a spin-3 chiral local field such as $W_3^\cL$, starting from a general ansatz built from $(\theta,\alpha)$ and their derivatives, is already a quite complicated and tedious task, while the derivation of $W_3^\cL$ above can be easily done in a few lines.

In addition, the procedure also provides a rather efficient way of computing the Poisson bracket between elements of $\Wc^\cL$. Indeed, rather than performing a canonical analysis, in terms of the coordinate fields $(\theta,\alpha)$ and their conjugate momenta, we can use the expression of the fields $W^\cL_s$ in terms of the parafermions $\Psi^\cL$, or alternatively the currents $\Jc_+$ and $\Jc_{\hf}$ (see Subsubsection \ref{SubSec:W}). The latter satisfy rather simple closed Poisson algebras, which can be used to compute the Poisson structure of $\Wc^\cL$. We expect that the Poisson bracket of $W^\cL_s(x)$ and $W^\cL_{s'}(y)$ takes the form of a linear combination of derived Dirac distributions $\p_x^k\delta(x-y)$, with $k\in\lbrace 0,\dots,s+s'-1\rbrace$, whose coefficients are differential polynomials in the fields $W^\cL_r(x)$, for $r\in\lbrace 2,\dots,s+s'-2 \rbrace$. The typical example of such a bracket is the one of $W_2^\cL=T^\cL$ with itself, which takes the form of the classical Virasoro algebra \eqref{Eq:VirL}.

\paragraph{The Lorentzian black hole $\bm\s$-model.} For future reference, we end this subsection with a brief discussion of a close cousin of the $SU(2)/U(1)$ gauged WZW theory, namely the Lorentzian black hole $\s$-model, which corresponds to an axially gauged $SL(2,\R)/\R$ WZW model~\cite{Witten:1991yr}. The degenerate $\Ec$-model interpretation of this theory is based on the double algebra $\df=\mathfrak{sl}(2,\R)\times \mathfrak{sl}(2,\R)$ and the gauge subalgebra $\hf = \R\, T_1 = \R\,(\td_1, -\td_1)$, where $\td_1 = \diag(1,-1)$ is the Cartan generator of $\mathfrak{sl}(2,\R)$. Compared to the $SU(2)/U(1)$ case considered earlier, we first note that this generator differs by a factor $\ri$ from its compact analogue defined in equation \eqref{Eq:BasisSu2}, as one can expect. The other difference is the minus sign in the second entry of $T_1=(\td_1,-\td_1)$, which encodes the fact that we are now gauging an axial symmetry rather than a vectorial one. One easily checks that this definition of $\hf$ yields an isotropic subalgebra of $\df$, as required. This model can be analysed following the exact same method as for the vector $SU(2)/U(1)$ theory above: we will skip all the details of this analysis and simply give the main end results, using the parallel with the compact case.

We find that the target space of the $\s$-model is still described by two physical fields $(\theta,\alpha)$, which, compared to the $SU(2)/U(1)$ case, are now valued in $\R$ rather than being angles. The metric and dilaton of this target space read
\begin{equation}\label{Eq:LorBH1}
\dd s^2 = \kay\,\bigl(\dd \theta^2 - \tanh^2(\theta)\,\dd \alpha^2 \bigr) \qquad \text{ and } \qquad \Phi = -\frac{1}{2} \log\bigl(\cosh^2(\theta)\bigr)\,.
\end{equation}
The model possesses an isometry, acting as translations of the coordinate $\alpha$. The corresponding conserved current $\Pi_\pm$ and dual field $\alpha^\cD(t,x)$ are given by
\begin{equation}
\Pi_\pm = - \tanh^2(\theta)\,\p_\pm\alpha = \pm\,\p_\pm\alpha^\cD \qquad \text{ and } \qquad \alpha^\cD(t,x) = -\int^x \tanh^2\bigl( \theta(t,y) \bigr) \, \p_t \alpha(t,y)\,\dd y\,.
\end{equation}
Moreover, we find non-local left-moving parafermions
\begin{equation}\label{Eq:ParaBH}
\psi^\cL = e^{\alpha-\alpha^\cD} \bigl( \p_+ \theta + \tanh(\theta) \,\p_+\alpha)\,, \;\;\qquad \widetilde{\psi}\,^\cL = e^{\alpha^\cD - \alpha} \bigl( \p_+ \theta -\tanh(\theta) \,\p_+\alpha)\,,
\end{equation}
as well as right-moving ones which we will not describe for conciseness. These are the analogues of the $SU(2)/U(1)$ parafermions \eqref{Eq:ParaSu2u1}: note that here, these fields are real and $\widetilde{\psi}\,^\cL$ is not the complex conjugate of $\psi^\cL$ (contrarily to $\psib^\cL$ in the compact case). As expected, the model also possesses local chiral fields, forming a $SL(2,\R)/\R$ coset $\Wc$-algebra. The first few of these fields read
\begin{subequations}\label{Eq:WBH}
\begin{align}
\kay^{-1}\,W_2^\cL & = \psi^\cL \, \widetilde{\psi}\,^\cL = (\p_+\theta)^2 - \tanh^2(\theta)\,(\p_+\alpha)^2 \,, \\[8pt]
\kay^{-1}\,W^\cL_3 &= \bigl( \psi^\cL \,\p_+ \widetilde{\psi}\,^\cL -  \widetilde{\psi}\,^\cL\,\p_+ \psi^\cL \bigr)/2 \\[3pt]
&= \tanh^2(\theta) \bigl( \tanh^2(\theta) + 1 \bigr) (\p_+\alpha)^3 + \tanh(\theta)\, \bigl( \p_+^2\theta\,\p_+\alpha - \p_+\theta\, \p_+^2 \alpha \bigr) - 2 \, (\p_+\theta)^2\;\p_+\alpha \,,  \notag \\[8pt]
\kay^{-1}\,W^\cL_4 &= \psi^\cL \,\p_+^{\hspace{1pt}2} \widetilde{\psi}\,^\cL +  \widetilde{\psi}\,^\cL\,\p_+^{\hspace{1pt} 2} \psi^\cL \,.
\end{align}
\end{subequations}

Finally, it will be useful for future reference to write the metric \eqref{Eq:LorBH1} in so-called conformal coordinates, \textit{i.e.} to bring it to the form of the flat Minkowski metric multiplied by a unique (non-constant) factor. This is done by introducing the new coordinate $r=-\log\bigl( \sinh(\theta) \bigr)$ and performing the change of variables $(\theta,\alpha) \mapsto (r,\alpha)$ in the metric \eqref{Eq:LorBH1}. We then simply find
\begin{equation}\label{Eq:LorBH2}
\dd s^2 = \kay \, \frac{\dd r^2 - \dd\alpha^2}{1+e^{2r}}\,.
\end{equation}

\subsection{An example based on a solvable Lie group}
\label{SubSec:Diam}

Up to a few exceptions, many of the conformal $\s$-models that are studied in the literature are based on simple or semi-simple Lie groups. Our goal in this subsection is to describe a conformal degenerate $\Ec$-model which, in contrast, is built from a solvable Lie group.

\paragraph{Double group and gauge subgroup.} Consider $6\!\times\!6$ matrices of the form
\begin{equation}\label{Eq:MatDiam}
M(\xi_i) =  \begin{pmatrix}
1 & e^{\ri \xi_1} (\xi_3+\ri\,\xi_4) & e^{\ri \xi_1} (\xi_5+\ri\,\xi_6) & 0 & 0 & \ri\, \xi_2 + \frac{1}{2}( \xi_3^2 + \xi_4^2 -  \xi_5^2 - \xi_6^2 ) \vspace{3pt} \\
0 & e^{\ri \xi_1} & 0 & 0 & 0 & \xi_3 - \ri\,\xi_4 \vspace{1pt}\\
0 & 0 & e^{\ri \xi_1} & 0 & 0 & -\xi_5+\ri\,\xi_6 \vspace{1pt}\\
0 & 0 & 0 & e^{-\ri \xi_1} & 0 & \xi_3 + \ri\,\xi_4 \vspace{1pt}\\
0 & 0 & 0 & 0 & e^{-\ri \xi_1} & \xi_5 + \ri\,\xi_6 \vspace{1pt}\\
0 & 0 & 0 & 0 & 0 & 1
\end{pmatrix}\,,
\end{equation}
parametrised by 6 numbers $\xi_1\in[0,2\pi]$ and $\xi_2,\dots,\xi_6\in\R$. We then define
\begin{equation}
D = \bigl\lbrace M(\xi_i),\,\;\xi_1\in[0,2\pi],\;\,\xi_2,\dots,\xi_6\in\R \bigr\rbrace\,,
\end{equation}
\textit{i.e.} the set formed by all these matrices. One checks that $D$ is stable under matrix inversion and multiplication and thus form a 6-dimensional Lie subgroup of $SL(6,\mathbb{C})$. We will take $D$ to be our $\Ec$-model double group in this subsection. The main difference with the other examples considered in this article is that this group is not semi-simple, but instead is solvable (this follows directly from the fact that it has a representation formed by triangular matrices).\\

The identity of the group $D$ corresponds to the point $\xi_i=0$, \textit{i.e.} $M(0)=\Id$. To obtain the Lie algebra $\df=\text{Lie}(D)$, one thus has to look at the matrices $M(\xi_i)$ for infinitesimal $\xi_i$'s. We find
\begin{equation}\label{Eq:TiDiam}
M(\xi_i) = \Id + \sum_{i=1}^6 T_i\, \xi_i + o(\xi_i)\,,
\end{equation}
where $T_1,\dots,T_6$ are $6\!\times\!6$ triangular matrices whose explicit expression we will not give here for conciseness. The family $\lbrace T_A \rbrace_{A=1}^6$ forms a basis of the Lie algebra $\df$ of $D$. The non-vanishing commutation relations in this basis read
\begin{equation}
[T_1,T_3] = -T_4\,, \quad [T_1,T_4] = T_3, \quad [T_1,T_5] = -T_6\,, \quad [T_1,T_6] = T_5, \quad [T_3,T_4]=-2T_2, \quad [T_5,T_6] = 2T_2\,.
\end{equation}
In particular, we note that the element $T_2\in\df$ is central, \textit{i.e.} commutes with all the other elements.

In order to see $\df$ as the double Lie algebra of an $\Ec$-model, we have to equip it with a non-degenerate invariant bilinear form $\psd$, with split signature. Here we will define this form by its entries $\eta_{AB}=\ps{T_A}{T_B}$ in the basis $\lbrace T_A \rbrace_{A=1}^6$. Namely, we let
\begin{equation}\label{Eq:FormDiam}
\bigl( \eta_{AB} \bigr)_{A,B=1,\dots,6} = \kay \text{\small $\begin{pmatrix}
0 & 1 & 0 & 0 & 0 & 0 \\
1 & 0 & 0 & 0 & 0 & 0 \\
0 & 0 & 2 & 0 & 0 & 0 \\
0 & 0 & 0 & 2 & 0 & 0 \\
0 & 0 & 0 & 0 & -2 & 0 \\
0 & 0 & 0 & 0 & 0 & -2 \\
\end{pmatrix}$}\,.
\end{equation}
One shows that this choice indeed defines an invariant bilinear form of split signature $(3,3)$.\\

Let us now discuss our choice of gauge subgroup $H \subset D$. We will take it to be the 1-parameter subgroup generated by $T_1$. Concretely, $H$ is then composed of the matrices $M(\xi_1,0,0,0,0,0)$ with $\xi_1\in[0,2\pi]$, \textit{i.e.} the matrices of the form \eqref{Eq:MatDiam} where all the parameters $\xi_i$ are set to zero except $\xi_1$. The corresponding Lie algebra $\hf=\text{Lie}(H)$ is then trivially given by
\begin{equation}
\hf = \R\,T_1\,, \qquad \text{ with } \qquad T_1 = \text{\footnotesize $\begin{pmatrix}
0 & 0 & 0 & 0 & 0 & 0 \\
0 & \ri & 0 & 0 & 0 & 0 \\
0 & 0 & \ri & 0 & 0 & 0 \\
0 & 0 & 0 & -\ri & 0 & 0 \\
0 & 0 & 0 & 0 & -\ri & 0 \\
0 & 0 & 0 & 0 & 0 & 0 \\
\end{pmatrix}$}\,.
\end{equation}
Since the entry $\eta_{11}$ in the bilinear form \eqref{Eq:FormDiam} is vanishing, we see that the subalgebra $\hf$ is isotropic. It thus possesses the right properties to be chosen as the gauge subalgebra of a degenerate $\Ec$-model.

\paragraph{$\bm\Ec$-operator.} We will define the operator $\Eh:\df\to\df$ of the degenerate $\Ec$-model by its action
\begin{equation}
\Eh(T_1)=0\,, \quad \Eh(T_2)=T_1\,, \quad \Eh(T_3)=T_3\,, \quad \Eh(T_4)=T_4\,, \quad \Eh(T_5)=-T_5\,, \quad \Eh(T_6)=-T_6
\end{equation}
on the basis $\lbrace T_A \rbrace_{A=1}^6$. One easily checks that $\Eh|_{\hf^\perp}^3=\Eh|_{\hf^\perp}$, $\Ker(\Eh)=\hf$ and $\Eh$ commutes with the adjoint action of $\hf$, ensuring that $\Eh$ is an appropriate choice of $\Eh$-operator.

In that case, the eigenspace decomposition $\hf^\perp=\hf\oplus V_+ \oplus V_-$ with $\hf=\Ker(\Eh)$ and $V_\pm=\Ker(\Eh\mp\Id)$ is given by $V_+ = \R\,T_3 \oplus \R\,T_4$ and $V_- = \R\,T_5 \oplus \R\,T_6$. In matrix terms, this gives
\begin{equation}\label{Eq:VDiam}
V_+ = \left\lbrace \text{\footnotesize $\begin{pmatrix}
0 & a & 0 & 0 & 0 & 0 \\
0 & 0 & 0 & 0 & 0 & \overline{a} \\
0 & 0 & 0 & 0 & 0 & 0 \\
0 & 0 & 0 & 0 & 0 & a \\
0 & 0 & 0 & 0 & 0 & 0 \\
0 & 0 & 0 & 0 & 0 & 0 \\
\end{pmatrix}$},\; a\in\mathbb{C} \right\rbrace\qquad \text{ and } \qquad V_- = \left\lbrace \text{\footnotesize $\begin{pmatrix}
0 & 0 & a & 0 & 0 & 0 \\
0 & 0 & 0 & 0 & 0 & 0 \\
0 & 0 & 0 & 0 & 0 & -\overline{a} \\
0 & 0 & 0 & 0 & 0 & 0 \\
0 & 0 & 0 & 0 & 0 & a \\
0 & 0 & 0 & 0 & 0 & 0 \\
\end{pmatrix}$},\; a\in\mathbb{C} \right\rbrace\,.
\end{equation}

The subspace $\hf'$, defined in Subsection \ref{SubSec:EModels} as the subspace of $\df$ which is isotropic, orthogonal to $V_\pm$ and pairing non-degenerately with $\hf$, is given in the present case by $\hf' = \R\,T_2$. One easily checks that $\hf\oplus\hf'$ is a subalgebra of $\df$ and that $[V_+,V_-]=\lbrace 0 \rbrace$. Thus, the degenerate $\Ec$-model considered here satisfies the strong conformal condition introduced in Subsection \ref{SubSec:Conf}.

\paragraph{Maximally isotroptic subalgebra and gauge symmetries.} To relate this degenerate $\Ec$-model to a $\s$-model, we have to choose a maximally isotropic subalgebra $\kf$ in $\df$. In terms of the basis $\lbrace T_A \rbrace_{A=1}^6$, we will take it to be
\begin{equation}
\kf = \R\,T_1 \, \oplus \, \R\,(T_3+T_5) \, \oplus \, \R\,(T_4+T_6)\,.
\end{equation}
One easily checks that $\kf$ is stable under commutation and is isotropic with respect to the bilinear form \eqref{Eq:FormDiam}. The corresponding subgroup $K\subset D$ can be seen as
\begin{equation}
K = \left\lbrace\;\; \text{\small $\begin{pmatrix}
1 & e^{\ri \zeta}\,a & e^{\ri \zeta}\,a & 0 & 0 & 0 \vspace{3pt} \\
0 & e^{\ri \zeta} & 0 & 0 & 0 & \overline{a} \vspace{1pt}\\
0 & 0 & e^{\ri \zeta} & 0 & 0 & -\overline{a} \vspace{1pt}\\
0 & 0 & 0 & e^{-\ri \zeta} & 0 & a \vspace{1pt}\\
0 & 0 & 0 & 0 & e^{-\ri \zeta} & a \vspace{1pt}\\
0 & 0 & 0 & 0 & 0 & 1
\end{pmatrix}$}\,, \; \zeta\in[0,2\pi]\, , \, a\in\mathbb{C} \right\rbrace\,.
\end{equation}

The resulting $\s$-model is described by a $D$-valued field $\ell(t,x)$, subject to a $K$-left and a  $H$-right gauge symmetries, acting as $\ell\mapsto k\ell h$, where $k\in K$ and $h\in H$ are arbitrary fields. Let us first treat the left symmetry. As explained above, elements of the double group $D$ can be seen as matrices $M(\xi_i)$, parametrised by 6 numbers $\xi_i$ as in equation \eqref{Eq:MatDiam}. Consider now the left multiplication $k\,M(\xi_i)$, where $k$ is valued in the group $K$ defined above. One shows that $k$ can be uniquely chosen so that it ``eliminates'' the parameters $(\xi_1,\xi_5,\xi_6)$, \textit{i.e.} such that $k\,M(\xi_i)$ takes the form $M(0,\xi'_2,\xi'_3,\xi'_4,0,0)$. In other words, any equivalence class in the left quotient $\KD$ contains a unique representative of the form $M(0,\xi'_2,\xi'_3,\xi'_4,0,0)$. We will use this to fix the $K$-left gauge symmetry of the model by choosing the field $\ell(t,x)$ to be of this form. More precisely, we will parametrise it as
\begin{equation}\label{Eq:lDiam}
\ell(t,x) = M \bigl(0\,,\,\alpha(t,x)\,,\,r(t,x)\cos(\beta(t,x))\,,\,r(t,x)\sin(\beta(t,x))\,,\,0\,,\,0 \bigr)
\end{equation}
in terms of three coordinate fields $(r,\alpha,\beta)$, valued in $\R_{\geq 0} \times \R \times [0,2\pi]$.

In the above equation, we have parametrised the third and fourth entries of $M$ using polar coordinates $(r,\beta)$, rather than Cartesian ones. This has to do with the right gauge symmetry of the model, with gauge group $H=\exp(\R\,T_1)$. Indeed, a straightforward analysis shows that the gauge transformation with parameter $h=\exp(\eta\,T_1)$ acts on the fields $(r,\alpha,\beta)$ by
\begin{equation}
r \longmapsto r, \qquad \alpha \longmapsto \alpha, \qquad \beta \longmapsto \beta + \eta\,.
\end{equation}
Thus, the fields $(r,\alpha)$ are physical degrees of freedom, while $\beta$ can be gauged away.

\paragraph{Action and target space geometry.} We now have all the data necessary to derive the $\s$-model realising the conformal degenerate $\Ec$-model considered above. We will not give the details of this computation here, as it is quite similar to the one presented in the previous subsections for gauged WZW models. The main intermediate step is the computation of the projectors $W^\pm_\ell$, characterised by equation \eqref{Eq:Wpm}. In the end, we find that the action of the $\s$-model is given by
\begin{equation}\label{Eq:ActionDiam}
S[r,\alpha] = 2\kay\,\iint_{\R\times \mathbb{S}^1} \dd t\,\dd x \; \bigl( \p_+ r \, \p_- r + r^{-2}\, \p_+\alpha\,\p_-\alpha \bigr)\,,
\end{equation}
in terms of the fields $(r,\alpha)$ introduced earlier in the parametrisation \eqref{Eq:lDiam}. As expected, the third field $\beta$ appearing in this equation does not enter the action in the end, as it is not gauge-invariant, contrarily to $(r,\alpha)$.

Let us quickly describe the target space of the model. It is 2-dimensional and has metric
\begin{equation}\label{Eq:MetDiam}
\dd s^2 = \kay \left(\dd r^2 + \frac{\dd\alpha^2}{r^2} \right)\,,
\end{equation}
with a singularity at $r=0$. One checks that this geometry defines a 1-loop conformal $\s$-model by showing that its Ricci tensor satisfies the equation \eqref{Eq:Ricci}, with the dilaton defined as
\begin{equation}
\Phi = - \log(r) + \text{constant}\,.
\end{equation}
To the best of our knowledge, this conformal target space cannot be obtained from a gauged WZW model. The metric \eqref{Eq:MetDiam} possesses an $\R$-isometry, acting by translation of the coordinate $\alpha$. This corresponds to a global symmetry of the $\s$-model \eqref{Eq:ActionDiam}, with conserved current
\begin{equation}
\Pi_\pm = \frac{\p_\pm \alpha}{r^2}\,, \qquad \text{ satisfying } \qquad \p_+\Pi_- + \p_-\Pi_+ = 0\,.
\end{equation}
This conservation equation ensures the existence of the so-called dual field
\begin{equation}\label{Eq:Dual2}
\alpha^\cD(t,x) = \int^x \frac{\p_t \alpha(t,y)}{r(t,y)^2}\,\dd y\,, \qquad \text{ such that } \qquad  \p_\pm\alpha^\cD = \pm\,\Pi_\pm\,.
\end{equation}
For more details and explanations, we refer to the similar discussion around equation \eqref{Eq:Dual}, in the context of the $SU(2)/U(1)$ gauged WZW model.

\paragraph{Parafermions and $\bm\Wc$-algebra.} The fact that the $\s$-model \eqref{Eq:ActionDiam} was constructed as a degenerate $\Ec$-model satisfying the strong conformal condition allows a simple description of its chiral fields. More precisely, we have non-local parafermionic chiral fields, as well as local ones, forming the $\Wc$-algebra of the theory. Their construction is very similar to the ones of the $SU(2)/U(1)$ gauged WZW model in the previous subsection. For this reason, we will not detail it here and will simply summarise the end results. For simplicity, we will also focus only on the left-moving sector.\\

The parafermionic field $\Psi^\cL$ is valued in the subspace $V_+$, defined as in equation \eqref{Eq:VDiam}. In terms of the $\s$-model fields, it takes the form
\begin{equation}
\Psi^\cL = \text{\small $\begin{pmatrix}
0 & \psib^\cL & 0 & 0 & 0 & 0 \\
0 & 0 & 0 & 0 & 0 & \psi^\cL \\
0 & 0 & 0 & 0 & 0 & 0 \\
0 & 0 & 0 & 0 & 0 & \psib^\cL \\
0 & 0 & 0 & 0 & 0 & 0 \\
0 & 0 & 0 & 0 & 0 & 0 \\
\end{pmatrix}$}, \qquad \text{ where } \qquad \psi^\cL = e^{\ri\,\alpha^\cD} \bigl( \p_+ r + \ri\, r^{-1}\, \p_+\alpha \bigr)
\end{equation}
and $\psib^\cL$ is the complex conjugate of $\psi^\cL$. In this formula, $\alpha^\cD$ is the dual field, introduced in equation \eqref{Eq:Dual2}. The non-local field $\psi^\cL$ is left-moving, \textit{i.e.} satisfy $\p_-\psi^\cL=0$ under the equations of motion derived from the action \eqref{Eq:ActionDiam}. Moreover, it satisfies a very simple Poisson algebra:
\begin{equation}
\bigl\lbrace \psi^\cL(x), \psi^\cL(y) \bigr\rbrace = 0 \qquad \text{ and } \qquad \bigl\lbrace \psi^\cL(x), \psib^\cL(y) \bigr\rbrace = -\frac{1}{\kay}\,\p_x\delta(x-y)\,.
\end{equation}
This Poisson algebra is a special case of the general one \eqref{Eq:PbParaBasisL}, for the particular choice of $\hf$, $\hf'$ and $V_+$ considered here. We note that, contrarily to the parafermionic algebra \eqref{Eq:PbParaSu2u1} of the $SU(2)/U(1)$ gauged WZW model, the present one does not contain any sign function $\epsilon(x-y)$, even if the field $\psi^\cL$ is non-local. This has to do with the rather simple algebraic structure underlying the theory: more precisely, the fact that the subspace $\hf'=\R\,T_2$ is central in $\df$ ensures that the $\epsilon(x-y)$ terms in the bracket \eqref{Eq:PbParaBasisL} vanish in the present case.\\

The construction of the $\Wc$-algebra is completely analogous to the one of the $SU(2)/U(1)$ gauged WZW model in Subsection \ref{SubSec:Su2u1}. Indeed, recall that the key ingredients in this construction are the multilinear forms on $V_+$ invariant under the adjoint action of $H$: as it turns out, the group $H \simeq U(1)$ and its action on $V_+$ for the present case are isomorphic to the ones for the $SU(2)/U(1)$ gauged WZW model. We will thus obtain the same expressions for the local chiral fields $W^\cL_s$ in terms of the parafermions $\psi^\cL$ and $\psib^\cL$: the only difference will then come from the explicit relation of these parafermions with the $\s$-model fields. The first two elements that we obtain by this construction are
\begin{equation}
W_2^\cL = T^\cL = \kay\, \psi^\cL \, \psib^\cL = \kay\left( \bigl(\p_+r\bigr)^2 +  \frac{\bigl(\p_+\alpha\bigr)^2}{r^2} \right)\,,
\end{equation}
\begin{equation}
W_3^\cL = \frac{\kay}{2\ri}\bigl( \psi^\cL \p_+ \psib^\cL - \psib^\cL \p_+ \psi^\cL \bigr) = \kay\left( \frac{(\p_+\alpha)^3}{r^4} + \frac{\p_+r\, \p_+^2 \alpha - \p_+^2 r\,\p_+\alpha}{r} \right)\,.
\end{equation}
For consistency, one easily checks that $\p_-W^\cL_2=\p_-W^\cL_3=0$.\\

Following the general discussion of Subsection \ref{SubSec:Conf}, the left-moving parafermions and $\Wc$-algebra described above are obtained respectively as the non-local and local observables in the Hamiltonian reduction of the $\df_+$--current algebra with respect to the subalgebra $\hf$, where $\df_+ = \hf\oplus\hf'\oplus V_+$. In the present case, we find that $\df_+$ is a 4-dimensional Lie algebra, generated by $T_1,\dots,T_4$. It is isomorphic to the so-called diamond algebra, or equivalently to a centrally extended 2d Poincaré algebra, which is a well-known example of a non-semi-simple Lie algebra admiting a non-degenerate invariant bilinear form (this Lie algebra also plays a crucial role in the Nappi-Witten model~\cite{Nappi:1993ie}). Similarly, the right-moving sector of the theory is described by the Hamiltonian reduction of a $\df_-$--current algebra by $\hf$, where $\df_- = \hf\oplus\hf'\oplus V_-$ is also isomorphic to the diamond algebra.

We note that, contrarily to the gauged WZW theory, the conformal $\Ec$-model described in this subsection does not fit into the GKO coset construction of Subsubsection \ref{SubSec:GKO}. More precisely, the double algebra $\df$ is not the direct sum of two orthogonal subalgebras and the decomposition $\df=\hf\oplus\hf'\oplus V_+\oplus V_-$ underlying the model is not of the form considered in this construction -- see equations \eqref{Eq:hGKO} and \eqref{Eq:DecoGKO} and the discussion around. This means that the chiral Poisson algebras of the model are obtained as Hamiltonian reductions of a current algebra but not as coset algebras. The structure underlying this theory then takes a form which is technically slightly different from the standard coset CFTs and in particular gauged WZW models. It could however happen that the target space \eqref{Eq:MatDiam} of the theory still arises from a gauged WZW model, based on another degenerate $\Ec$-model (we will see such phenomema happening in the next subsection, for other examples): to the best of our knowledge, this is not the case here.

\subsection[The conformal point of the \texorpdfstring{$SL(2,\R)$}{SL(2,R)} Klim\v{c}\'{i}k model]{The conformal point of the \texorpdfstring{$\bm{SL(2,\R)}$}{SL(2,R)} Klim\v{c}\'{i}k model}\label{SubSec:BYB}

Our final example will be the conformal point of the $SL(2,\R)$ Klim\v{c}\'{i}k (or Bi-Yang-Baxter) model. The general Klim\v{c}\'{i}k model was introduced in~\cite{Klimcik:2008eq,Klimcik:2014bta} and is an integrable double deformation of the Principal Chiral Model on a simple group $G$, generalising a construction of Fateev~\cite{Fateev:1996ea} for $G=SU(2)$. The conformal limit of this theory has been studied in details in~\cite{Fateev:1996ea,Bazhanov:2018xzh,Kotousov:2022azm}, especially for $G=SU(2)$. In this subsection, we revisit the conformal point for $G=SL(2,\R)$, using the formalism of degenerate $\Ec$-models\footnote{The Klim\v{c}\'{i}k model was reinterpreted as a non-degenerate $\Ec$-model in~\cite{Klimcik:2016rov}. However, in this subsection, we will use a different formulation, this time as a degenerate $\Ec$-model, which is better suited for our purpose. Up to differences in reality conditions, we expect this formulation to be related to a particular limit of the one considered in~\cite{Klimcik:2019kkf}.} and the general approach developed in this article, guided by the results of~\cite{Bazhanov:2018xzh,Kotousov:2022azm}\footnote{The references~\cite{Bazhanov:2018xzh,Kotousov:2022azm} were mostly considering the $SU(2)$ case and were motivated by the study of integrable structures at the conformal point. In particular, the formalism of affine Gaudin models used in~\cite{Kotousov:2022azm} is closely related to the one of $\Ec$-models and provided strong inspirations for the present work. Here, we will consider $G=SL(2,\R)$ rather than $G=SU(2)$, to avoid certain issues with reality conditions (however, most of the results of~\cite{Bazhanov:2018xzh,Kotousov:2022azm} can be readily transposed to this case).}. In addition to providing another explicit illustration of this approach, this will also allow us to point out some of its subtleties and open questions, such as the existence of different target space domains in conformal limits of $\Ec$-models or the fact that the $\Ec$-model $\Wc$-algebra introduced in Subsection \ref{SubSec:Conf} does not always encode all the local chiral fields of the corresponding $\s$-model. To keep the length of the subsection contained, we will mostly focus on these aspects and will not give many details on the construction of the theory as a conformal $\Ec$-model or the structure of its chiral Poisson algebras.

Moreover, we will end this subsection by considering a one-parameter deformation of this theory, which is based on the same underlying degenerate $\Ec$-model but with a deformed choice of maximally isotropic subalgebra (corresponding to a so-called TsT transformation). This will lead to a more complicated target space, defining a non-trivial conformal $\s$-model with two parameters. As we will see, although the theories considered in this subsection initially arise from a quite distinct construction, the target spaces that we will obtain will in fact be related to gauged WZW models on $SL(2,\R)/\R$ and $\bigl( SL(2,\R) \times \R \bigr)/\R$. Using Subsection \ref{SubSec:gWZW}, the latter can also be obtained from degenerate $\Ec$-models, which are however different from the one initially considered for the Klim\v{c}\'{i}k model (including a different choice of double algebra and of gauged subalgebra). This shows that various, seemingly unrelated, conformal $\Ec$-models can in fact lead to the same theory.

\paragraph{The degenerate $\bm{\Ec}$-model.} We denote by $\kappa(\cdot,\cdot)=-\Tr(\cdot\,\cdot)$ the minimal bilinear form on $\slf(2,\R)$ (where the trace is taken in the fundamental representation). We will define a degenerate $\Ec$-model with double group $D=SL(2,\R)^{\times 4}$. We equip the Lie algebra $\df=\slf(2,\R)^{\times 4}$ with the bilinear form
\begin{equation}\label{Eq:FormBYB}
\psb{(X_1,\dots,X_4)}{(Y_1,\dots,Y_4)} = \frac{\kay}{1+\nu^2} \Bigl( \nu^2 \, \kappa(X_1,Y_1) + \kappa(X_2,Y_2) - \nu^2 \, \kappa(X_3,Y_3) - \kappa(X_4,Y_4) \,\Bigr) \,,
\end{equation}
defined in terms of parameters $\kay>0$ and $\nu>0$. It is non-degenerate, invariant and of split signature.

We will denote pure tensor products in $\df \simeq \R^4\otimes\slf(2,\R)$ by $(a_1,a_2,a_3,a_4)\otimes X$, with $a_i\in\R$ and $X\in\slf(2,\R)$ (for instance, in these notations, the element $(X,0,2X,0)$ in  $\slf(2,\R)^{\times 4}$ corresponds to $(1,0,2,0)\otimes X$). We then let
\begin{equation}
\hf = \slf(2,\R)_{\diag} =(1,1,1,1) \otimes \slf(2,\R) \, ,\;\; \qquad \hf' = (1,1,-1,-1) \otimes \slf(2,\R)\,, 
\end{equation}
\begin{equation}
V_+ =(1,-\nu^2,0,0) \otimes \slf(2,\R) \qquad \text{ and } \qquad V_- =(0,0,1,-\nu^2) \otimes \slf(2,\R)\,.
\end{equation}
This specifies the decomposition $\df=\hf\oplus\hf'\oplus V_+ \oplus V_-$ underlying the degenerate $\Ec$-model, where
$\hf$ is the gauge isotropic subalgebra, $\hf'$ its dual space and $V_\pm$ the eigenspaces of the $\Eh$-operator (we refer to Section \ref{Sec:D} for details). One checks that this decomposition satisfies the strong conformal condition introduced in Subsection \ref{SubSec:Conf}, ensuring that the degenerate $\Ec$-model will be conformal.

\paragraph{Maximally isotropic subalgebra and physical fields.} To build a $\s$-model corresponding to this $\Ec$-model, the next step is to pick a maximally isotropic subalgebra $\kf$ inside of $\df$. For the general Klim\v{c}\'{i}k model on a group $G$, it is built from the choice of an $R$-matrix, \textit{i.e.} a skew-symmetric endomorphism of $\text{Lie}(G)$ satisfying the classical Yang-Baxter equation. In the case at hand, our choice for $\kf$ can be written explicitly as
\begin{equation}\label{Eq:kBYB}
\kf = \left\lbrace (a_1,a_2,-a_1,-a_2)\otimes \begin{pmatrix}
1 & 0 \\ 0 & -1
\end{pmatrix} + (a_3,a_4,0,0)\otimes \begin{pmatrix}
0 & 1 \\ 0 & 0
\end{pmatrix} + (0,0,a_5,a_6)\otimes \begin{pmatrix}
0 & 0 \\ 1 & 0
\end{pmatrix}\,, a_i\in\R\, \right\rbrace\,.
\end{equation}
This is clearly a 6-dimensional subspace of $\df$ and one easily checks that it is isotropic and stable under the commutation relations of $\df \simeq \R^4 \otimes \slf(2,\R)$.\\

We denote by $K$ and $H=SL(2,\R)_{\diag}$ the subgroups of $D=SL(2,\R)^{\times 4}$ with Lie algebras $\kf$ and $\hf$. Following the general construction recalled in Subsubsection \ref{SubSec:Lag}, the $\s$-model that we are building is described by a $D$-valued field $\ell(t,x)$, with gauge symmetries $\ell\mapsto k\ell h$, $(k,h)\in K \times H$, so that the target space is the double quotient $\KDH$. In what follows, we will focus on a particular domain in this target space. At the level of the Lie algebra, the above choice of $\kf$ and $\hf$ induces a decomposition
\begin{equation}
\df = \kf \oplus  \bigl( (1,0,1,0) \otimes \slf(2,\R) \bigr) \oplus \hf\,.
\end{equation}
``Exponentiating'' this result, we find that there is a neighbourhood of the identity in the group $D$ which takes the factorised form
\begin{equation}
K \cdot \bigl( (g,\Id,g,\Id), \, g\in SL(2,\R) \bigr\rbrace \cdot H\,,
\end{equation}
which we will call the main cell of $D$. Taking the left and right quotients by $K$ and $H$ respectively, we obtain a subdomain of the target space described by the choice of $g\in SL(2,\R)$. In terms of the $\s$-model, focusing on this domain amounts to requiring the $D$-valued field $\ell(t,x)$ to take the form
\begin{equation}
\ell(t,x) = \bigl( g(t,x), \Id, g(t,x), \Id \bigr)\,,
\end{equation}
where $g(t,x)$ is valued in $SL(2,\R)$. This corresponds to starting from $\ell$ in the main cell of $D$ and imposing a complete gauge-fixing of both the $K$- and $H$-gauge symmetries. All the degrees of freedom in $g(t,x)$ are thus physical ones and the resulting theory can be seen as a $\s$-model on $SL(2,\R)$.

To further extract these degrees of freedom, we have to parametrise $g(t,x)$ by making a choice of coordinates on $SL(2,\R)$. Here, we will pick some which are adapted to our model and in particular will make its symmetries manifest. The price to pay for this choice is that these coordinates will not be global ones: to cover the whole $SL(2,\R)$ manifold, one would have to consider different sets of such coordinates on various patches. For simplicity, we will not describe all of them in this article and will focus only on one patch, thus restricting to a specific subdomain in the target space of the $\s$-model. In this domain, we will parametrise the $SL(2,\R)$-valued field $g(t,x)$ as
\begin{equation}\label{Eq:gBYB}
g(t,x) = \dfrac{1}{\sqrt{1+e^{2r(t,x)}}} \begin{pmatrix}
e^{ -\alpha(t,x) + \frac{1}{2}(\nu-\nu^{-1})\chi(t,x)} & -e^{ r(t,x) -\frac{1}{2}(\nu+\nu^{-1})\chi(t,x)} \vspace{8pt}\\
e^{r(t,x) + \frac{1}{2}(\nu+\nu^{-1})\chi(t,x)} & e^{ \alpha(t,x) - \frac{1}{2}(\nu-\nu^{-1})\chi(t,x)}
\end{pmatrix}\,,
\end{equation}
in terms of three scalar fields $\bigl( r(t,x), \alpha(t,x), \chi(t,x) \bigr)$ valued in $\R$ (the dependence on $\nu$, which is one of the constant parameter entering the bilinear form \eqref{Eq:FormBYB}, was introduced for later convenience). One easily checks that the above matrix is of determinant $1$ and thus belongs to $SL(2,\R)$.

\paragraph{Action and target space geometry.} Given the setup described in the previous paragraphs, one can derive the action of the $\s$-model following the construction of Subsubsection \ref{SubSec:Lag}. We will skip all the details of this computation and give only the end result, written in terms of the three fields $\bigl( r(t,x), \alpha(t,x), \chi(t,x) \bigr)$ introduced above. We find that the action reads
\begin{equation}\label{Eq:BYB}
S[r,\alpha,\chi] = 2\kay\, \iint_{\R \times \mathbb{S}^1} \dd t\,\dd x \left( \frac{\p_+ r\,\p_- r - \p_+\alpha\,\p_-\alpha}{1+e^{2r}} - \p_+\chi\,\p_-\chi \right)\,.
\end{equation}
We note that in the derivation of this formula, we discarded a total derivative $B$-field. From the action \eqref{Eq:BYB}, one directly reads off the metric of the corresponding $3$-dimensional target space:
\begin{equation}\label{Eq:ConfBYB}
\dd s^2 = \kay\left( \frac{\dd r^2 - \dd \alpha^2}{1+e^{2r}} - \dd\chi^2 \right)\,.
\end{equation}
We see in particular that $\chi$ completely decouples, so that the theory in fact becomes a free field $\chi$ plus a $\s$-model with 2-dimensional target space, described by the coordinate fields $(r,\alpha)$. We recognise in this target space the 2d Lorentzian black hole geometry \eqref{Eq:LorBH2}, which can be obtained from a $SL(2,\R)/\R$ axially gauged WZW model~\cite{Witten:1991yr}, as reviewed at the end of Subsection \ref{SubSec:Su2u1}. 

\paragraph{Chiral fields.} The fact that the theory \eqref{Eq:BYB} was built as a degenerate $\Ec$-model satisfying the strong conformal condition means that it possesses many chiral fields, following the results of Subsection \ref{SubSec:Conf}. We will not describe these fields and their Poisson algebras in details but will comment on some related aspects, as they illustrate some of the natural limits, open questions and perspectives of the conformal $\Ec$-model formalism. For simplicity, we will focus on left-moving fields.\\

The first type of left-moving fields that one can obtain from the $\Ec$-model approach are non-local parafermions. In the present case, applying the general construction of Subsubsection \ref{SubSec:Para} in fact leads to complicated parafermions, whose expression in terms of the $\s$-model fields $(r,\alpha,\chi)$ is quite involved. This does not reflect the simple form of the action \eqref{Eq:BYB}. In particular, since the boson $\chi$ completely decouples in this action, the most obvious left-moving field of the theory is simply the derivative $\p_+\chi$. In addition, more complicated but still rather simple left-moving fields are given by the parafermions of the Lorentzian black hole $\s$-model in terms of $(r,\alpha)$ -- see the end of Subsection \ref{SubSec:Su2u1} and more precisely equation \eqref{Eq:ParaBH}, up to the change of coordinate $(\theta,\alpha)\mapsto(r,\alpha)$. These are not the chiral fields that naturally arise from the construction of Subsubsection \ref{SubSec:Para}. However, this does not mean that they are not describable from the $\Ec$-model perspective, as we now explain.

Recall that the approach of Subsubsection \ref{SubSec:Para} relied on a simple choice of gauge-fixing, in which the components of the $\Ec$-model current $\Jc_+$ became the left-moving parafermions. For the present model, there exists in fact another type of gauge-fixing condition, in which $\Jc_+$ simply yields the free boson $\p_+\chi$ and the black hole parafermions. This gauge-fixing condition, which we will not write down here, can be readily obtained from the results of~\cite[Subsection 6.2]{Kotousov:2022azm}, replacing the group $SU(2)$ considered there with the one $SL(2,\R)$ considered here. We expect it to be mostly adapted to the particular choice of maximally isotropic subalgebra $\kf$ considered here\footnote{More concretely, the gauge-fixing condition of~\cite[Subsection 6.2]{Kotousov:2022azm} can be naturally expressed in terms of the $R$-matrix underlying the Klim\v{c}\'{i}k model, which is the main ingredient used to construct the subalgebra $\kf$.}, but not necessarily to other potential choices. In contrast, the general approach of Subsubsection \ref{SubSec:Para} did not rely at all on the choice of $\kf$. This suggests that there might exist a more refined construction than this general one, which could take into account some specificity of the choice of $\kf$ and thus the particular form of the $\s$-model realising the underlying conformal $\Ec$-model. A more thorough analysis of these aspects is out of the scope of this paper but constitutes a natural perspective for future developments.\\

In addition to non-local parafermions, the other left-moving fields obtained from the $\Ec$-model formalism are the local ones, forming the $\Wc$-algebra of the theory. Applying the general construction of Subsubsection \ref{SubSec:W} in the present case, we obtain the $\slf(2,\R) \times \slf(2,\R) / \slf(2,\R)_{\diag}$ coset $\Wc$-algebra (see~\cite{Semikhatov:2001zz} for a detailed description of this $\Wc$-algebra at the quantum level). The first few fields obtained this way (which have spins $2$, $4$ and $6$) can be easily written explicitly and are the analogues of those discussed in~\cite[Subsections 5.3 and 6.2]{Kotousov:2022azm}, replacing the group $SU(2)$ by $SL(2,\R)$. Here also, this construction coming from the $\Ec$-model formalism does not capture completely the simplicity of the $\s$-model \eqref{Eq:BYB}. For instance, the derivative $\p_+\chi$ is an obvious spin-1 local left-moving field of the theory, which however does not appear in the coset $\Wc$-algebra. Moreover, the latter also does not describe all the left-moving local densities \eqref{Eq:WBH} built from the Lorentzian black hole fields, which form a $SL(2,\R)/\R$ coset $\Wc$-algebra (see the end of Subsection \ref{SubSec:Su2u1}). In other words, the complete set of local left-moving fields of the theory \eqref{Eq:BYB} is composed by the Heinsenberg algebra of $\p_+\chi$ and the $SL(2,\R)/\R$ coset $\Wc$-algebra built from the fields $(r,\alpha)$, but the $\Ec$-model construction only yields a subalgebra thereof, corresponding to a $\slf(2,\R) \times \slf(2,\R) / \slf(2,\R)_{\diag}$ coset (we refer to~\cite[Subsections 5.3 and 6.2]{Kotousov:2022azm} for a detailed discussion of the $SU(2)$ analogue of this statement).

This is again a specificity that comes with the particular choice of isotropic subalgebra $\kf$ made here. Another choice for this subalgebra should lead to the $SL(2,\R) \times SL(2,\R) / SL(2,\R)_{\diag}$ gauged WZW model (following the construction of Subsection \ref{SubSec:WZW}). For that case, we expect the $\Ec$-model formalism to capture all the local chiral fields. It would be interesting to explore these aspects in more details in the future and in particular see if the $\Wc$-algebra construction of Subsection \ref{SubSec:W} can be refined to take into account the specificity of the choice of $\kf$.

\paragraph{Non-conformal Klim\v{c}\'{i}k model and its conformal limit.} As mentioned earlier, the $\s$-model \eqref{Eq:BYB} can be thought of as the UV limit of a non-conformal theory, the Klim\v{c}\'{i}k model on $SL(2,\R)$. Similarly to the conformal point, this theory can be realised as a degenerate $\Ec$-model: more precisely, it shares the same double Lie algebra $\df$ and isotropic subalgebras $\hf$ and $\kf$ as the conformal point but differs in the choice of subspaces $V_\pm$ (or equivalently of $\Eh$-operator). The resulting $\s$-model can be described in terms of three coupling constants $(\kay,\nu,\lambda)$: the first two are the same parameters ($\kay,\nu)$ as in the conformal theory above, while $\lambda \geq 0$ is a new parameter encoding the perturbation away from the conformal point. Letting $\mu=(1-\nu^2)/(1+\nu^2)$ and using the coordinates $(r,\alpha,\chi)$ introduced earlier, this non-conformal $\s$-model corresponds to the metric
\begin{align}\label{Eq:NonConfBYB}
&\hspace{-35pt}  \dd s^2 = \kay\left(\frac{1-\lambda^2}{1-\lambda^2\mu^2}\right)^{1/2} \left[ \frac{1-\lambda^2\mu^2}{(1+e^{2r})(1+\lambda^2\,e^{-2r})}(\dd r^2 - \dd \alpha^2) - \frac{1+\lambda^2\mu^2\,e^{-2r}}{1+\lambda^2\,e^{-2r}} \dd\chi^2 \right. \\ 
&\hspace{125pt} \left. - \frac{4\lambda^2\,e^{-2r}}{1+\lambda^2\,e^{-2r}} \left( \frac{\nu^2}{(1+\nu^2)^2} \frac{\dd\alpha^2}{1+e^{2r}} + \frac{\mu\,\nu}{1+\nu^2} \,\dd\alpha\,\dd\chi \right) \right]\,. \notag
\end{align}
As expected, one shows that this theory admits a non-trivial RG-flow. The above parametrisation has been chosen so that the coupling constants $(\kay,\nu)$ are in fact RG-invariants, while the new one $\lambda$ runs. More precisely, we find the following expressions for the RG-flow of $\lambda$ and the dilaton of the model:
\begin{equation}
\frac{\dd\lambda}{\dd \tau} = -\frac{2\hbar\lambda}{\kay}\left(\frac{1-\lambda^2}{1-\lambda^2\mu^2}\right)^{1/2} \qquad \text{ and } \qquad \Phi = r - \frac{1}{2} \log \left( \frac{1+e^{2r}}{1+\lambda^2\,e^{-2r}} \right) + \text{constant}\,.
\end{equation}
In particular, one sees that the UV limit $\tau\to+\infty$ of the theory corresponds to $\lambda\to 0$. This is consistent with the previous results of this subsection. Indeed, it is clear that setting $\lambda = 0$ into the equation \eqref{Eq:NonConfBYB} yields the metric \eqref{Eq:ConfBYB} of the conformal model introduced earlier. The latter thus arises in the UV limit of the full non-conformal Klim\v{c}\'{i}k model.\\

The main goal of this paragraph is to explain that this UV limit is in fact a bit more subtle than that.\footnote{This is inspired by various already established results in the literature. We refer for instance to the lecture notes~\cite{Lukyanov:2019asr} and to~\cite{Bazhanov:2017nzh,Kotousov:2022azm} for a more detailed discussion of similar ideas.} To understand why, let us quickly describe the geometry of the non-conformal target space \eqref{Eq:NonConfBYB}. Its submanifolds corresponding to fixed values of the coordinate $r$ are infinitely-extended flat 2-dimensional planes, described by the two coordinates $(\alpha,\chi)\in\R^2$. In contrast, let us consider a curve $\mathcal{C}$ in this target space corresponding to fixed values of the coordinates $(\alpha,\chi)$ and $r$ varying from $-\infty$ to $+\infty$. The length of this curve with respect to the metric \eqref{Eq:NonConfBYB} can be computed by integrating $\sqrt{\dd s^2|_{\mathcal{C}}}$ over $r\in\R$. We find that it is finite for $\lambda$ strictly positive, but behaves as $\propto|\log(\lambda)|$ when $\lambda$ approaches $0$. The length of $\mathcal{C}$ thus diverges logarithmically in the UV-limit $\lambda\to 0$. Schematically, this means that the target space of the theory ``decompactifies'' along the $r$-direction in this limit.

This introduces some subtleties in the way we take the UV-limit of the metric \eqref{Eq:NonConfBYB}. Indeed, one can consider various such limits focusing on different ``domains'' along the $r$-direction. Earlier, we simply took the limit $\lambda\to 0$ while keeping the coordinate $r$ fixed and in particular finite: we then recovered the metric \eqref{Eq:ConfBYB} of the Lorentzian black hole times a decoupled line. Alternatively, one can consider a shifted coordinate $\tilde{r} = r - a \log (\lambda)$ and take the limit keeping $\tilde{r}$ finite: schematically, when $\lambda$ approaches $0$, this new coordinate $\tilde{r}$ will probe different domains along the decompactifying curve $\mathcal{C}$, depending on the value of the parameter $a$. For instance, for $a<0<1$, we find that the limit $\lambda\to 0$, $\tilde{r}$ fixed, of the metric \eqref{Eq:NonConfBYB} simply yields the flat metric
\begin{equation}\label{Eq:DomainAs}
\dd s^2_{\text{as}} = \kay\,(\dd \tilde{r}^2 - \dd \alpha^2 - \dd \chi^2)\,.
\end{equation}
This corresponds to staying in a region away from the tips of the curve $\mathcal{C}$ while taking the limit. In this region, which we call the asymptotic domain of the theory, the geometry approaches the flat one when $\lambda\to 0$ (the corresponding $\s$-model is thus asymptotically free in this domain).

Another interesting case is to take $a=1$. Doing so, we find that the limit of the metric \eqref{Eq:NonConfBYB} when $\lambda\to 0$ and $\tilde{r}$ stays fixed takes the form
\begin{equation*}
\dd s^2_{\text{II}} = \frac{\kay}{1+e^{-2\tilde{r}}} \left( \dd \tilde{r}^2 - \left( 1 + \frac{4\nu^2 e^{-2\tilde{r}}}{(1+\nu^2)^2} \right) \dd \alpha^2 - \left( 1 + \frac{(1-\nu^2 )^2 e^{-2\tilde{r}}}{(1+\nu^2)^2}\right) \dd \chi^2  - \frac{4\nu(1-\nu^2 ) e^{-2\tilde{r}}}{(1+\nu^2)^2}\, \dd\alpha\,\dd\chi \right)\,.
\end{equation*}
Performing a well-chosen rotation of $(\alpha,\chi)$ to new coordinates $(\tilde{\alpha},\tilde{\chi})$, one brings this metric to
\begin{equation}\label{Eq:DomainII}
\dd s^2_{\text{II}} = \kay \left( \frac{\dd\tilde{r}^2-\dd\tilde{\alpha}^2}{1+e^{-2\tilde{r}}} - \dd\tilde{\chi}^2 \right)\,.
\end{equation}
We thus find a second domain in which the target space flows to a Lorentzian black hole times a line.

It is natural to wonder what is the interpretation of these different domains in the formalism of conformal degenerate $\Ec$-models. The conformal theory in the first domain (corresponding to the limit with $r$ fixed) was initially introduced in this subsection as a specific conformal $\Ec$-model. The latter can in fact be seen as the naive $\lambda\to 0$ limit of the $\Ec$-model underlying the non-conformal theory (in particular, this limit does not change the double algebra $\df$, its bilinear form $\psd$ or the isotropic subalgebra $\kf$, but only the subspaces $V_\pm$). It would be interesting to understand, if it exists, the $\Ec$-model interpretation of the other two domains \eqref{Eq:DomainAs} and \eqref{Eq:DomainII}. Some elements in the analysis of~\cite{Bazhanov:2018xzh,Kotousov:2022azm}, which concerned the $SU(2)$ analogue of the present story, could suggest that the $\Ec$-model current $\Jc$ will have a non-degenerate limit in these domains only if one performs a well-chosen conjugation before we take $\lambda\to 0$. This offers a potential approach to explore these questions.

\paragraph{Deformed conformal $\bm\s$-model.} We end this subsection by describing a deformation of the theory \eqref{Eq:BYB} which still defines a conformal $\s$-model but which corresponds to a more complicated target space. It is obtained by considering the exact same underlying degenerate $\Ec$-model, but deforming the maximally isotropic subalgebra $\kf$, initially chosen as in equation \eqref{Eq:kBYB}. We now take it to be
\begin{align}
\hspace{-30pt} \kf_\rho &= \left\lbrace a_1(1,0,-\cos 2\rho,-\nu\sin 2\rho )\otimes \begin{pmatrix}
1 & 0 \\ 0 & -1
\end{pmatrix} + a_2 (0,1,\nu^{-1}\sin 2\rho,-\cos 2\rho)\otimes \begin{pmatrix}
1 & 0 \\ 0 & -1
\end{pmatrix} \right. \\
&\left. \hspace{130pt} +\; (a_3,a_4,0,0)\otimes \begin{pmatrix}
0 & 1 \\ 0 & 0
\end{pmatrix} + (0,0,a_5,a_6)\otimes \begin{pmatrix}
0 & 0 \\ 1 & 0
\end{pmatrix}\,,\; a_i\in\R\, \right\rbrace\,, \notag
\end{align}
where $\rho$ is a constant deformation parameter. It is clear that $\kf_\rho$ reduces to the subalgebra $\kf$ defined in equation \eqref{Eq:kBYB} when $\rho=0$. Moreover, one checks that it still defines a subalgebra of $\df$, maximally isotropic with respect to the bilinear form \eqref{Eq:FormBYB}. We expect that this deformation corresponds to applying a so-called TsT transformation~\cite{Lunin:2005jy} to the undeformed model.\footnote{The deformed theory would then correspond to a specific conformal point among the very general class of integrable models built in the reference~\cite{Delduc:2017fib}, which describe deformations of the Klim\v{c}\'{i}k model for any group $G$, including a Wess-Zumino term and TsT transformations. In the case of $G=SU(2)$, which is the compact analogue of the setup $G=SL(2,\R)$ considered here, this deformed model becomes equivalent to the integrable $\s$-model introduced by Lukyanov in~\cite{Lukyanov:2012zt}. We finally note that the general class of models of~\cite{Delduc:2017fib} (including TsT transformations) has been realised in the language of degenerate $\Ec$-models in~\cite{Klimcik:2019kkf}, at least for the case of compact groups (the $\Ec$-model considered here should thus be a special case of this construction, up to differences in reality conditions).} \\

The explicit $\s$-model corresponding to the deformed isotropic subalgebra $\kf_\rho$ can be obtained from the same procedure as the undeformed one \eqref{Eq:BYB}, which we do not detail here. In order to present the end result in the simplest form, we first have to redefine the coordinate fields $(\alpha,\chi)$ appearing in the parametrisation \eqref{Eq:gBYB} of the field $g(t,x)$. More precisely, we perform the following rescaling and $r$--dependent shift
\begin{equation}
\alpha \longmapsto \frac{\alpha}{\cos\rho} + \frac{\nu\tan\rho}{1+\nu^2} \bigl( 2r - \log(1+e^{2r}) \bigr)\,, \qquad \chi \longmapsto \frac{\chi}{\cos\rho} + \frac{(1-\nu^2)\tan\rho}{2(1+\nu^2)} \bigl( 2r - \log(1+e^{2r}) \bigr)\,.
\end{equation}
As one should expect, these new coordinates $(\alpha,\chi)$ coincide with the initial ones in the undeformed case $\rho=0$. In these terms, the deformed metric then takes the form
\begin{equation}\label{Eq:BYBtst-Met}
\dd s^2 = \kay\left( \frac{\dd r^2}{1+e^{2r}} - \frac{\dd \alpha^2}{1 + \cos^2(\rho)\,e^{2r}} - \frac{(1 + e^{2r})\,\dd\chi^2}{1 + \cos^2(\rho)\,e^{2r}} \right)\,.
\end{equation}
It reduces to the metric \eqref{Eq:ConfBYB} when $\rho=0$ and corresponds to a non-trivial deformation of the model. In particular, the field $\chi$ is not a decoupled free field when $\rho\neq 0$, so that the target space is a genuine 3-dimensional manifold and not the direct product of lower-dimensional ones as is the undeformed case $\rho=0$. Moreover, it is important to note that this deformed model also comes with a non-trivial $B$-field, which takes the form
\begin{equation}\label{Eq:BYBtst-B}
B = \frac{2\kay\,\tan(\rho)}{1+\cos^2(\rho)\,e^{2r}}\, \dd\chi \wedge \dd \alpha\,.
\end{equation}
As expected, it vanishes when $\rho=0$.\footnote{Note that, along the procedure that we followed to obtain the $\s$-model, we have in fact discarded a part of the $B$-field which is exact and which thus contributes a total derivative to the Lagrangian.} This $B$-field is essential to obtain a conformal $\s$-model. Indeed, one checks that the 1-loop RG-flow of the theory is trivial once we include the non-trivial contribution of the torsion deriving from $B$ in the $\beta$-functions of the model (using the general results of~\cite{Friedan:1980jf,Fradkin:1985ys,Callan:1985ia}), as well as the contribution coming from the dilaton
\begin{equation}
\Phi = r - \frac{1}{2} \log\bigl( 1 + \cos^2(\rho)\,e^{2r} \bigr) +  \text{constant}\,.
\end{equation}

The metric \eqref{Eq:BYBtst-Met} and B-field \eqref{Eq:BYBtst-B} in fact coincide with the ones of an axially gauged WZW model on $(SL(2,\R) \times \R) / \R$, similar to the 3-dimensional black hole geometry considered in~\cite{Horne:1991gn}. This is analogous to the relation described earlier in this subsection between the undeformed conformal model and the $SL(2,\R) / \R$ gauged WZW theory with an additional free boson. In particular, we see from this relation that the conformal degenerate $\Ec$-model considered here will not directly capture all the details of the chiral structure of the theory, as was the case already for the undeformed model. For instance, the gauged WZW formulation makes obvious the existence of local spin-$(\pm 1)$ chiral fields
\begin{equation}
W_1^\cLR = \frac{(1 + e^{2r} )\p_\pm \chi \pm \tan(\rho)\,\p_\pm \alpha}{1 + \cos^2(\rho)\,e^{2r}}\,,
\end{equation}
satisfying $\p_- W_1^\cL = \p_+ W_1^\cR = 0$ and reducing to the chiral free bosons $\p_\pm\chi$ in the undeformed case $\rho=0$. These particularly simple chiral densities are not contained in the $\slf(2,\R) \times \slf(2,\R) / \slf(2,\R)_{\diag}$ coset $\Wc$-algebra arising from the $\Ec$-model. Similarly, the gauged WZW formulation leads to the construction of rather simple chiral parafermionic fields, which are not staightforwardly obtained from the $\Ec$-model under consideration here. This gives another example where the general constructions of Subsection \ref{SubSec:Conf} fail to capture the finer details of the chiral structures that arise from specific choices of maximally isotropic subalgebra.

The above results illustrate that different, seemingly unrelated, conformal $\Ec$-models can in fact produce the same theory, in particular with one of these $\Ec$-models taking the form of a gauged WZW model (as in Subsection \ref{SubSec:gWZW}). It is not clear to us whether such a phenomenon is an accidental low-dimensional isomorphism or the sign of a deeper connection. It would be interesting to explore these aspects further and in particular to develop a more systematic approach to determine whether two conformal $\Ec$-models correspond to the same $\s$-model.

\section{Conclusion and perspectives}
\label{Sec:Conclusion}

We conclude by discussing some natural perspectives and potential extensions of the results presented in this paper.

\subsection[More general conformal \texorpdfstring{$\Ec$}{E}-models]{More general conformal \texorpdfstring{$\bm{\Ec}$}{E}-models}
\label{SubSec:GenConf}

A degenerate $\Ec$-model is characterised by the decomposition $\df = \hf \oplus \hf' \oplus  V_+ \oplus V_-$ of its double algebra. The main result of this article is that a simple algebraic hypothesis on the commutation relations of the subspaces $V_\pm$ and $\hf\oplus\hf'$ ensures the 1-loop conformal invariance of the model as well as the decoupling of its classical observables into two chiral sectors. We called this hypothesis the \textit{strong conformal condition} (see Subsection \ref{SubSec:Conf} for details). The use of the adjective ``strong'' to qualify this condition is due to our expectation that it is not the most general one ensuring the above-mentioned properties of conformality and chirality. The goal of this subsection is to discuss briefly some perspectives about these more general cases. We present it in the context of degenerate $\Ec$-models: however, the discussion also applies directly to the non-degenerate case by setting $\hf=\hf'=\lbrace 0 \rbrace$.

Recall the general form \eqref{Eq:Rg}--\eqref{Eq:S} of the 1-loop RG-flow of $\Ec$-models~\cite{Valent:2009nv,Sfetsos:2009dj,Avramis:2009xi,Sfetsos:2009vt,Severa:2018pag}. The triviality of this flow translates into a collection of complicated algebraic constraints, involving adjoint operators and projectors on the subspaces $V_\pm$ and $\hf$. The strong conformal condition was introduced in the main text as an easy way to satisfy those. A thorough analysis of the general solutions to these constraints is beyond the scope of this paper and constitutes a natural perspective for future work. As a first illustration, we exhibit one such solution here, which generalises the strong conformal condition but still takes a rather simple form. Namely, let us suppose that the subspaces $V_\pm$ admit an orthogonal decomposition
\begin{subequations}\label{Eq:GenCond}
\begin{equation}
V_\pm = V_\pm' \oplus V_\pm''
\end{equation}
such that
\begin{equation}
\bigl[ V_+', V_-' \bigl]\, = \bigl[ V_+'', V_-'' \bigr] = \lbrace 0 \rbrace \qquad \text{ and } \qquad \bigl[ V_\pm', V_\mp'' \bigl]\, \subset V_\mp'' \,.
\end{equation}
\end{subequations}
It is a straightforward algebraic exercise to show that this configuration ensures the vanishing of the RG-flow \eqref{Eq:Rg}--\eqref{Eq:S}, as required. Moreover, if one considers the subcase $V_\pm'' = \lbrace 0 \rbrace$ and further requires that $\hf\oplus\hf'$ forms a subalgebra, this condition reduces to the strong conformal one.

In addition to the triviality of the RG-flow, the strong conformal condition was also useful in the main text to study the classical chiral fields of the theory. Similarly, the more general configuration \eqref{Eq:GenCond} also ensures the existence of such fields. This is shown by projecting the equations of motion \eqref{Eq:EoMpm} on $V_\pm'$ and $V_\pm''$. For instance, we find that the $V_\pm'$--valued component $\Jc_\pm'$ of the current $\Jc_\pm$ is annihilated by $\p_\mp$ up to the adjoint action of the $\hf$-valued gauge field $\Ac_\mp$. This is the exact analogue of the situation studied in Subsubsection \ref{SubSec:Para} for the strong conformal condition. In particular, as explained there, it ensures the existence of non-local gauge-invariant chiral fields obtained by dressing $\Jc_\pm'$ by the adjoint action of a well-chosen $H$-valued quantity, namely the path-ordered exponential of the gauge field. The case of the $V_\pm''$--valued component $\Jc_\pm''$ is slightly more subtle: we find that it is annihilated by $\p_\mp$ up to the adjoint action of $\Ac_\mp + \Jc'_\mp \in \hf \oplus V_\mp'$. One then also obtains non-local chiral fields by taking appropriate conjugations of $\Jc_\pm''$, but this time by quantities which are not directly built from the gauge field only. Moreover, there should exist a procedure similar to the one of Subsubsection \ref{SubSec:W} which also allows the construction of local chiral fields from this setup.\footnote{More precisely, the construction of local chiral fields from the component $\Jc_\pm'$ follows exactly the $\Wc$-algebra procedure developed in Subsubsection \ref{SubSec:W}. In contrast, the construction of local fields involving the other component $\Jc_\pm''$ would require a more involved analysis.}

Let us finally mention that the strong conformal condition also allowed us in the main text to determine the Poisson brackets obeyed by the chiral fields of the conformal theory, taking the form of parafermionic algebras and coset-type $\Wc$-algebras. For the more general case considered here, this question seems to not be as straightforward to study and would require a more in-depth analysis (in particular for the chiral fields built from $\Jc_\pm''$). It would be interesting to explore this in the future.\\

The condition \eqref{Eq:GenCond} discussed above provides a first example of more general configurations useful to study conformal degenerate $\Ec$-models. It is a quite natural perspective of the present work to study further these more general cases and explore the panorama of conformal $\s$-models obtained by such constructions. We hope that this panorama will cover a large class of theories, including interesting examples beyond the class of gauged WZW models. For instance, based on some preliminary results that we will not discuss here, we expect the condition \eqref{Eq:GenCond} to be relevant for the description of the Guadagnini-Martellini-Mintchev conformal theory~\cite{Guadagnini:1987ty} as a degenerate $\Ec$-model.

\subsection[Open questions on the conformal limits of \texorpdfstring{$\Ec$}{E}-models]{Open questions on the conformal limits of \texorpdfstring{$\bm\Ec$}{E}-models}
\label{SubSec:Open}

The Subsection \ref{SubSec:BYB} was devoted to the conformal limit of the $SL(2,\R)$ Klim\v{c}\'{i}k/bi-Yang-Baxter model~\cite{Klimcik:2008eq} and its interpretation in the language of conformal $\Ec$-models, based on the works~\cite{Bazhanov:2018xzh,Kotousov:2022azm}. This raised various subtleties and open questions in the treatment of $\Ec$-models satisfying the strong conformal condition. We briefly summarise these points in this conclusion, as they form natural subjects for future investigations.

For instance, this theory gives an example of conformal degenerate $\Ec$-model where the parafermionic algebra built in Subsubsection \ref{SubSec:Para} does not entirely reflect the simple chiral structure of the $\s$-model and the coset $\Wc$-algebra of Subsubsection \ref{SubSec:W} fails to capture all of its local chiral fields (we refer to Subsection \ref{SubSec:BYB} for more details about these statements). We stress here that this property depends on the choice of maximally isotropic subalgebra $\kf \subset \df$ necessary to built the $\s$-model from the underlying $\Ec$-model. Indeed, the constructions of Subsubsections \ref{SubSec:Para} and \ref{SubSec:W} rely only on the $\Ec$-model current, independently of the choice of $\kf$. Different choices for the latter lead to various realisations of the current in terms of the $\s$-model fields: for certain specific choices, there can exist additional structures in the chiral algebras of the $\s$-model which are not directly detectable from the current formulation. It would be interesting to explore these aspects further and understand whether one can also take into account the specificity of the choice of $\kf$ in our description of conformal $\Ec$-models.

Another subtlety discussed in Subsection \ref{SubSec:BYB} is the potential existence of several target space domains in the conformal limit of the $SL(2,\R)$ Klim\v{c}\'{i}k model. The target space of this $\s$-model possesses a bounded direction which is decompactified in the conformal limit, \textit{i.e.} whose length with respect to the target space metric becomes infinite.  Schematically, depending on which region of this decompactifying direction we probe while taking the limit, one obtains different domains in the target space of the conformal model, which become separated by an infinite distance one from another. In the case of the $SL(2,\R)$ Klim\v{c}\'{i}k model, we more precisely obtain three such domains: two of them form regions in which the target space stays curved in the conformal limit while the last one is a region where the metric becomes flat. In this so-called asymptotic domain, the $\s$-model then essentially flows to a free theory. As mentioned in Subsection \ref{SubSec:BYB}, taking the naive conformal limit of the degenerate $\Ec$-model describing the Klim\v{c}\'{i}k theory yields a conformal $\Ec$-model which captures only one of the (non-asymptotic) domains. A natural question here is thus whether there exists an $\Ec$-model interpretation of the other ones: we refer to the paragraphs around equation \eqref{Eq:DomainII} for more details and a preliminary discussion of potential directions to answer this question. More generally, it would be interesting to investigate this phenomenom for a broader class of $\Ec$-models.

\subsection[Quantisation of conformal degenerate \texorpdfstring{$\Ec$}{E}-models]{Quantisation of conformal degenerate \texorpdfstring{$\bm\Ec$}{E}-models}
\label{SubSec:Quant}

The most natural perspective of the present work is the quantisation of degenerate $\Ec$-models satisfying the strong conformal condition. Indeed, the results presented here on the chiral structure of these models pave the way for their exact quantisation using standard operator techniques of two-dimensional conformal field theory. For instance, the classical $\Wc$-algebra built in Subsubsection \ref{SubSec:W} is expected to quantise in the form of a vertex operator algebra, describing the quantum local chiral fields of the theory. In many cases, for instance when the Lie algebras $\df_\pm$ and $\hf$ are semi-simple (or more generally reductive), these quantum $\Wc$-algebras have been studied in great detail in the literature (see e.g.~\cite{Goddard:1984vk,Goddard:1986ee,Bouwknegt:1992wg} for a non-exhaustive list of references). Similarly, it is natural to consider the quantisation of the non-local parafermionic fields of Subsubsection \ref{SubSec:Para}. For semi-simple and reductive Lie algebras, this has been studied in~\cite{Lepowsky:1984,Fateev:1985mm,Ninomiya:1986dp,Gepner:1987sm,Karabali:1989dk}. It would be interesting to apply these various results to the models considered in this paper as well as extend them to more exotic choices for the Lie algebras $\df_\pm$ and $\hf$, which have been less explored so far.

The construction of these algebras of quantum chiral fields is the starting point for the quantisation of the model. Further steps include for instance the study of representations and characters of these algebras, the determination of the spectrum of states/fields of the model, the computation of its correlation functions and the study of its modular invariance. We hope that many of these aspects can be investigated exploiting the Lie-theoretic formulation underlying conformal degenerate $\Ec$-models.

\subsection[Quantum corrections to conformal \texorpdfstring{$\Ec$}{E}-models geometry]{Quantum corrections to conformal \texorpdfstring{$\bm\Ec$}{E}-models geometry}

The conformal invariance of the $\Ec$-models considered in this article was only established at 1-loop, \textit{i.e.} at first order in the quantum perturbative expansion of the model. It would be natural to extend these results to higher-loops. As already established in a variety of cases, this generally requires the addition of quantum $\alpha'$-corrections to the geometry of the model, \textit{i.e.} to its metric and B-field\footnote{Note that the full treatment of the conformal invariance of these models would also require the description of diffeomorphism/dilaton terms in their RG-flow. For simplicity, we did not address this subject in the present paper. It should however be kept in mind for future investigations.}. It would be particularly interesting to understand whether such corrections can be understood in a systematic way for conformal $\Ec$-models and can be phrased in the Lie-theoretic language used above. Relevant for these questions is the work~\cite{Hassler:2020wnp}, which studies the 2-loop RG-flow of non-degenerate $\Ec$-models, generalising the 1-loop results of~\cite{Valent:2009nv,Sfetsos:2009dj,Avramis:2009xi,Sfetsos:2009vt}.

In this context, it might be useful to develop a ``target space approach'', by seeing the requirement of conformal invariance of the $\s$-models as gravity equations for their target space geometry, subject to $\alpha'$-corrections. Of particular interest here is the formulation of gravity in the language of double field theory~\cite{Siegel:1993th,Hull:2009mi,Hohm:2010jy,Hohm:2010pp}, especially on group manifolds~\cite{Blumenhagen:2014gva,Blumenhagen:2015zma,Hassler:2016srl,Hassler:2017yza}, which is closely related to $\Ec$-models.

In another direction, it also seems worthwhile to follow approaches based on the worldsheet theory rather than that of the target space. For instance, in the work~\cite{Dijkgraaf:1991ba}, it was argued that the all-loop quantum corrections to the euclidean black hole $\s$-model can be inferred using conformal field theory techniques and in particular the determination and the geometric interpretation of the vertex operators of the underlying $SL(2,\R)/U(1)$ gauged WZW model. A similar approach was applied to general $G/H$ gauged WZW models in~\cite{Bars:1992sr}.\footnote{See also~\cite{Tseytlin:1992ri} for another derivation of these results, based on the quantum effective action of the worldsheet theory.} It is quite remarkable that such methods, which essentially start with the (quotiented) current algebras underlying the theory, allows one to extract exact results on the quantum geometry of the target space. Since the gauged WZW theories belong to the class of conformal $\Ec$-models studied in this paper (see~\cite{Klimcik:1999ax} and Subsections \ref{SubSec:gWZW} and \ref{SubSec:Su2u1}) and in light of the algebraic nature of the works~\cite{Dijkgraaf:1991ba,Bars:1992sr}, it is very appealing to explore similar approaches for more general conformal $\Ec$-models.

\subsection[Integrable \texorpdfstring{$\s$}{sigma}-models, affine Gaudin models and their quantisation]{Integrable \texorpdfstring{$\bm\s$}{sigma}-models, affine Gaudin models and their quantisation}
\label{SubSec:Int}

One of the main long-term motivation for the present paper, which we develop in this subsection, is the study of integrable $\s$-models and their quantisation. These models are characterised by the existence, at least at the classical level, of an infinite number of conserved Poisson-commuting charges. The latter are often built from a Lax connection, formed by a pair of matrices $\Lc_t(z\,;t,x)$ and $\Lc_x(z\,;t,x)$ valued in a complex Lie algebra $\g^\C$, built from the fields of the model and depending meromorphically on an auxiliary complex variable $z$, called the spectral parameter. The main property of this Lax connection is that the equations of motion of the theory are equivalent to its flatness $[\p_t + \Lc_t(z), \p_x + \Lc_x(z) ] = 0$, ensuring the construction of an infinite number of (non-local) conserved charges from the monodromy of $\Lc_x(z)$. Moreover, if $\Lc_x(z)$ satisfies a certain Poisson algebra called the Maillet bracket~\cite{Maillet:1985fn,Maillet:1985ek}, these charges are pairwise Poisson-commuting, thus proving the integrability of the theory.

In the last decades, it was shown that many examples of integrable $\s$-models can be recast as $\Ec$-models~\cite{Sfetsos:2015nya,Klimcik:2015gba,Klimcik:2016rov,Klimcik:2017ken,Hoare:2017ukq,Severa:2017kcs,Demulder:2017zhz,Klimcik:2019kkf,Demulder:2019bha,Demulder:2019vvh,Klimcik:2020fhs,Demulder:2020dlo,Hoare:2020mpv,Lacroix:2020flf,Klimcik:2021bjy,Klimcik:2021bqm,Shah:2021xbf,Hoare:2022vnw,Liniado:2023uoo}. In particular, a systematic approach to the question of the integrability of non-degenerate $\Ec$-models was initiated in~\cite{Severa:2017kcs}. The main proposal of this reference is to consider Lax matrices $\Lc_x(z)$ built from the $\Ec$-model current as $\Lc_x(z) = p_z(\Jc)$, where $p_z : \df \to \g^\C$ is a linear operator depending meromorphically on $z$. The time-evolution of $\Jc$, and thus of $\Lc_x(z)$, is dictated by the choice of Hamiltonian of the theory, characterised by the operator $\Ec:\df\to\df$. The work~\cite{Severa:2017kcs} derived a purely algebraic compatibility condition on $\Ec$ and $p_z$ which ensures that this time-evolution takes the form of a flatness equation, thus guaranteeing the existence of an infinite number of conserved charges. An extension of this result for degenerate $\Ec$-models as well as a further algebraic condition on $p_z$ ensuring the existence of a Maillet bracket for $\Lc_x(z)$ were recently established in~\cite{Klimcik:2021bjy,Klimcik:2021bqm}.

To summarise, given a degenerate $\Ec$-model with double algebra $(\df,\psd)$, current $\Jc\in\df$ and degenerate $\Ec$-operator $\Eh : \df \to \df$, \cite{Severa:2017kcs,Klimcik:2021bjy,Klimcik:2021bqm} provide simple conditions on a linear operator $p_z : \df \to \g^{\mathbb{C}}$ which ensure that the model is integrable with Lax matrix $\Lc_x(z) = p_z(\Jc)$. A natural question at this point is whether there exist a systematic way of constructing $(\df,\psd,\Eh,p_z)$ such that these conditions are automatically satisfied. The recent results of~\cite{Lacroix:2020flf,Liniado:2023uoo} indicate that such a construction is provided by the formalisms of 4d Chern-Simons theory~\cite{Costello:2019tri} and affine Gaudin models~\cite{Levin:2001nm,Feigin:2007mr,Vicedo:2017cge}, which are known to be deeply connected one to another~\cite{Vicedo:2019dej,Levin:2022dnq}. Here, we will use the language of affine Gaudin models (AGM), which is more directly related to the one used in this paper. The main defining data of an AGM is the choice of a real simple Lie algebra $\g$ and a rational function $\vp(z)$ of the spectral parameter, called the twist function. To this data is associated a double algebra $(\df,\psd)$, naturally built from the poles of $\vp(z)\dd z$ in $\CP$. For instance, if $\vp(z)\dd z$ has only $N$ simple real poles, then $\df=\g^{\times N}$ and the invariant non-degenerate bilinear form $\psd$ is defined on each copy of $\g$ as a multiple of the Killing form, with the proportionality factor being the residue of $\vp(z)\dd z$ at the corresponding pole. If we promote one of these simple poles to a singularity of order $m\in \Z_{\geq 1}$, then we replace the corresponding copy of $\g$ in $\df$ by the so-called Takiff algebra $\mathsf{T}^m\g = \g \otimes \R[\epsilon] / \epsilon^m \R[\epsilon]$. Finally, for a pair of complex conjugate poles, one considers the corresponding complexified Takiff algebra $\mathsf{T}^m\g^\C$.

The main result of the AGM formalism can be seen as the construction of a degenerate $\Ec$-model with this double algebra $(\df,\psd)$ which is automatically integrable. This model is described in terms of a $\df$-valued current $\Jc$ and a Hamiltonian density quadratic in $\Jc$, with coefficients explicitly built as rational functions of the poles and zeroes of the twist function $\vp(z)\dd z$~\cite{Vicedo:2017cge,Delduc:2019bcl,Arutyunov:2020sdo}. This specific choice ensures the integrability of the theory: more precisely, the AGM construction also naturally provides an explicit linear operator $p_z : \df \to \g^\C$ such that $\Lc_x(z) = p_z(\Jc)$ satisfies a flatness equation and a Maillet bracket.\footnote{These flatness equation and Maillet bracket are built-in in the AGM construction and are thus not checked using the general integrability conditions for $\Ec$-models derived in~\cite{Severa:2017kcs,Klimcik:2021bjy,Klimcik:2021bqm}. However, we expect that the operators $p_z$ obtained from AGMs automatically satisfy these conditions. This has been partially checked in~\cite{Lacroix:2020flf}.} As explained above, this ensures the existence of an infinite number of non-local Poisson-commuting conserved charges. In addition, the AGM formalism also guarantees the existence of an infinite family of local higher-spin conserved charges~\cite{Lacroix:2017isl}, which Poisson-commute pairwise and with the non-local ones, forming another manifestation of the integrability of this theory. \\

One of the main open challenge in the study of integrable $\s$-models is the question of their ``first-principle'' quantum integrability. By that, we mean essentially the following two steps: (i) define the non-commutative algebra of observables of the quantum theory and (ii) construct an infinite family of observables in this algebra which are pairwise commuting (and which reduce to the Poisson-commuting charges of the classical model in the classical limit). For a generic classically integrable $\s$-model, both of these steps are in general quite challenging questions. A potential simplification of the step (i), inspired by the works~\cite{Zamolodchikov:1989hfa,Sasaki:1987mm,Bazhanov:1994ft,Bazhanov:1996dr,Bazhanov:1998dq} and which makes the link with the present paper, is to consider the conformal limit of the theory. In this limit, the integrable structure is expected to decouple according to the chiral decomposition of the conformal model: at the classical level, we more precisely expect half of the Poisson-commuting charges to be built from left-moving fields and the other half from right-moving ones. As sketched in Subsection \ref{SubSec:Quant}, the operator formalism of two-dimensional CFTs offers a rigorous and powerful handle on the quantisation of these chiral fields, for instance by constructing the  quantum algebra of local chiral observables of the theory in the form of a vertex operator algebra (as well as, in some cases, non-local chiral algebras such as parafermionic ones). In this context, we can now pass to the step (ii) in the investigation of ``first-principle'' quantum integrability, namely the construction of an infinite family of commuting operators in these chiral algebras. This is a hard but well-defined mathematical question.

The study of integrable structures in CFTs was initiated in~\cite{Zamolodchikov:1989hfa,Sasaki:1987mm,Bazhanov:1994ft,Bazhanov:1996dr,Bazhanov:1998dq} for the case of the quantum KdV theory, where the underlying vertex operator algebra is the Virasoro one. It has since been extended to various other theories and $\Wc$-algebras. In the case of integrable $\s$-models, first progresses on the explicit construction of quantum integrable structures in their conformal limit were obtained in various cases in~\cite{Bazhanov:2013cua,Litvinov:2016mgi,Bazhanov:2017nzh,Litvinov:2018bou,Bazhanov:2018xzh,Lukyanov:2019asr,Litvinov:2019rlv,Alfimov:2020jpy,Kotousov:2022azm}.\footnote{The study of these integrable structures is often done using the notion of asymptotic domain. As explained in Subsection \ref{SubSec:Open}, this is a region of the target space which becomes flat in the conformal limit and where the $\s$-model thus flows to a free theory. This allows the construction of a free-field realisation of the $\Wc$-algebra and an algebraic characterisation of the local commuting operators using the so-called Wakimoto/bosonic screening charges~\cite{Semikhatov:2001zz,Bazhanov:2013cua,Litvinov:2016mgi,Bazhanov:2017nzh,Litvinov:2018bou,Bazhanov:2018xzh,Lukyanov:2019asr,Litvinov:2019rlv,Alfimov:2020jpy,Kotousov:2022azm}, which schematically encode the perturbation from the free theory in the asymptotic domain to an interacting $\s$-model. The perspectives discussed in Subsection \ref{SubSec:Open} might thus also play a role in the study of integrable structures. Let us finally note that the $\Wc$-algebras and commuting local charges that arise in these conformal $\s$-models can often be studied from the alternative points of view of Toda-like dual theories (see for instance~\cite{Fateev:1996ea,Bazhanov:2013cua,Fateev:2018yos,Fateev:2019xuq} and references therein) and integrable models of boundary interactions~\cite{Lukyanov:2003nj,Lukyanov:2012wq}.} The first important step in these works is the identification of the quantum chiral algebras underlying these conformal limits. One of the motivation for the present paper was to initiate a systematic analysis of this question, based on the formalism of $\Ec$-models. Indeed, as explained above, a very large class of classical integrable $\s$-models can be interpreted as affine Gaudin models and thus as $\Ec$-models: based on the results of this paper and their potential extensions, we expect that the chiral sectors of their conformal limits will be reinterpreted as so-called ``chiral AGMs'', built from chiral current algebras and their quotients.\footnote{As explained above, the currents appearing in AGMs are generally based on Takiff Lie algebras, which do not belong to the class of semi-simple Lie algebras. In view of the general perspectives discussed here, we note that the quantisation of such currents was studied in~\cite{Babichenko:2012uq,Quella:2020uhk}.} Such a construction would then provide an appropriate and rigorous starting point for a systematic investigation of the quantum integrability of these CFTs. Indeed, although a complete answer to this question is still open at the moment, first progresses concerning the construction and the diagonalisation of commuting charges in quantised chiral AGMs were obtained in~\cite{Feigin:2007mr,Frenkel:2016gxg,Lacroix:2018fhf,Lacroix:2018itd,Gaiotto:2020fdr,Gaiotto:2020dhf,Kotousov:2021vih,Wu:2021jir,Franzini:2022duf}. The application of these ideas to a specific integrable $\s$-model (the Klim\v{c}\'{i}k model~\cite{Klimcik:2008eq,Klimcik:2014bta}) was the subject of the recent work~\cite{Kotousov:2022azm}: the present paper can thus be seen as a first step towards a more systematic application of this quantisation programme to a larger class of theories.\\

In addition to integrability, let us note that the formalism of $\Ec$-models has also been quite fruitful for the study of Poisson-Lie symmetries and dualities at the classical level~\cite{Klimcik:1995ux,Klimcik:1995dy,Klimcik:1996nq,Klimcik:1996np}. Similarly to the discussion above for integrability, we hope that the chiral operators approach advocated in this paper can help to shed some light on the quantum nature of these properties, which have been so far less explored. For conciseness, we will not discuss these perspectives further in this conclusion.

\section*{Acknowledgements} 

The author would like to thank F. Delduc, S. Driezen, M. Gaberdiel, F. Hassler, O. Hul\'{i}k, C. Klim\v{c}\'{i}k, M. Magro, D. Thompson and B. Vicedo for useful and interesting discussions as well as F. Delduc for valuable comments on the draft. This work is supported by Dr. Max R\"ossler, the Walter Haefner Foundation and the ETH Z\"urich Foundation.

\appendix

\section{Dirac bracket of the parafermionic fields}
\label{App:Dirac}

\paragraph{Setup.} In this appendix, we explain the main steps in the computation of the Dirac bracket of the fields $\Psi^\cL(x)$ and $\Psi^\cR(x)$. Let us briefly recall the situation. We start with the $\df$-valued current $\Jc(x)$, which we decompose as $\Jc = \Jc_{\hf}+\Jc'+\Jc_+-\Jc_-$, along the direct sum $\df=\hf\oplus\hf'\oplus V_+ \oplus V_-$. In particular, the component $\Jc'\in\hf'$ is identified as the generator of the $H$-gauge symmetry, so that the corresponding first-class constraint is $\Jc' \approx 0$. We are interested in the Poisson brackets of the fields $\Psi^\cL$ and $\Psi^\cR$ built from $\Jc_\pm$ as in equation \eqref{Eq:Psi}. As observed in the main text, this construction is such that $\Psi^\cL$ and $\Psi^\cR$ coincide with the fields $\Jc_+$ and $\Jc_-$ in the gauge where the current $\Jc_\hf$ is set to zero. We will denote equalities valid in this gauge using the symbol `$\equiv$', in a similar way that we used the weak symbol `$\approx$' for equalities that hold under the first-class constraints. By construction, in this gauge, we then have
\begin{equation}
\Jc_{\hf} \equiv 0\,, \qquad \Jc' \equiv 0\,, \qquad \Jc_+ \equiv \Psi^\cL \qquad \text{ and } \qquad \Jc_- \equiv \Psi^\cR\,.
\end{equation}
The second equality above simply encodes the initial $\dim\hf$ constraints associated with the gauge symmetry, while the first one encodes the additional $\dim\hf$ gauge-fixing conditions that we impose to reach the specific gauge we are interested in. In this language, our goal can now be rephrased as determining the Poisson brackets of the components $\Jc_+$ and $\Jc_-$ in this gauge.

\paragraph{Second-class constraints.} It will be useful to slightly rephrase the above discussion. Recall that under the strong conformal condition discussed in Subsection \ref{SubSec:Conf}, the subspace $D\hf = \hf \oplus \hf'$ is a subalgebra of $\df$. We denote by $\Jc_{D\hf}=\Jc_{\hf}+\Jc'$ the component of the current $\Jc$ along this subalgebra. By construction, the initial first-class constraints combined with the gauge-fixing conditions imposed above amount to
\begin{equation}
\Jc_{D\hf} \equiv 0\,.
\end{equation}
This can be seen as a system of $\dim D\hf = 2\dim\hf$ constraints, which crucially satisfies the so-called second-class property, as we will now explain.

The restricted bilinear form $\psd|_{D\hf}$ is invariant and non-degenerate on the subalgebra $D\hf$: the corresponding split quadratic Casimir $\Cdh$ was defined in equation \eqref{Eq:CasDeco}. The $D\hf$-valued field $\Jc_{D\hf}$ then satisfies the Poisson bracket
\begin{equation}\label{Eq:PbJDh}
\bigl\lbrace \Jc_{D\hf}(x)\ti{1}, \Jc_{D\hf}(y)\ti{2} \bigr\rbrace  = \bigl[ \Cdh, \Jc_{D\hf}(x)\ti{1} \bigr] \,\delta(x-y) - \Cdh\,\p_x\delta(x-y) \equiv  - \Cdh\,\p_x\delta(x-y)\,.
\end{equation}
In particular, we find that the Poisson bracket of the $2\dim\hf$ constraints contained in $\Jc_{D\hf}\equiv 0$ does not vanish: in fact, it even defines an invertible kernel $- \Cdh\,\p_x\delta(x-y)$. This is the aforementioned second-class property. In contrast, recall that the constraints contained in $\Jc' \approx 0$ and generating the gauge symmetry were first-class, meaning that they had vanishing Poisson brackets. The fact that the system of constraints becomes second-class when adding $\Jc_{\hf}\equiv 0$ is what characterises this equality as an appropriate choice of gauge-fixing condition: more explicitly, the invertibility of the above Poisson brackets translates at the infinitesimal level the property that the condition $\Jc_{\hf}\equiv 0$ is always attainable by a gauge transformation and furthermore that it completely fixes the symmetry.

\paragraph{The Dirac bracket.} Following Dirac's theory of constrained Hamiltonian systems~\cite{dirac1964lectures,Henneaux:1992ig}, the Poisson bracket of observables in the gauge $\Jc_{D\hf}\equiv 0$ can be computed using the so-called Dirac bracket. The latter is built from the initial current bracket $\lbrace\cdot,\cdot\rbrace$ by the following construction:
\begin{equation}\label{Eq:Dirac0}
\lbrace \Oc, \Oc' \rbrace^* \equiv \lbrace \Oc,\Oc' \rbrace + \int_0^{2\pi} \int_0^{2\pi} \Bigl\langle \bigl\lbrace \Jc_{D\hf}(x)\ti{1}, \Oc\bigr\rbrace\, \bigl\lbrace \Jc_{D\hf}(y)\ti{2}, \Oc'\bigr\rbrace, \, \mathsf{K}(x,y)\ti{12}\, \Bigr\rangle\ti{12}\;\dd x\, \dd y\,.
\end{equation}
Here, $\Oc$ and $\Oc'$ are any two observables of the model, $\lbrace\cdot,\cdot\rbrace^*$ denotes the Dirac bracket and $\mathsf{K}(x,y)\ti{12}$ is a $(D\hf\otimes D\hf)$-valued distribution, depending on the space coordinates $x,y\in[0,2\pi]$. This distribution is built as the ``inverse'' of the kernel defined by the Poisson bracket $\bigl\lbrace \Jc_{D\hf}(x)\ti{1}, \Jc_{D\hf}(y)\ti{2} \bigr\rbrace$ of the constraints (this is where the second-class property of these constraints plays a crucial role). This construction ensures that the Dirac bracket is compatible with $\Jc_{D\hf} \equiv 0$, \textit{i.e.} that
\begin{equation}\label{Eq:DiracConst}
\lbrace \Oc, \Jc_{D\hf} \rbrace^* \equiv 0\,
\end{equation}
for any observable $\Oc$. It is a standard result~\cite{dirac1964lectures,Henneaux:1992ig} that the Dirac bracket \eqref{Eq:Dirac0} is skew-symmetric and satisfies the Leibniz and Jacobi identities, making it a well-defined Poisson bracket. Moreover, the above property \eqref{Eq:DiracConst} ensures that this is the appropriate Poisson structure induced on observables in the gauge where we imposed $\Jc_{D\hf} \equiv 0$, as wanted.

This property \eqref{Eq:DiracConst} is in fact enough to concretely determine the inverse kernel $\mathsf{K}(x,y)\ti{12}$. Indeed, using equation \eqref{Eq:PbJDh}, we get
\begin{align*}
\lbrace \Oc, \Jc_{D\hf}(z) \rbrace^* &\equiv \lbrace \Oc, \Jc_{D\hf}(z) \rbrace + \int_0^{2\pi} \int_0^{2\pi} \Bigl\langle \bigl\lbrace \Jc_{D\hf}(x)\ti{1}, \Oc\bigr\rbrace\, \bigl\lbrace \Jc_{D\hf}(y)\ti{2}, \Jc_{D\hf}(z)\ti{3} \bigr\rbrace,\, \mathsf{K}(x,y)\ti{12}\, \Bigr\rangle\ti{12}\;\dd x\, \dd y \\
&\equiv \lbrace \Oc, \Jc_{D\hf}(z) \rbrace - \int_0^{2\pi} \int_0^{2\pi} \Bigl\langle \bigl\lbrace \Jc_{D\hf}(x)\ti{1}, \Oc\bigr\rbrace\, \mathsf{C}^{D\hf}\ti{23},\, \mathsf{K}(x,y)\ti{12}\, \Bigr\rangle\ti{12} \p_y\delta(y-z)\;\dd x\, \dd y\,\\
&\equiv \lbrace \Oc, \Jc_{D\hf}(z) \rbrace - \int_0^{2\pi} \Bigl\langle  \bigl\lbrace \Oc,\Jc_{D\hf}(x)\ti{1}\bigr\rbrace\,,\, \p_z \mathsf{K}(x,z)\ti{13}\, \Bigr\rangle\ti{1}\; \dd x\,,
\end{align*}
where to obtain the last equality we performed the integral over $y$ using the derived Dirac distribution, applied the identity $\langle \mathsf{C}^{D\hf}\ti{23}, X\ti{2} \rangle\ti{2} = X\ti{3}$ for  $X\in D\hf$ and used the skew-symmetry of the Poisson bracket $\lbrace \cdot,\cdot\rbrace$. For the above expression to vanish for all $\Oc$, as required by the construction of the Dirac bracket, we thus need the $(D\hf\otimes D\hf)$-valued kernel $\mathsf{K}(x,y)$ to satisfy
\begin{equation}
\int_0^{2\pi} \bigl\langle  F(x)\ti{1}\,,\, \p_y \mathsf{K}(x,y)\ti{12}\, \bigr\rangle\ti{1} \dd x = F(y)
\end{equation}
for all $D\hf$-valued function $F$. The solution to this condition is given by
\begin{equation}
\mathsf{K}(x,y)\ti{12} = -\frac{1}{2}\, \Cdh\,\epsilon(x-y)\,,
\end{equation}
where $\epsilon(x-y)$ is the sign function, defined by $\epsilon(z)=-1$ if $-2\pi < z < 0$ and $\epsilon(z)=+1$ if $0 < z < 2\pi$, such that $\p_z \epsilon(z) = 2\delta(z)$. Given this expression for $\mathsf{K}$, and using the identity $\langle X\ti{1}Y\ti{2}, \mathsf{C}^{D\hf}\ti{12} \rangle\ti{12} = \ps{X}{Y}$ for $X,Y\in D\hf$, we thus find that the Dirac bracket \eqref{Eq:Dirac0} explicitly reads
\begin{equation}\label{Eq:Dirac}
\lbrace \Oc, \Oc' \rbrace^* \equiv \lbrace \Oc,\Oc' \rbrace - \frac{1}{2} \int_0^{2\pi} \int_0^{2\pi} \Bigl\langle \bigl\lbrace \Jc_{D\hf}(x), \Oc\bigr\rbrace\,, \bigl\lbrace \Jc_{D\hf}(y), \Oc'\bigr\rbrace \Bigr\rangle\,\epsilon(x-y)\,\dd x\, \dd y\,.
\end{equation}
This is the expression that we will use in the rest of this appendix.

\paragraph{Dirac bracket of $\bm{\Jc_+}$ with $\bm{\Jc_-}$.} Let us now compute the Dirac bracket of $\Jc_+$ with $\Jc_-$. Applying the formula \eqref{Eq:Dirac}, we get
\begin{align}
\bigl\lbrace \Jc_+(x)\ti{1} , \Jc_-(y)\ti{2} \bigr\rbrace^* &\equiv \bigl\lbrace \Jc_+(x)\ti{1} , \Jc_-(y)\ti{2} \bigr\rbrace \label{Eq:DiracPM0} \\ 
& \hspace{18pt}- \frac{1}{2} \int_0^{2\pi} \int_0^{2\pi} \Bigl\langle \bigl\lbrace \Jc_{D\hf}(x')\ti{3}, \Jc_+(x)\ti{1} \bigr\rbrace\,, \bigl\lbrace \Jc_{D\hf}(y')\ti{3}, \Jc_-(y)\ti{2} \bigr\rbrace \Bigr\rangle\ti{3}\,\epsilon(x'-y')\,\dd x'\, \dd y'\,. \notag
\end{align}
To compute this, we will need two main ingredients. The first one is the bracket of $\Jc_+$ with $\Jc_-$ with respect to the initial current Poisson structure $\lbrace\cdot,\cdot\rbrace$. This is obtained by projecting the current Poisson algebra \eqref{Eq:PbJ} of $\Jc$ onto $V_+ \otimes V_-$. Using the fact that $V_+$ and $V_-$ are orthogonal and pair non-degenerately with themselves with respect to the bilinear form $\psd$, combined with $[V_+,V_-]=\lbrace 0 \rbrace$ by the strong conformal condition, we find that this projection in fact vanishes. Thus, we have
\begin{equation}\label{Eq:PbJpJm}
\bigl\lbrace \Jc_+(x)\ti{1} , \Jc_-(y)\ti{2} \bigr\rbrace = 0\,.
\end{equation}

The second ingredient that we need is the bracket of $\Jc_\pm$ with the second-class constraint $\Jc_{D\hf}$. This is obtained from the current algebra \eqref{Eq:PbJ} by projecting onto $D\hf \otimes V_\pm$. Using in particular that $\langle D\hf, V_\pm \rangle = 0$, $[ D\hf, D\hf ] \subset D\hf$ and $[ D\hf, V_\pm ] \subset V_\pm$ -- see equation \eqref{Eq:ComHVpm}, we simply find
\begin{equation}\label{Eq:PbJDhPm}
\bigl\lbrace \Jc_{D\hf}(x)\ti{1} , \Jc_\pm(y)\ti{2} \bigr\rbrace = -\bigl[ \Cdh, \Jc_\pm(x)\ti{2} \bigr]\, \delta(x-y)\,.
\end{equation}

Reinserting the equations \eqref{Eq:PbJpJm} and \eqref{Eq:PbJDhPm} in the formula \eqref{Eq:DiracPM0} and performing the integrals over $x'$ and $y'$ using the Dirac distributions, we obtain
\begin{align}
\bigl\lbrace \Jc_+(x)\ti{1} , \Jc_-(y)\ti{2} \bigr\rbrace^* &\equiv - \frac{1}{2} \,\Bigl\langle \bigl[ \mathsf{C}^{D\hf}\ti{31}, \Jc_+(x)\ti{1} \bigr], \bigl[ \mathsf{C}^{D\hf}\ti{32}, \Jc_-(y)\ti{2} \bigr] \Bigr\rangle\ti{3} \,\epsilon(x-y) \notag \\
&\equiv - \frac{1}{2} \,\Bigl[ \bigl[ \mathsf{C}^{D\hf}\ti{12}, \Jc_+(x)\ti{1} \bigr], \Jc_-(y)\ti{2} \bigr] \Bigr] \,\epsilon(x-y)\,. \label{Eq:DirPmAux}
\end{align}
This expression is valued in $[ D\hf, V_+] \otimes [ D\hf, V_-] \subset V_+ \otimes V_-$, as expected. In fact, it turns out to identically vanish. To prove it, let us consider two arbitrary elements $Z_+ \in V_+$ and $Z_- \in V_-$. Using the ad-invariance of $\psd$ and the property $\langle \mathsf{C}^{D\hf}\ti{12}, X\ti{1}Y\ti{2} \rangle\ti{12} = \ps{\pi_{D\hf}X}{Y}$ for $X,Y\in \df$, we find
\begin{equation}
\bigl\langle \bigl\lbrace \Jc_+(x)\ti{1} , \Jc_-(y)\ti{2} \bigr\rbrace^*, Z_+ \otimes Z_- \bigr\rangle\ti{12} \equiv -\frac{1}{2}\, \bigl\langle \pi_{D\hf}\, \bigl[ Z_+, \Jc_+(x) \bigr] , \bigl[ Z_-, \Jc_-(y) \bigr] \bigr\rangle\, \epsilon(x-y)\,.
\end{equation}
Recall that under the strong conformal condition, $[V_\pm,V_\pm] \subset D\hf \oplus V_\pm$. Since $\langle V_+, D\hf\oplus V_-\rangle = 0$, we see that the projector $\pi_{D\hf}$ can be removed in the above equation as the component of $[ Z_+, \Jc_+(x) ]$ which is not in $D\hf$ is valued in $V_+$ and is thus orthogonal to $[ Z_-, \Jc_-(x) ]$. We then get
\begin{align*}
\bigl\langle \bigl\lbrace \Jc_+(x)\ti{1} , \Jc_-(y)\ti{2} \bigr\rbrace^*, Z_+ \otimes Z_- \bigr\rangle\ti{12} & \equiv - \frac{1}{2}\, \bigl\langle  \bigl[ Z_+, \Jc_+(x) \bigr] , \bigl[ Z_-, \Jc_-(y) \bigr] \bigr\rangle\, \epsilon(x-y)\\
&\equiv -\frac{1}{2}\, \bigl\langle  \bigl[ \bigl[ Z_+, \Jc_+(x) \bigr] ,  Z_- \bigr], \Jc_-(y)  \bigr\rangle\, \epsilon(x-y)\,,
\end{align*}
where we used the ad-invariance of $\psd$ in the last equality. Using the Jacobi identity and the fact that $[V_+,V_-]=\lbrace 0 \rbrace$, we then find that the above quantity vanishes. Since this is true for all $Z_+\in V_+$ and $Z_-\in V_-$ and $\psd$ is non-degenerate on $V_+$ and $V_-$, this simply means that
\begin{equation}
\bigl\lbrace \Jc_+(x)\ti{1} , \Jc_-(y)\ti{2} \bigr\rbrace^* \equiv 0\,.
\end{equation}
Since the gauge-fixing of $\Jc_+$ and $\Jc_-$ coincides with the fields $\Psi^\cL$ and $\Psi^\cR$, this can equivalently be written
\begin{equation}
\bigl\lbrace \Psi^\cL(x)\ti{1} , \Psi^\cR(y)\ti{2} \bigr\rbrace = 0\,.
\end{equation}
Here, we have used the standard bracket and a strong equality, since the fields $\Psi^\cL$ and $\Psi^\cR$ are gauge-invariant (see the discussion below equation \eqref{Eq:Psi}): indeed, the Dirac bracket coincides with the initial one for gauge-invariant observables.

\paragraph{Dirac bracket of $\bm{\Jc_\pm}$ with itself.} We now turn to the computation of the Dirac bracket of $\Jc_\pm$ with itself. Using the equation \eqref{Eq:PbJDhPm} and following a computation similar to the derivation of equation \eqref{Eq:DirPmAux} in the previous paragraph, we get
\begin{equation}\label{Eq:DirPpMm}
\bigl\lbrace \Jc_\pm(x)\ti{1} , \Jc_\pm(y)\ti{2} \bigr\rbrace^*\equiv \bigl\lbrace \Jc_\pm(x)\ti{1} , \Jc_\pm(y)\ti{2} \bigr\rbrace - \frac{1}{2} \,\Bigl[ \bigl[ \mathsf{C}^{D\hf}\ti{12}, \Jc_\pm(x)\ti{1} \bigr], \Jc_\pm(y)\ti{2} \bigr] \Bigr] \,\epsilon(x-y)\,.
\end{equation}
We now need to compute the first term in the right-hand side of this equation. This is done by extracting the $V_\pm \otimes V_\pm$ component of the current Poisson algebra \eqref{Eq:PbJ}, yielding\footnote{An efficient way to do this computation is as follows. First recall that $\df_\pm = D\hf \oplus V_\pm$ is a subalgebra of $\df$ and that $\pm\psd|_{\df_\pm}$ is a non-degenerate invariant bilinear form on $\df_\pm$. The component $\Jc_\pm \pm \Jc_{D\hf}$ of $\Jc$ along $\df_\pm$ thus satisfies the closed current Poisson algebra of $\df_\pm$. From there, it is easy to extract the bracket of $\Jc_\pm$ with itself since we already know all the brackets involving $\Jc_{D\hf}$.}
\begin{equation}
\bigl\lbrace \Jc_\pm(x)\ti{1} , \Jc_\pm(y)\ti{2} \bigr\rbrace \equiv \left( \bigl[ \Cpm, \Jc_\pm(x)\ti{1} \bigr] \pm \bigl[ \Cdh, \Jc_\pm(x)\ti{2} \bigr] \right) \delta(x-y) \mp \Cpm \, \p_x\delta(x-y)\,.
\end{equation}
Reinserting this expression in the Dirac bracket \eqref{Eq:DirPpMm} and translating the result in terms of the gauge-invariant field $\Psi^\cLR$ coinciding with $\Jc_\pm$ in the considered gauge, we finally obtain
\begin{align}
\bigl\lbrace \Psi^\cL(x)\ti{1}, \Psi^\cL(y)\ti{1} \bigr\rbrace &= \left( \bigl[ \Cp , \Psi^\cL(x)\ti{1}  \bigr] + \bigl[ \Cdh, \Psi^\cL(x)\ti{2}  \bigr] \right)\,\delta(x-y) \\
& \hspace{30pt} - \Cp \,\p_x\delta(x-y) - \frac{1}{2} \Bigl[ \bigl[ \Cdh , \Psi^\cL(x)\ti{1}  \bigr]  , \Psi^\cL(y)\ti{2}  \Bigr] \, \epsilon(x-y)\,,  \notag \\[5pt]
\bigl\lbrace \Psi^\cR(x)\ti{1}, \Psi^\cR(y)\ti{1} \bigr\rbrace &= \left( \bigl[ \Cm, \Psi^\cR(x)\ti{1}  \bigr] - \bigl[ \Cdh, \Psi^\cR(x)\ti{2}  \bigr] \right)\,\delta(x-y) \\
& \hspace{30pt} + \Cm \,\p_x\delta(x-y) - \frac{1}{2} \Bigl[ \bigl[ \Cdh , \Psi^\cR(x)\ti{1}  \bigr]  , \Psi^\cR(y)\ti{2}  \Bigr] \, \epsilon(x-y)\,.  \notag
\end{align}

\providecommand{\href}[2]{#2}\begingroup\raggedright

\end{document}